\documentclass[sigconf, nonacm]{acmart}

\usepackage{graphicx}
\usepackage{subcaption}
\usepackage{enumitem}
\usepackage{algorithm}
\usepackage{algpseudocode}
\usepackage{multirow}
\usepackage{amsmath}

\AtBeginDocument{%
  }

\begin{document}
\title{Bitcoin under Volatile Block Rewards: How Mempool Statistics Can Influence Bitcoin Mining}


\author{Roozbeh Sarenche}
\affiliation{%
             \institution{COSIC, KU Leuven}
             \country{Belgium}}
\email{roozbeh.sarenche@esat.kuleuven.be}

\author{Alireza Aghabagherloo}
\affiliation{%
             \institution{COSIC, KU Leuven}
             \country{Belgium}}
\email{alireza.aghabagherloo@esat.kuleuven.be}

\author{Svetla Nikova}
\affiliation{%
             \institution{COSIC, KU Leuven}
             \country{Belgium}}
\email{svetla.nikova@esat.kuleuven.be}

\author{Bart Preneel}
\affiliation{%
             \institution{COSIC, KU Leuven}
             \country{Belgium}}
\email{bart.preneel@esat.kuleuven.be}

\begin{abstract}

The security of Bitcoin protocols is deeply dependent on the incentives provided to miners, which come from a combination of block rewards and transaction fees. As Bitcoin experiences more halving events, the protocol reward converges to zero, making transaction fees the primary source of miner rewards. This shift in Bitcoin’s incentivization mechanism, which introduces volatility into block rewards, leads to the emergence of new security threats or intensifies existing ones. Previous security analyses of Bitcoin have either considered a fixed block reward model or a highly simplified volatile model, overlooking the complexities of Bitcoin's mempool behavior.

In this paper, we present a reinforcement learning-based tool for analyzing mining strategies under a more realistic volatile reward model. The tool leverages the Asynchronous Advantage Actor-Critic (A3C) algorithm to derive near-optimal strategies while interacting with an environment that simulates the behavior of the Bitcoin mempool during any specified period, enabling analysis based on actual historical patterns. It supports the evaluation of adversarial mining strategies, such as selfish mining and undercutting, both before and after the difficulty adjustment, offering insights into the effects of mining attacks in both the short and long term.

We revisit the Bitcoin security threshold presented in the WeRLman paper~\cite{bar2022werlman} and demonstrate that the implicit predictability of valuable transaction arrivals in this model leads to an underestimation of the reported threshold. Additionally, we show that, while adversarial strategies like selfish mining under the fixed reward model incur an initial loss period of at least two weeks, the transition toward a transaction-fee era incentivizes mining pools to abandon honest mining for \emph{immediate} profits. This incentive is expected to become more significant as the protocol reward approaches zero in the future.


\end{abstract}





\maketitle

\section{Introduction}
\label{sec:introduction}

As of December 2024, Bitcoin~\cite{nakamoto2008bitcoin} holds a market capitalization of nearly 1.8 trillion dollars~\cite{market_cap}, surpassing the annual Gross Domestic Product (GDP) of over 94\% of countries~\cite{gdp}. While Bitcoin has progressed without major security attacks thus far, its steady growth over the past decade and a half does not guarantee future security, highlighting the need to address potential vulnerabilities.
Protecting Bitcoin’s economy requires ensuring the resilience of its incentivization mechanism. If consensus nodes find their rewards insufficient to cover costs, they may abandon the chain, risking a halt in blockchain operations. Bitcoin's incentivization comes from two sources: the \emph{protocol reward} for mining a block and the \emph{transaction fee} paid by users. In Bitcoin’s early years, the protocol reward significantly outweighed transaction fees. However, due to the halving mechanism, which reduces the protocol reward every four years, transaction fees are expected to match and eventually replace the protocol reward as the sole source of incentivization. In fact, there have already been periods, such as in December 2023 and April 20, 2024, when transaction fees equaled or even exceeded the protocol reward~\cite{explorer}.

As the balance between transaction fees and the protocol reward changes over time, the incentivization dynamics of Bitcoin change as well, potentially leading to the emergence of new security threats or intensifying existing ones. One consequence of transitioning to a transaction-fee-driven era is that the total reward per block can no longer be considered \emph{fixed}. Block rewards become \emph{volatile}, as the inclusion of transactions with varying fee levels causes fluctuations in the total reward per block. 
This volatility in block rewards alters the profitability of various mining strategies. For example, under the fixed reward model, once a miner receives a new block that extends the longest adopted fork in its view, the miner is incentivized to accept the new block and start mining on top of it~\cite{sapirshtein2017optimal}. Since block rewards are fixed, there is no advantage in taking the risk of orphaning an already-mined block. However, in a volatile block reward model, a miner might find it more profitable to attempt to orphan a newly mined block with significantly high transaction fees to steal those transactions~\cite{bar2022werlman, carlsten2016instability}, a strategy referred to as \emph{undercutting}. Additionally, a miner may engage in \emph{selfish mining} by creating a private fork of blocks to take advantage of including all available transactions from the mempool in its private fork.
These examples highlight the need to revisit the security analyses conducted under the fixed reward model~\cite{zur2020efficient, sapirshtein2017optimal, grunspan2018profitability} to better understand Bitcoin’s future security in the context of volatile block rewards.

Some research has analyzed Bitcoin's security under the volatile block reward model in a simplified scenario. Carlsten et al.~\cite{carlsten2016instability} raised concerns about Bitcoin security as the protocol reward diminishes to zero, introducing the undercutting attack and showing how selfish mining could become more threatening in a transaction-fee-driven era. While the paper highlights Bitcoin's instability under the volatile reward model, the analysis relies on simplified assumptions, such as unlimited block space, allowing blocks to collect all available transaction fees. Additionally, the analysis is based on a basic selfish mining strategy, overlooking the optimal strategy and other actions that adversarial miners might use in the volatile block reward model.
To improve mining analysis in the transaction-fee setting, the authors in~\cite{bar2022werlman} introduced the WeRLman framework, which combines deep reinforcement learning with a Monte Carlo Tree Search (MCTS) tool, inspired by the AlphaGo Zero implementation~\cite{silver2017mastering}. They assumed occasional \emph{whale} transactions with significantly high fees that can result in volatility of block rewards. The study showed that the minimum hash rate required for a profitable deviation from honest mining, referred to as the \emph{security threshold}, decreases as Bitcoin undergoes more halvings. While the WeRLman paper made progress in analyzing Bitcoin security, its mempool implementation and treatment of transaction fees do not accurately reflect real-world conditions. In WeRLman, a block can include either a normal transaction or a whale transaction, resulting in blocks with only two distinct fee levels. This assumption simplifies the complex mempool environment to consider solely the count of whale transactions, all sharing the same fee. Moreover, WeRLman is designed to derive a near-optimal mining strategy only in a setting where the mining difficulty has been adjusted according to the adopted strategy. However, in practice, if a miner deviates from the honest strategy, it takes an epoch of at least two weeks for the difficulty to be adjusted, a relatively long period that WeRLman overlooks in its analysis.


In this paper, we enhance the analysis of mining strategies in Bitcoin by incorporating a more realistic volatile block reward model that obtains the near-optimal strategies both before and after the difficulty adjustment. We begin our analysis by revisiting the Bitcoin security threshold presented in the WeRLman paper~\cite{bar2022werlman}. We demonstrate that, due to the design of the WeRLman environment, an adversary in this setting can predict the arrival of the next whale transaction before deciding on its action—an ability we refer to as \emph{predictive capability}. We argue that the predictive capability strengthens the adversary compared to a real-world adversary, resulting in an underestimation of the security threshold. We modify the WeRLman environment to transform it into a non-predictable setting, where the adversary decides its action solely based on the mempool status at the moment of block generation, without any foresight regarding the types of subsequent transactions. The Bitcoin security threshold obtained under the non-predictable WeRLman model, while still lower than that in the fixed block reward model, is higher than what can be achieved under the original WeRLman environment.

We proceed by introducing a simplified model to analyze mining strategies, in which the volatility of block rewards arises from the varying generation times of blocks. We argue that block reward volatility is not solely attributed to the infrequent arrival of whale transactions. The inherent competition among transactions to be included in the chain quickly causes blocks with longer generation times to include higher transaction fees, resulting in block reward volatility even in the absence of whale transactions. To this end, we build our simplified model upon the observed positive correlation between block generation time and its fee. Compared to the WeRLman environment, our simplified environment includes a greater number of blocks with varying fee levels, allowing us to analyze more fine-grained strategies. We employ a Markov Decision Process (MDP)-based tool to analyze the profitability of mining strategies under our simplified volatile model.
Our MDP tool estimates a more accurate security threshold for Bitcoin in the transaction-fee era compared to the non-predictable WeRLman model. However, this threshold serves as an \emph{upper bound} on the actual value, primarily due to the limited number of states the MDP can handle and the restriction of decision-making to block mining events.

\begin{figure}[t]
    \centering
    \includegraphics[height=1.8in]{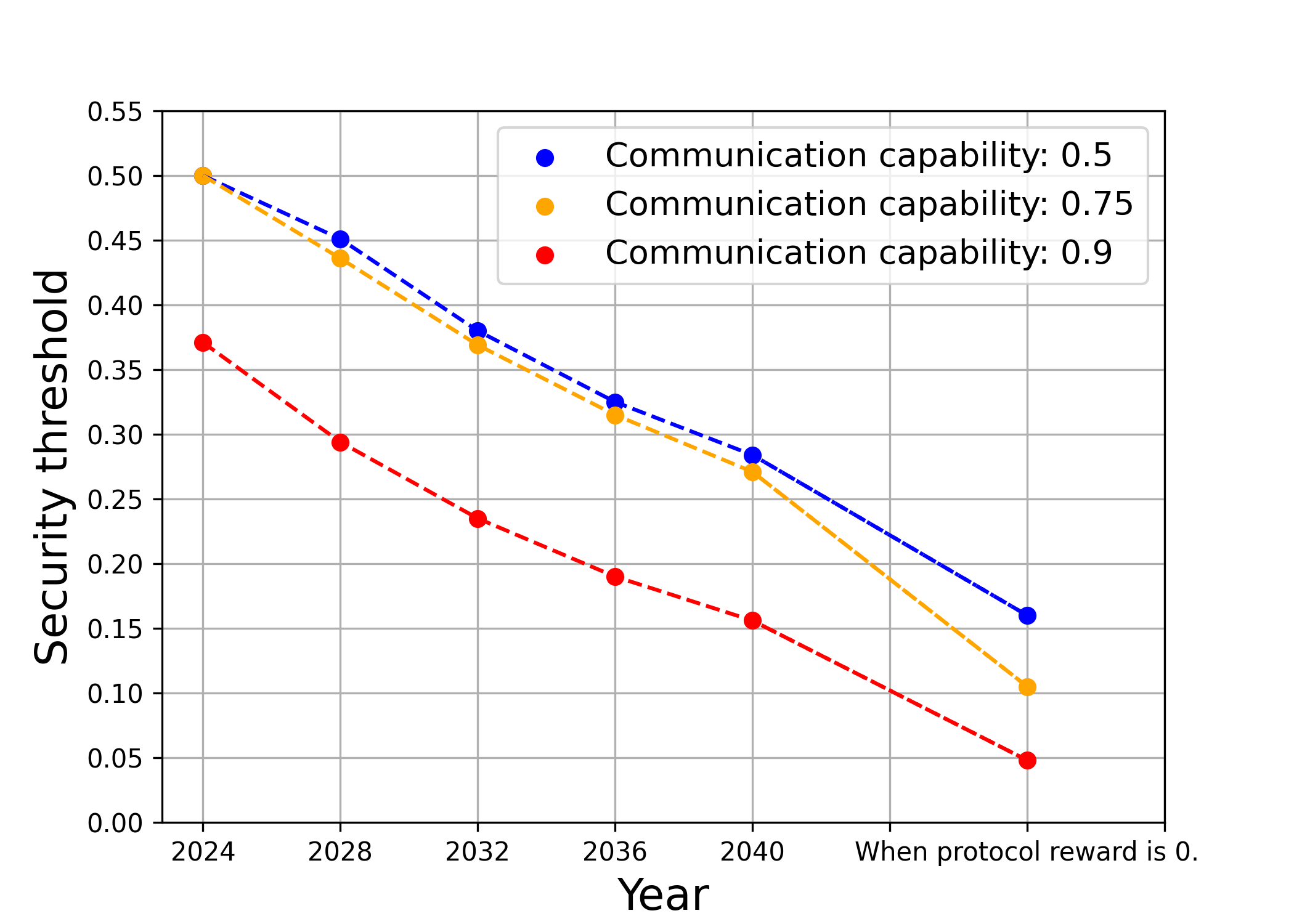}
    \caption{Bitcoin security threshold before DAM.}
    \label{fig:security_threshold_before_DAM}
\end{figure}

In the next step, we analyze the profitability of mining strategies before difficulty adjustment. An interesting observation under the volatile reward model is that a deviant strategy, such as a selfish mining attack, can become profitable immediately after launch without needing to wait for difficulty adjustment. As discussed in the literature~\cite{grunspan2018profitability, negy2020selfish, grunspan2023profit, sarenche2024deep}, under the fixed reward model assumption, a selfish miner incurs losses during the initial epoch of attack when the mining difficulty has not yet adjusted, the duration of which is at least two weeks. 
However, under the volatile reward model, this loss period is shortened or may disappear entirely, making selfish mining a more significant threat in the transaction-fee era. Using our MDP-based tool within the simplified model, we analyze the Bitcoin security threshold prior to difficulty adjustment, which has not been explored in the WeRLman paper. Figure~\ref{fig:security_threshold_before_DAM} illustrates the Bitcoin security threshold in the upcoming year as Bitcoin undergoes more halvings, based on different adversarial communication capabilities\footnote{Communication capability is defined as the ratio of honest miners who receive the adversarial block first in the case of a block race.}. Miners with a mining share above this threshold are not only incentivized to deviate from the honest strategy, but also do not face any initial loss. For a typical communication capability of $50\%$, when the protocol reward approaches 0, the Bitcoin security threshold reduces to $16\%$, which is within the range of current mining pool shares.

Due to the limited state space in the simplified model and the simplifying assumptions made for implementation, the simplified model calculates a lower bound on mining strategy profitability. To conduct a more precise analysis of mining strategies under the volatile reward model, we propose an enhanced model that improves both the implementation of mempool behavior and the training approach. Specifically, we apply the Asynchronous Advantage Actor-Critic (A3C) reinforcement learning algorithm~\cite{mnih2016asynchronous} to derive a near-optimal mining strategy. In this algorithm, multiple agents (representing a payoff-maximizing miner) interact with their own instance of the environment (the Bitcoin blockchain) to train a global policy simultaneously. To better model the Bitcoin mempool in the A3C environment, we analyze historical Bitcoin mempool data to estimate the transaction fee arrival patterns. In our implementation, transactions are categorized into different groups based on the fee paid per unit of weight. We estimate how the total transaction weight grows over time for each category, incorporating these estimates into our environment. Our A3C implementation considers a larger set of actions and states compared to models designed for the fixed reward system, enabling it to obtain near-optimal policies both before and after difficulty adjustments under the volatile reward model. The undercut action is modeled as a continuous action in our A3C model, helping the adversary determine when and for how long to undercut the tip of the canonical chain.

To summarize, our key contributions are:

\begin{itemize} [leftmargin=*]
    \item Revisiting the security threshold of difficulty-adjusted Bitcoin mining from the WeRLman paper~\cite{bar2022werlman}, based on a real-world adversary without predictive capability (Section~\ref{sec: revisiting_werlman}).
    \item Introducing a simplified volatile reward model that includes a greater number of block reward levels and an MDP-based tool to achieve a more precise security threshold (Section~\ref{sec:simplified_volatile_reward_model}).
    \item Analyzing the security threshold of Bitcoin mining under the volatile model before difficulty adjustment to highlight the security threat regarding the immediate profitability of adversarial strategies (Section~\ref{sec:pre_DAM_selfish}).
    \item Implementing an A3C-based tool to derive a near-optimal mining strategy under a realistic mempool model (Section~\ref{sec:Enhanced_volatile_reward}).
\end{itemize}

To present the results in this paper, we have run our implementation using real-world mempool statistics~\cite{mempool} from different periods to demonstrate the threat of adversarial deviation from the honest strategy if the same mempool patterns reoccur.

\section{Preliminaries} \label{sec:preleminaries}
\subsection{System Model and Definitions}
\label{sec:system_model}
The honest strategy in the Bitcoin protocol is defined as follows:
\begin{definition} [Honest strategy] \label{def:honest strategy}
    A mining node should always mine on top of the longest chain~\footnote{In Proof-of-Work blockchains, the longest chain is a chain with the highest accumulated difficulty.} available in its view. Once a new block is mined, the mining node should immediately publish the block to all the other mining nodes.
\end{definition}
The mining nodes in our scheme are categorized into two groups:
\begin{itemize} [leftmargin=*]
    \item \emph{Honest} mining nodes: These nodes, which are denoted by $\mathcal{H}$, follow the honest strategy defined in Definition~\ref{def:honest strategy}.
    \item \emph{Adversarial} mining nodes: In our paper, we assume that the \emph{adversary}, denoted by $\mathcal{A}$, controls a subset of mining nodes referred to as adversarial nodes. These adversarial nodes can arbitrarily deviate from the honest strategy. Namely, they can choose any fork rather than the longest chain to mine on top of it or withhold the mined blocks.
\end{itemize}
Depending on the context, we may refer to adversarial nodes as an attacker, a selfish miner, or an undercutter. In selfish mining, an attacker withholds a newly mined block instead of immediately broadcasting it to the network. In undercutting, the attacker attempts to orphan the tip of the canonical chain, forcing the transactions it contains back into the mempool. A brief discussion of mining attacks that an adversary can conduct in our model, including selfish mining and undercutting, is presented in Appendix~\ref{appendix:mining attacks}.

Although the longest chain fork choice rule outputs a deterministic chain in most cases, situations may arise in which there are multiple forks with the same height in the view of a mining node (the same-height fork race), making the chain selection somewhat subjective. Based on how mining nodes select the winning chain in the presence of multiple same-height forks, the honest nodes are divided into $2$ sub-groups: 
\vspace{-3 pt}
\begin{itemize} [leftmargin=*]
    \item Altruistic mining nodes: These nodes always choose a fork they have heard about earlier in the case of a same-height fork race.
    \item Petty-compliant mining nodes: These nodes choose a fork that leads to a greater profit for them in the case of a same-height fork race.
\end{itemize}
\vspace{-3 pt}
In a same-height fork race, petty-compliant nodes choose the fork with the lower total transaction fees included. The rationale behind this choice is that the fork with fewer transaction fees leaves more fees in the transaction pool, allowing the next block mined on this fork to include a higher amount of fees. Assuming all mining nodes share the same view of the transaction pool, in a same-height fork race, the winning chain for a petty-compliant node is the one whose tip block (the highest block) was mined earlier and, as a result, includes lower fees.

Another factor, besides mining share, that can affect the adversary's profitability in following a deviant strategy instead of honest mining is its connectivity to other mining nodes. We define adversarial \emph{communication capability} as follows:
\begin{definition}[Adversarial communication capability]
    The adversarial communication capability is defined as the ratio of altruistic nodes that receive the adversarial fork earlier than the honest fork in the case of a same-height fork race.
\end{definition}
The adversarial communication capability specifies the altruistic node ratio that selects the adversarial chain as the winning one in a same-height fork race.

In our model, time is assumed to be divided into rounds, denoted by $r$. In each round, a mining node can calculate multiple mining (hash) queries, the number of which is proportional to its mining share. A time step, denoted by $t$, represents the set of consecutive rounds after which a block is mined in the system. At each time step $t$, a block is generated by either the adversary or the honest miners. However, depending on the adversary's strategy, the block generated at time step $t$ might not be immediately added to the canonical chain. Instead, it could be either added to the canonical chain or eventually identified as an orphan block (excluded from the canonical chain) at a future time step.
\begin{definition}[Block ratio] \label{def:block_ratio}
    The block ratio of adversary $\mathcal{A}$ following strategy $\pi$ is defined as follows:
    \begin{equation}
        \begin{split}
             & \texttt{BR}^{\mathcal{A}}_{T}(\pi) = \frac{\sum_{t=0}^{T}{N_\mathcal{A}(t; \pi)}}{\sum_{t=0}^{T}{N_\mathcal{A}(t; \pi) + N_\mathcal{H}(t; \pi)}}, 
             \;\;\;\; \text{and} \;\;\;\; \\
             & \texttt{BR}^{\mathcal{A}}(\pi) =\lim_{T\to\infty} {\texttt{BR}^{\mathcal{A}}_{T}(\pi)}
             \enspace,
        \end{split}
    \end{equation}
    where $N_\mathcal{A}(t; \pi)$ and $N_\mathcal{H}(t; \pi)$ denote the number of adversarial blocks and the number of honest blocks, respectively, added to the canonical chain at time step $t$ under policy $\pi$.
\end{definition} 

\begin{definition}[Time-averaged profit] \label{def:time_avg_profit}
    The time-averaged profit (per-round profit) of adversary $\mathcal{A}$ following strategy $\pi$ is defined as follows:
    \begin{equation}
        \begin{split}
            & \texttt{Profit}^{\mathcal{A}}_{T}(\pi) = {\frac{\sum_{t=1}^{T} {R_\mathcal{A}(t; \pi)-C_\mathcal{A}(t; \pi)}}{T}},
            \;\;\;\; \text{and} \;\;\;\; \\
            & \texttt{Profit}^{\mathcal{A}}(\pi) =\lim_{T\to\infty} {\texttt{Profit}^{\mathcal{A}}_{T}(\pi)} \enspace,
        \end{split}
    \end{equation}
    where $R_\mathcal{A}(t; \pi)$ and $C_\mathcal{A}(t; \pi)$ denote the revenue achieved and the mining cost incurred by adversary $\mathcal{A}$ at time step $t$ under policy $\pi$, respectively.
\end{definition}
The revenue a miner receives at a given time step equals the total rewards from its blocks added to the canonical chain at that time. Each block reward consists of the protocol reward and the transaction fees included in the block.

\begin{definition}[Security threshold] \label{def:security_threshold}
    For a given security parameter $\epsilon$, the security threshold of a mining strategy $\pi$ is defined as the minimum mining share required for the time-averaged profit of strategy $\pi$ to exceed the time-averaged profit of the honest strategy by at least $\epsilon$. The security threshold within an environment is defined as the minimum mining share for which the profitability of the corresponding optimal strategy exceeds the time-averaged profit of the honest strategy by at least $\epsilon$.
\end{definition}

\subsection{Overview of Selfish Mining Profitability}
\label{sec:background_selfish_mining_profitability}
In Bitcoin, a difficulty epoch is defined as the period during which 2016 blocks are appended to the canonical chain. At the end of each epoch, the Difficulty Adjustment Mechanism (DAM) recalibrates the difficulty target for the subsequent epoch based on the hash power estimated from the previous epoch.
Adversarial behaviors that destroy a portion of the network's mining power can cause DAM to underestimate the active mining power. This underestimation lowers mining difficulty, potentially making such behaviors profitable. In the following, we briefly review the well-established fact that, under the fixed reward model, adversarial behaviors like selfish mining are unprofitable before difficulty adjustment~\cite{grunspan2018profitability, sarenche2024selfish, negy2020selfish}. However, they can become profitable over the long term after applying DAM.
\subsubsection{Non-profitability of Pre-DAM Selfish Mining Under the Fixed Block Reward Model}
\label{sec:non_profit_before_DAM}
Under a fixed block reward model, selfish mining cannot be more profitable than honest mining during the first epoch of the attack and before the DAM is applied~\cite{grunspan2018profitability}.
To compare profits from honest and selfish mining before difficulty adjustment, let $e$ denote an epoch, $\lambda$ the mining rate, $\alpha$ the mining share of adversary $\mathcal{A}$, and $R$ the fixed reward per block. We define the adversary's \emph{total} and \emph{canonical} block generation rates as the average number of rounds in which a new adversarial block is mined, and the average number of rounds in which a new adversarial block is added to the canonical chain, respectively. Under honest mining, the adversary's canonical block generation rate is $\lambda \alpha$, yielding a time-averaged profit of $\lambda \alpha R$. If the adversary $\mathcal{A}$ starts selfish mining at the beginning of epoch $e$, its canonical block generation rate would be less than or equal to that under honest mining ($\le \lambda \alpha$). This is because the difficulty of epoch $e$ is already set before the start of the epoch, and the adversary's strategy during the epoch cannot influence the epoch difficulty. Since both the epoch difficulty and the adversary's mining share are fixed, the adversary's total block generation rate remains the same under both honest and selfish mining. However, under selfish mining, some of the adversarial blocks may become orphaned, leading to a decrease in the adversary's canonical block generation rate. Therefore, the time-averaged profit from selfish mining during epoch $e$ (before a DAM is applied) is $\le \lambda \alpha R$. Intuitively, before a DAM is applied, the adversary cannot mine blocks faster, regardless of the strategy chosen~\cite{sarenche2024selfish}.

Note that while selfish mining may potentially increase the adversary's block ratio during epoch $e$, this increase is due to orphaned honest blocks and does not reflect an increase in the adversary's time-averaged profit.

\subsubsection{Profitability of Long-Range Selfish Mining Under the Fixed Block Reward Model} \label{sec:profit_long_range_selfish_mining}
The adversary needs to sustain the selfish mining attack for a longer period to achieve profitability. In Bitcoin, DAM is responsible for adjusting the block generation rate to ensure that it takes, on average, 10 minutes to mine a new block. This implies that the ideal epoch duration is two weeks.
Assume the adversary starts selfish mining at the beginning of epoch $e$. As a result of selfish mining, some of the blocks (both adversarial and honest) mined during epoch $e$ may become orphaned. The generation of these orphan blocks extends the duration of epoch $e$ because to add 2016 blocks to the canonical chain, more than 2016 blocks need to be mined. At the end of epoch $e$, which exceeds two weeks, the DAM decreases the mining difficulty for the following epoch to restore the canonical block generation rate to one block per 10 minutes. This reduction in difficulty leads to a higher block generation rate for the adversary, indicating that starting from epoch $e+1$, the adversary can mine new blocks in fewer rounds and consequently increase its time-averaged profit~\cite{sarenche2024deep}. 
This implies that under the fixed block reward model, selfish mining requires at least one difficulty epoch to become profitable. In practice, the attack duration must be extended further to ensure profitability. This is because the selfish miner incurs losses during the initial epoch due to orphaned blocks, necessitating a longer attack period to compensate for these losses.

\subsection{Asynchronous Advantage Actor-Critic (A3C)} \label{sec: A3C_intro}
Reinforcement learning (RL), a branch of artificial intelligence (AI), aims to develop autonomous agents that learn optimal policies to maximize long-term rewards through iterative interactions with their environment~\cite{alzubaidi2021review}.
Asynchronous Advantage Actor-Critic (A3C)~\cite{mnih2016asynchronous} is a policy gradient method within the RL domain that simultaneously optimizes both the policy and value function estimation. The policy, denoted by $\pi$, guides the agent's actions in the environment by assigning probabilities to each possible action in a given state, indicating the likelihood of the agent choosing that particular action. The value function at a given state $s$, denoted by $V(s; \pi)$, estimates the expected cumulative reward that an agent can obtain starting from state $s$ and following a particular policy $\pi$ thereafter, formally defined as follows:
\begin{equation}
    V(s; \pi) = \mathbb{E}\left[ \sum_{k=0}^{\infty} \gamma^k r_{t+k} \mid s_t = s, \pi \right] \enspace,
\end{equation}
where $s_t$ is the state visited at step $t$, $r_t$ is the reward agent receives at step $t$, and $\gamma$ is the discount factor.
A3C uses multiple agents that interact with their own instances of the environment in parallel. These agents run independently and asynchronously to explore different parts of the state space simultaneously, reducing the correlation between experiences and improving exploration. Unlike many reinforcement learning algorithms, such as deep Q-learning (DQN), which rely on experience replay (storing and reusing past experiences), A3C updates the network in real-time as experiences are generated. The diversity and independence of experiences gathered by multiple agents compensate for the lack of experience replay. 
A3C uses 1-to-$n$-step returns to update both the policy and the value function. These updates occur after every $T$ steps or when a terminal state is reached. The value loss for the trajectory over the time steps $t \in \{0, 1, \cdots, T-1\} $, denoted by $L_v$, is defined as the Mean Squared Error (MSE) between the value function and the expected return as follows:
\begin{equation}
    L_v (\theta_v) = \frac{1}{2} \mathbb{E}_t\Big[\big( R_t - V(s_t; \theta_v))^2\Big] \enspace,
\end{equation}
where $R_t$ is the expected return in step $t$ with respect to the updating time step $T$ and is defined as follows:
\begin{equation}
    R_t = \sum_{k=0}^{T-1- t} {\gamma^k r_{t+k}} + \gamma^{T-t} V(s_{T}; \theta_v) \enspace.
\end{equation}
The main goal of the policy update in A3C is to improve the policy $\pi$ such that actions with higher advantages are more likely to be chosen. The policy loss function for the trajectory over the time steps $t \in \{0, 1, \cdots, T-1\} $, denoted by $L_\pi$ is formulated as follows:
\begin{equation}
   L_\pi (\theta_\pi)=  \mathbb{E}_t\Big[\log \pi(a_t|s_t) \cdot A_t \Big] \enspace ,
\end{equation}
where $\pi(a_t|s_t)$ represents the probability of taking action $a_t$ at state $s_t$ according to the current policy, and $A_t$ is the advantage function defined as $R_t - V(s_t)$. Note that while the policy parameters ($\theta_\pi$) and value function parameters ($\theta_v$) are typically treated as separate for generality, in practice, some of these parameters are often shared. 

\section{Revisiting the WeRLman Model} \label{sec: revisiting_werlman}
The WeRLman paper~\cite{bar2022werlman} studied selfish mining under the volatile block reward model, assuming that mining difficulty is well-adjusted during the attack. The WeRLman authors showed that the security threshold of Bitcoin mining, defined in Definition~\ref{def:security_threshold}, is lower in the volatile reward model compared to the fixed block reward model. This is because, in the volatile block reward model, some blocks may occasionally be mined that include significantly higher rewards due to the inclusion of valuable transaction fees. Once such a block is mined, the adversary is incentivized to orphan it rather than extend it, in an attempt to steal the valuable transactions contained within. 
Although transitions to the volatile reward model can reduce the mining power threshold for profitable selfish mining, we argue that, due to the WeRLman environment design, it underestimates the mining power threshold.

\noindent\textbf{The WeRLman environment.} In the WeRLman environment, two types of transactions exist in the mempool: normal transactions with a normalized fee of $1$ and whale transactions with a normalized fee of $1+F$. The model assumes that, at each event of mining a block $B$ with a new height, a single transaction is added to the mempool. This transaction implicitly represents the transactions that arrive in the mempool between the mining time of block $B$ and the next block. More specifically, once a block is mined, the WeRLman environment samples a new transaction and adds it to the mempool. This new transaction is either normal, with probability $1-p$, or a whale, with probability $p$. This implementation assumption in the WeRLman model ensures a constant transaction arrival rate but introduces a significant bottleneck that benefits the adversary. In the WeRLman model, at each decision-making point, the adversary knows whether the next transaction added to the mempool will be a whale or a normal transaction before deciding its next action. We refer to this property of knowing the type of the next transaction in advance as \emph{predictive capability}.

Our analysis shows that the implicit predictive capability in the WeRLman model allows the adversary to make more profitable decisions regarding whether to publish or withhold its blocks, thereby increasing its profitability compared to a real-world miner. Note that, in reality, miners do not have knowledge of future transactions in advance. While mining a new block, miners frequently update the contents of their blocks to include the highest-fee transactions available. Therefore, once a block is mined, it includes the latest available whale transactions, and the miner is unaware of the arrival time of the next whale transaction. To determine Bitcoin's security threshold in the presence of real-world adversaries, we modified the transaction sampling method in the WeRLman environment to eliminate the effect of predictive capability. We refer to this modified version of the WeRLman environment as the \emph{non-predictable} WeRLman, which is explained below.

\noindent\textbf{The non-predictable WeRLman environment.} The only difference between the non-predictable WeRLman environment and the original version is that, in the non-predictable environment, the mempool includes only the transactions that arrived up to the block generation event, excluding future transactions. At each event of mining a block $B$ with a new height, the non-predictable environment samples a single transaction that resembles those transactions that arrived between the generation of $B$'s parent block and the generation of $B$, allowing $B$ to include the transaction. The implementation details of the non-predictable WeRLman environment are provided in Appendix~\ref{appendix:non_predictable_werlman}.

\begin{figure}[t]
    \centering
    \includegraphics[height=2in]{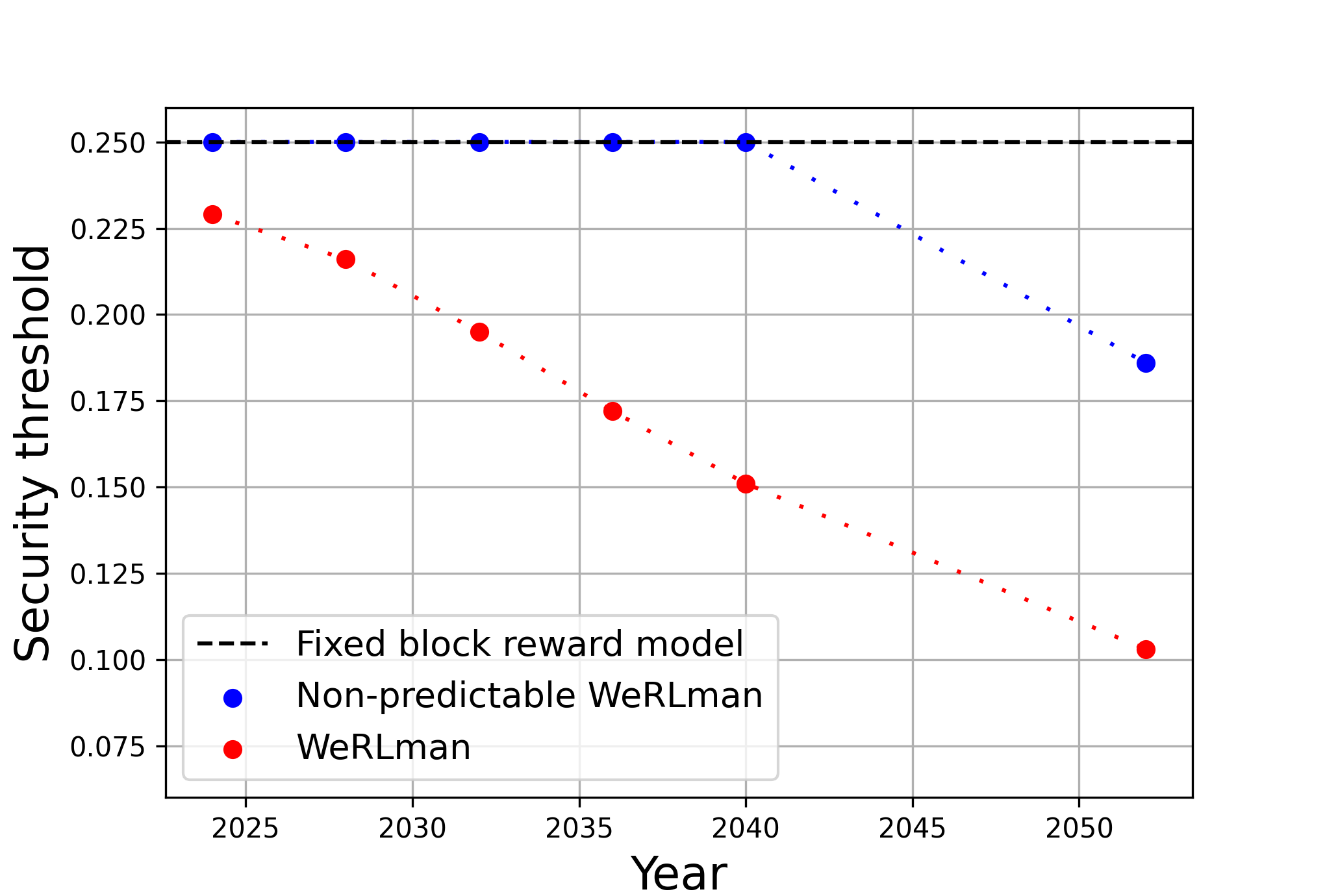}
    \caption{Bitcoin security threshold in upcoming years.}
    \label{fig:comparison_werlman}
\end{figure}

The authors of the WeRLman paper implemented an MDP tool to analyze the security threshold under the WeRLman environment. For a given adversary with specific mining power, this tool determines the optimal strategy for the adversary within the WeRLman environment. The security threshold is then identified as the minimum mining power at which the optimal strategy deviates from the honest strategy. This MDP tool can also be adapted to derive the optimal strategy for an adversary operating under the non-predictable WeRLman environment. In Figure~\ref{fig:comparison_werlman}, we compare the security threshold for a profitable deviation of honest mining in the WeRLman environment and the non-predictable WeRLman environment. To depict Figure~\ref{fig:comparison_werlman}, we use the same extra fee values $F$ as estimated in the WeRLman paper\footnote{In the WeRLman paper, the authors analyzed Bitcoin blockchain data to estimate the extra fee value $F$ for a whale transaction, assuming whale transactions arrive with a frequency of $p=0.001$. Based on the current protocol reward of 3.125 BTC, and the protocol rewards after 1, 2, 3, 4, and 7 halvings, the estimated values of $F$ are $0.26$, $0.45$, $0.74$, $1.14$, $1.58$, and $3.2$, respectively.}. Note that as Bitcoin experiences more halving events, the protocol reward decreases, causing the ratio of a whale-included block reward to a normal block reward, i.e., $1+F$, to increase.

As can be seen in Figure~\ref{fig:comparison_werlman}, the security threshold in the non-predictable environment is higher than in the original WeRLman environment. 

\subsection{Rationale for Higher Security Threshold in Non-Predictable WeRLman}
To explain the reason behind the higher security threshold in the non-predictable environment, we define three adversarial strategies: $\pi_1^\text{WeRLman}$, $\pi_1^\text{Non-predictable}$, and $\pi_2^\text{Non-predictable}$. The first strategy applies only in the original WeRLman environment, as it leverages predictive capability, while the other two are suited for the non-predictable WeRLman environment, where no predictive capability exists. We analyze the profitability of these strategies to identify the factors limiting the adversary's profitability in the non-predictable setting. These three strategies are precisely defined in the Appendix~\ref{appendix:adversarial_strategies}, and their main ideas are described as follows:

\noindent\textbf{Strategy $\pi_1^\text{WeRLman}$.} This strategy applies exclusively to the WeRLman environment. Once the adversary mines a normal block and knows that the next transaction added to the mempool is a whale transaction, it withholds the block and creates a private fork\footnote{By following this strategy, the adversary creates a private fork that increases its chances of winning the whale transaction available in the mempool.}. In other scenarios, the adversary follows honest behavior.

\noindent\textbf{Strategy $\pi_1^\text{Non-predictable}$.} This strategy is suitable for a non-predictable WeRLman environment. Once the adversary mines a normal block, it withholds the block and creates a private fork. In other scenarios, the adversary follows honest behavior.

\noindent\textbf{Strategy $\pi_2^\text{Non-predictable}$.} This strategy is suitable for a non-predictable WeRLman environment. Once the honest miners mine a whale-included block, the adversary does not accept the block but continues mining on the parent of the published block to undercut it. In other scenarios, the adversary follows honest behavior.

\begin{figure}[b]
    \vspace{-5 pt}
    \centering
    \includegraphics[height=2in]{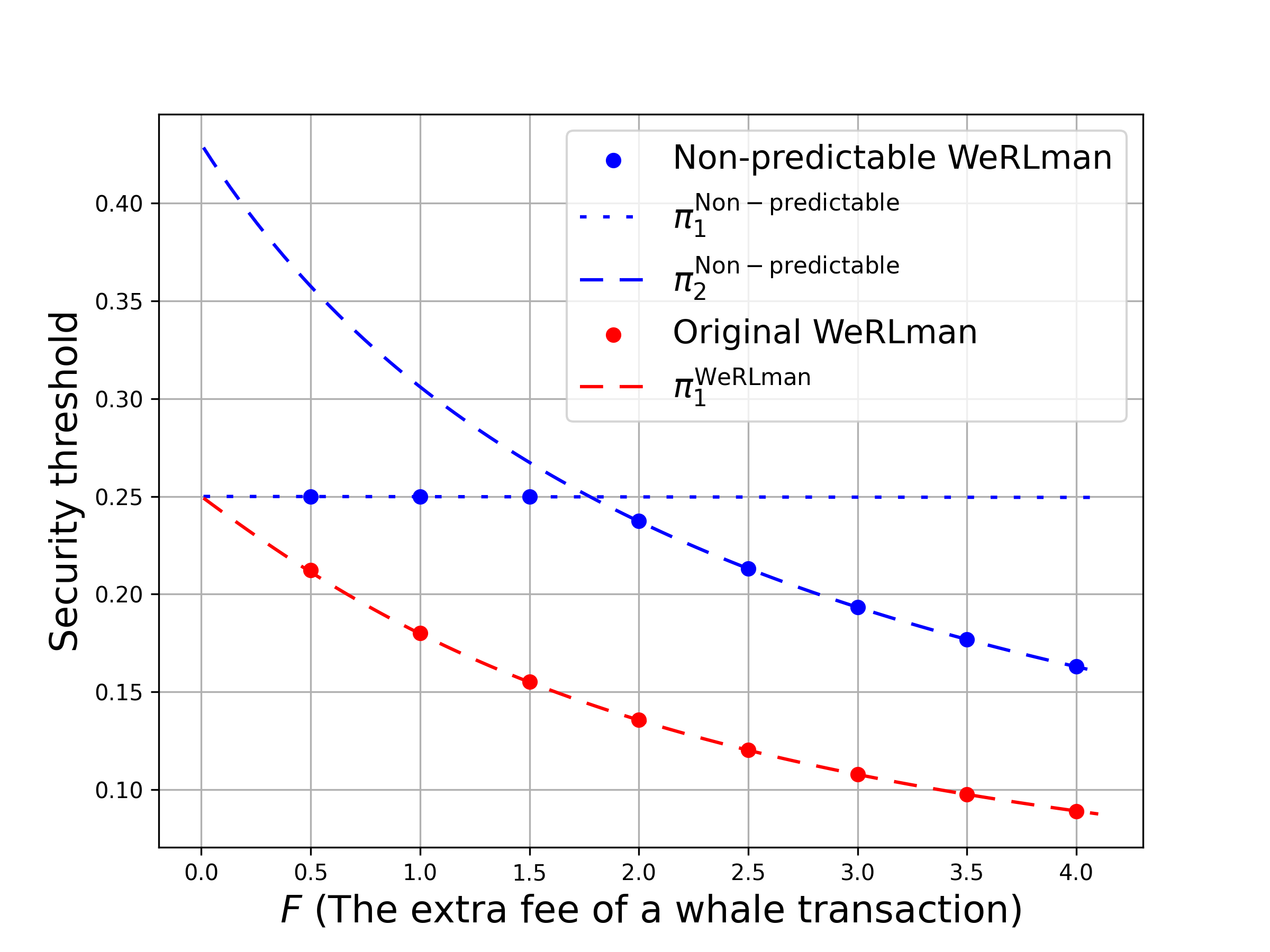}
    \caption{Security threshold of adversarial strategies}
    \label{fig:strategies_security_thereshold}
\end{figure}

In the Appendix~\ref{appendix:adversarial_strategies}, we theoretically analyze the three introduced strategies and determine their profitability. In Figure~\ref{fig:strategies_security_thereshold}, we compare the security threshold under the original WeRLman and non-predictable WeRLman environments with the security thresholds of the introduced strategies: $\pi_1^\text{WeRLman}$, $\pi_1^\text{Non-predictable}$, and $\pi_2^\text{Non-predictable}$. This comparison is based on different values of $F$, a whale transaction frequency of $p=0.001$, and a security parameter $\epsilon = 10^ {-6}$. The security thresholds for the original WeRLman and non-predictable WeRLman environments are obtained using the MDP tool from the WeRLman paper, while the security thresholds for the introduced strategies are derived from the theoretical analysis presented in Section~\ref{appendix:adversarial_strategies}. The comparison shows that the security threshold in the original WeRLman environment overlaps with the security threshold of strategy $\pi_1^\text{WeRLman}$, while the security threshold in the non-predictable WeRLman environment overlaps with the minimum of the security thresholds of $\pi_1^\text{Non-predictable}$ and $\pi_2^\text{Non-predictable}$. We can conclude that the strategy $\pi_1^\text{WeRLman}$ is the deviating strategy that determines the security threshold in the original WeRLman environment. This is because $\pi_1^\text{WeRLman}$ is a near-optimal strategy for miners whose mining share is close to the security threshold in the original WeRLman model.
However, since strategy $\pi_1^\text{WeRLman}$ is not applicable to the non-predictable environment, the security threshold in the non-predictable environment cannot be as low as that in the original WeRLman environment.


Consider the scenario in which the adversary has mined a normal block. Under the WeRLman environment, the adversary has access to additional information about the next transaction type and can choose to withhold its normal block only in scenarios where the next transaction is known to be a whale (strategy $\pi_1^\text{WeRLman}$). If the honest miners mine the next block and include the whale transaction, the withheld adversarial block can be used to orphan the honest block and revive the whale transaction. In other words, the adversary under the WeRLman model can withhold its normal block only if it knows that winning a fork race will guarantee the inclusion of the valuable whale transaction.
However, under the non-predictable environment, if an adversary with limited mining power mines a new block, it would typically publish the block immediately to secure the reward, as it has no information about the next transaction type. This means that for adversaries with limited mining power, following strategy $\pi_1^\text{Non-predictable}$ is not profitable under the realistic non-predictable environment.

\section{Simplified Volatile Reward Model}
\label{sec:simplified_volatile_reward_model}
As can be seen in Figure~\ref{fig:comparison_werlman}, the non-predictable WeRLman model, despite showing a reduction in the security threshold in the long term (after 2040), cannot precisely assess Bitcoin's threshold in the current and near-future years. This imprecision occurs because the WeRLman environment analyzes adversarial strategies by considering only two types of transactions---normal and whale transactions---while in practice, transactions can have varying fee levels.

In this section, we introduce a new volatile reward model, referred to as the simplified environment. This environment enhances the WeRLman environment by including a greater number of transaction types, each with different fee levels, to provide a more accurate simulation of the real-world mempool. Besides, we present a Markov Decision Process (MDP)-based tool that calculates an optimal strategy under the simplified environment. Our MDP tool is capable of obtaining an \emph{upper bound} for Bitcoin's security threshold in the transaction-fee era. The looseness of this upper bound is due to several reasons. First, the MDP tool can handle only a limited number of states, requiring simplifications in the mempool model, such as assuming a linear relationship between the block's transaction fee and block generation time. Second, in the designed MDP tool, the adversary's decision-making is restricted to block mining events. This means the adversary cannot adjust its strategy between two block mining events based on updates to the mempool status, such as the arrival of a fee-valuable transaction.

\subsection{Block Generation Time Variability and Reward Volatility}
In the WeRLman model, the volatility of block rewards arises from the inclusion of whale transactions. However, it is important to emphasize that the primary source of volatility in block rewards during the transaction fee era is not the occasional arrival of whale transactions that offer significantly higher fees compared to others. Instead, the main factor driving this volatility is the competition among transactions to be included in a block as quickly as possible. This competition forces transactions to adjust their fees based on the state of the mempool. Over time, as more valuable transactions accumulate in the mempool, new transactions need to offer higher fees to have a chance of inclusion in the next block. Therefore, during periods of high demand, block rewards can rise significantly due to the presence of higher-fee transactions in the mempool. 

An important factor that increases transaction inclusion demand is extended periods during which no blocks are generated. Due to the random nature of Bitcoin’s mining process, block generation times can vary widely, ranging from a few seconds to over an hour. When block generation times are longer, miners have more opportunities to include higher-fee transactions in their blocks, thus increasing block rewards. This phenomenon is evident in the Bitcoin blockchain, where blocks with significantly higher transaction fee rewards compared to their surrounding blocks often correspond to those with unusually long block generation times. In our simplified model, instead of focusing on the concept of whale transactions, we concentrate on block generation time to analyze Bitcoin mining strategies.


\subsection{Simplified Mempool Environment} \label{sec:simplified_memppol}
In a volatile reward model, miners' profits depend not only on the number of blocks they mine but also on the transaction fees included in those blocks.
To reflect the effect of transaction fees on the block rewards, our simplified environment encompasses an equation that calculates the transaction fee included in a block as a function of block generation time. We refer to this equation as the \emph{time-fee} equation. In the simplified mempool environment, we adopt the regression technique introduced in~\cite{tsabary2018gap} to model mempool behavior using a \emph{linear} time-fee equation. This equation is expressed as $\texttt{fee}(t) = \texttt{fee}_0 + r_{\texttt{fee}} t$, where $\texttt{fee}_0$ and $r_{\texttt{fee}}$ denote the \emph{base fee} and the \emph{fee increase rate}, respectively. The base fee, corresponding to the intercept of the linear equation, represents the average amount of transaction fees remaining in the pool immediately after a block is mined. The fee increase rate, corresponding to the slope, captures the rate at which the collected transaction fees of a block increase over time.
Since the simplified model uses an MDP-based tool to analyze selfish mining profitability, it is necessary to simplify the time-fee equation to fit within a limited number of states. To this end, we divide the linear time-fee equation into $M \geq 2$ discrete time steps. A detailed discussion of the discrete linear time-fee equation used in the simplified model is provided in Appendix~\ref{appendix:simplified_memppol}.

\subsection{MDP-Based Tool}\label{sec:MDP}
In this section, we introduce an MDP-based tool to analyze the profitability of selfish mining under the simplified environment introduced in Section~\ref{sec:simplified_memppol}. Our MDP-based tool, presented in~\cite{our_implementation}, is designed to calculate a lower bound for selfish mining profitability under the volatile reward model, enabling us to determine the security threshold both before and after the difficulty adjustment.

In our simplified model, we consider two forks: an honest fork and an adversarial fork. The honest fork is publicly known to all miners, while the adversarial fork is known only to the selfish miner, who can strategically decide when to publish the blocks in the adversarial fork. The actions available to the adversary are similar to those introduced in~\cite{sapirshtein2017optimal} and include: \texttt{override}, \texttt{adopt}, \texttt{match}, and \texttt{wait}. Each state in our simplified environment is represented by five elements: $(l_\mathcal{A}, l_\mathcal{H}, T_\texttt{total}, T_\texttt{last}, \texttt{fork})$. Here, $l_\mathcal{A}$ and $l_\mathcal{H}$ denote the lengths of the adversarial and honest forks, respectively, and can range from 0 to $\texttt{maxForkLen}$, where $\texttt{maxForkLen}$ represents the maximum fork length in this environment. If one of the forks reaches $\texttt{maxForkLen}$, the selfish miner must either publish its private fork or adopt the honest fork. $T_\texttt{total}$ represents the total sum of block generation time steps for all the blocks in the adversarial fork. Given the assumption of a linear time-fee equation, the reward of the adversarial fork can be calculated using $T_\texttt{total}$ without the need to store the block generation time for each block separately, as follows: $l_\mathcal{A} (R + \texttt{fee}_0) + r_\texttt{fee}T_\texttt{total},$ where $R$ is the protocol reward per block.
$T_\texttt{last}$ represents the time steps elapsed since the mining of the last adversarial block and is used to update the value of $T_\texttt{total}$ at each event of adversarial block generation. \texttt{fork} can take values 0, 1, or 2, indicating that the latest mined block is adversarial, the latest block is honest, or the action \texttt{match} is currently in progress, respectively.

\subsection{Difficulty-Adjusted Selfish Mining in the Simplified Environment}\label{sec:result_simplified_model_after_DAM}
We begin by discussing the objective function that the MDP tool aims to maximize. Let $\mathbf{t}^B_\texttt{ideal}$ denote Bitcoin's ideal average block generation time\footnote{$\mathbf{t}^B_\texttt{ideal}=10$ minutes.}. At each time step in the simplified environment, a new block (either honest or adversarial) is mined, with its block generation time following an exponential distribution process. Let $N_\mathcal{A}(t; \pi)$, $N_\mathcal{H}(t; \pi)$, and $R_\mathcal{A}(t; \pi)$ represent the number of adversarial blocks, honest blocks, and the total reward included in adversarial blocks, respectively, added to the canonical chain at time step $t$ under strategy $\pi$. 
In difficulty-adjusted selfish mining, the average canonical block generation time remains $\mathbf{t}^B_\texttt{ideal}$. Thus, the expected elapsed time up to step $T$ is $\mathbf{t}^B_\texttt{ideal}\sum_{t=1}^{T}{(N_\mathcal{A}(t; \pi) + N_\mathcal{H}(t; \pi))}$. By normalizing $\mathbf{t}^B_\texttt{ideal}$ to $1$ and ignoring mining costs\footnote{Mining costs remain unchanged unless the adversary turns off some miners.}, the time-averaged profit of selfish mining after difficulty adjustment, i.e., the objective function, is given by:
\begin{equation} \label{eq: mining_time_averaged_profit}
    \texttt{Profit}^\mathcal{A}(\pi) = \lim_{T\to\infty} {\frac{\sum_{t=1}^{T}{R_\mathcal{A}(t; \pi)}}{\sum_{t=1}^{T}{N_\mathcal{A}(t; \pi)+N_\mathcal{H}(t; \pi)}}} \enspace.
\end{equation}
Note that although the average canonical block generation time under difficulty-adjusted selfish mining remains equal to $\mathbf{t}^B_\texttt{ideal}$, the average total block generation time (considering both orphan and canonical blocks) is affected by the adversary's strategy. Therefore, we cannot assume that the total block generation time follows the initial rate $\lambda$. Ensuring the correct block generation time is crucial, as it impacts the transaction fee reward. 
To maintain a constant transaction arrival rate, we adopt the same heuristic as the WeRLman paper. At each step of mining a new block $B$, we sample an exponential random variable with mean $\lambda$ to determine the block generation time of $B$, only if $B$ is a block with a new height\footnote{The block height is the distance of the block from the genesis block.} in the system. However, if there already exists a block in the opposite fork with the same height as $B$, the generation time is set to $0$. This approach ensures the transaction arrival rate remains consistent regardless of the adversary's strategy.

\noindent\textbf{Security threshold (comparison with WeRLman model).} We provide a comparison between the security threshold in our simplified environment and the non-predictable WeRLman environment. The reason our results should not be compared with the security threshold of the original WeRLman environment is that, unlike the original environment, our simplified environment does not grant the adversary predictive capabilities. These predictive capabilities could be incorporated into our model by allowing the adversary to know the block generation time of the next block; however, this would not reflect real-world practices.

To ensure a fair comparison, we adjust the parameters in our model so that they align with the WeRLman model. Let $F$ and $p$ denote the additional fee value of a whale transaction and the probability of a whale transaction in the WeRLman model, respectively. Also, let $M$ denote the number of discrete time points in our model, $t_{M-1}$ the greatest time point in our model, $\lambda$ the block generation rate, and $\texttt{fee}(t)$ the time-fee equation. We set $t_{M-1}$ such that the probability of the block generation time being equal to or greater than $t_{M-1}$ matches $p$. Specifically, $t_{M-1} = \frac{-\texttt{Ln}(p)}{\lambda}$. This ensures that the probability of mining a block with the highest possible block generation time (the block with the highest fee) in our model matches the probability of mining a whale-included block in the WeRLman model. 
Additionally, we configure the time-fee equation $\texttt{fee}(t)$ to satisfy $\frac{\texttt{fee}(t_{M-1})}{\texttt{fee}(t_\texttt{average})} = 1 + F$,
where $t_\texttt{average}$ is the average block generation time according to the discrete time division in our model. This ensures that, in both models, the ratio of the fee of the most valuable block to the fee of a normal block is equal to $1 + F$.

\begin{figure}[t]
    \centering
    \includegraphics[height=2.3in]{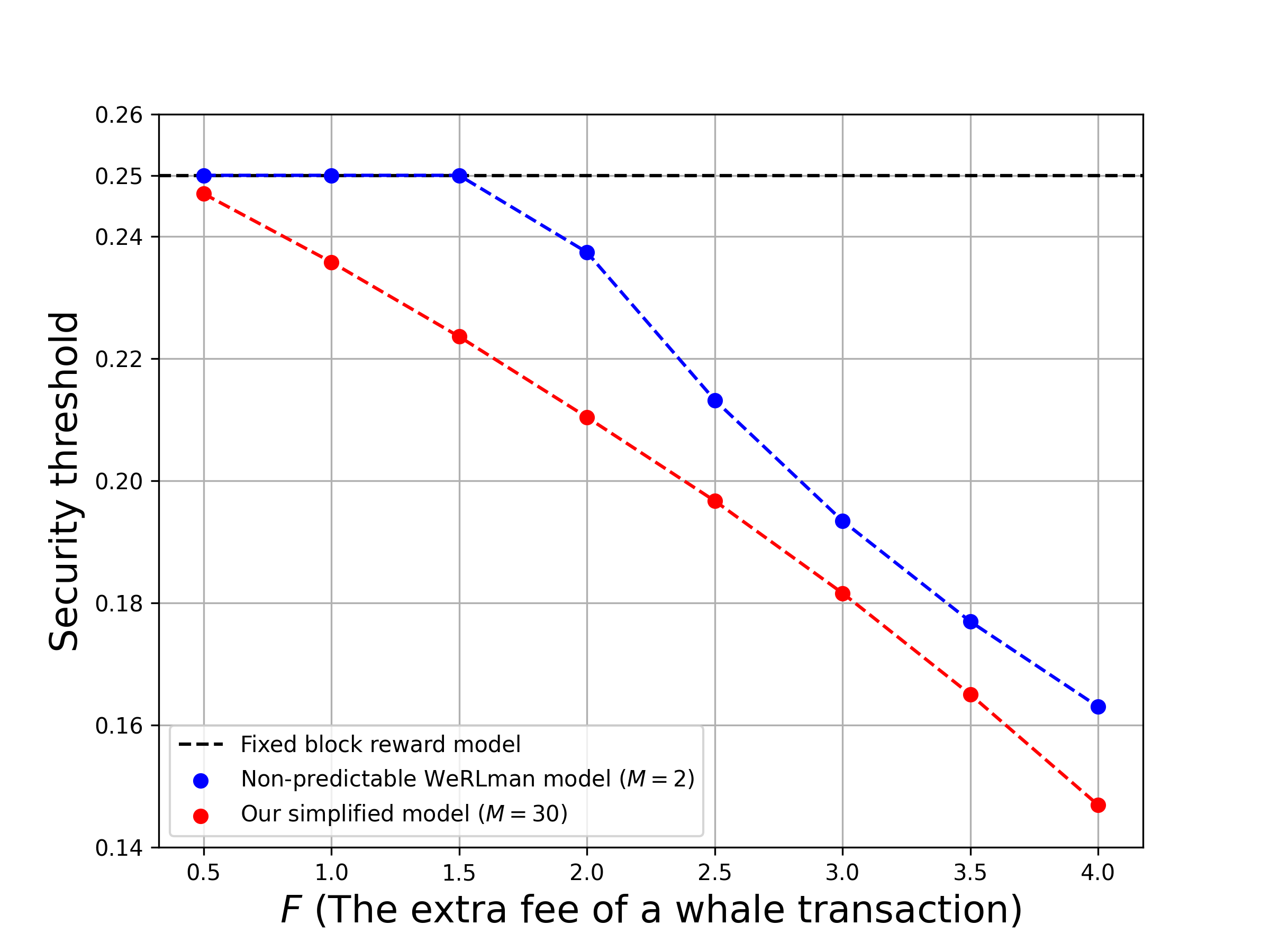}
    \caption{Security threshold comparison under our simplified model and the non-predictable WeRLman model.}
    \label{fig:comparison_werlman_our_model_MDP}
\end{figure}

By setting the number of discrete time points $M=2$, our model exactly matches the non-predictable WeRLman model, as both models only include blocks with two distinct fee levels. Increasing the number of discrete time points allows our model to represent a wider variety of blocks with different rewards, enabling a more precise analysis of profitability in the volatile block reward setting. 
In Figure~\ref{fig:comparison_werlman_our_model_MDP}, we compare the security thresholds in the non-predictable WeRLman model and our simplified model, assuming a protocol reward of $0$ and a probability of mining the most valuable block of $p=0.001$. To obtain the security threshold for our model, we used the MDP-based tool introduced in Section~\ref{sec:MDP}, with the number of time points set to $M=30$. 
As shown in Figure~\ref{fig:comparison_werlman_our_model_MDP}, considering an environment with higher block reward volatility decreases the security threshold. This occurs because, in a more volatile model, blocks with rewards lower than a normal 10-minute block exist, allowing the adversary to take the risk of withholding these blocks to orphan valuable honest blocks. Additionally, a greater number of blocks with rewards exceeding the normal reward exist, resulting in more frequent attractive opportunities to deviate from the honest strategy.

\section{Pre-DAM Selfish Mining Profitability under Volatile Reward Model} \label{sec:pre_DAM_selfish}
In this section, we demonstrate that selfish mining before difficulty adjustment can be profitable under the volatile reward model, enabling miners to deviate from the honest strategy for immediate gain. We first discuss the intuition of how selfish mining, which is unprofitable under the fixed reward model, becomes immediately profitable and thus more threatening in the volatile reward model.

\subsection{Selfish Mining Profit Lag under the Fixed Block Reward Model} \label{sec:profit_lag}
\begin{figure}[t]
    \centering
    \includegraphics[height=2in]{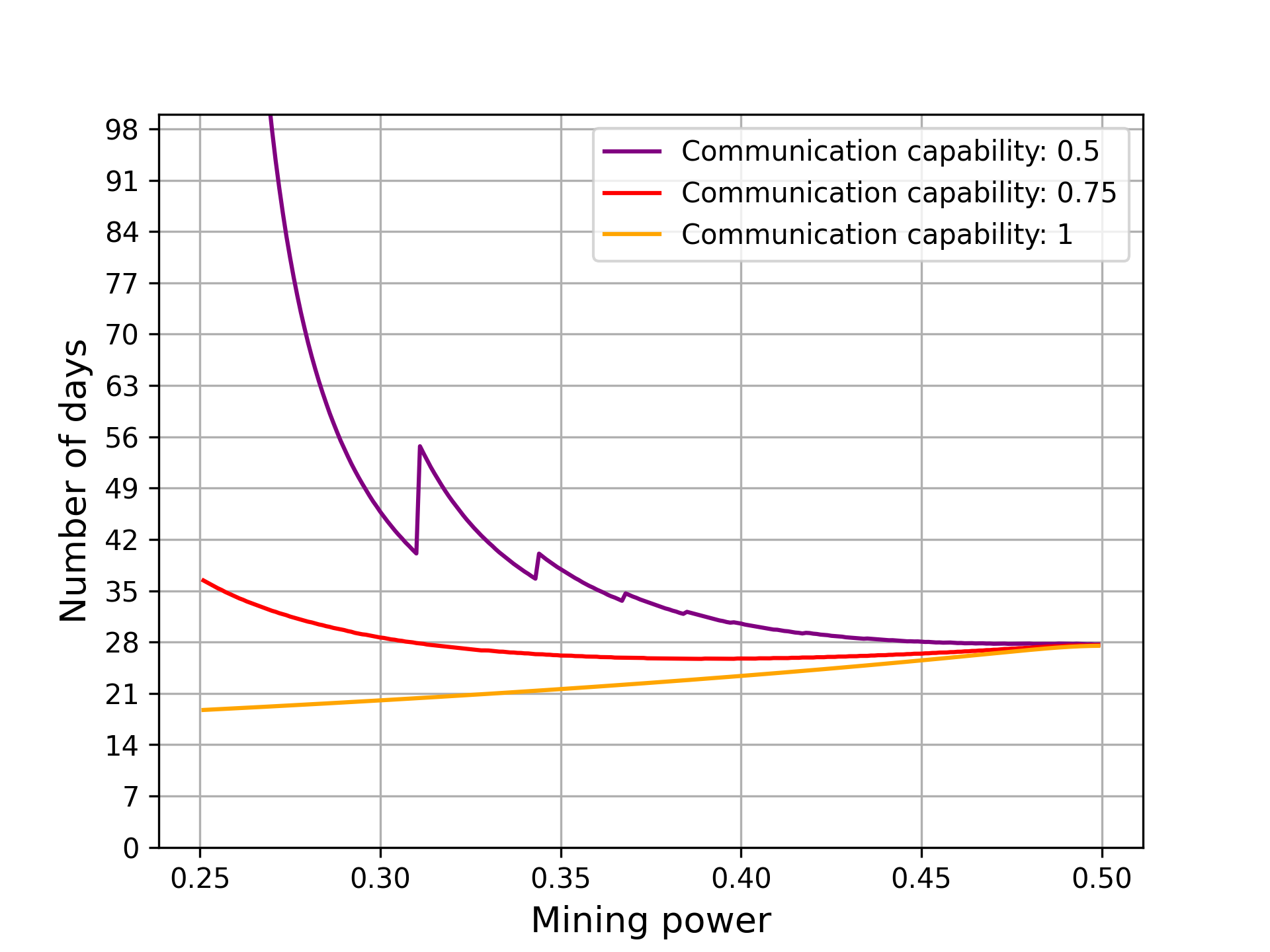}
    \caption{The minimum number of days required for a selfish mining attack to become profitable under the fixed reward model.}
    \label{fig:minimum_day}
\end{figure}

Since Bitcoin's introduction, it has progressed without experiencing long-range or detectable selfish mining attacks. We argue that, in addition to the high hash power required for a successful attack, another contributing factor is the prolonged time required for selfish mining to become profitable under the fixed block reward model. As discussed in Section~\ref{sec:non_profit_before_DAM}, selfish mining during the first difficulty epoch incurs financial losses for the adversary and must be sustained for several days to become profitable. Figure~\ref{fig:minimum_day} shows the minimum number of days required for a selfish mining attack to exceed the profitability of honest mining under the fixed block reward model, a period referred to as the profit lag in~\cite{grunspan2023profit}. The figure considers adversaries with communication capabilities of $\gamma = 0.5$, $0.75$, and $1$ who follow the optimal selfish mining strategy from~\cite{sapirshtein2017optimal}, which maximizes block ratio. The main takeaway is that, under the fixed block reward model, selfish mining demands a significant time investment (more than one ideal epoch or two weeks) to become profitable.

\subsection{Profitability of Pre-DAM Selfish Mining under the Volatile Block Reward Model} \label{sec: profit_selfish_mining_before_DAM_volatile_reward}
Under the volatile block reward model, selfish mining can yield immediate profits by shortening or eliminating the initial loss period, making it a more significant threat to the Bitcoin network.
The primary reason for the immediate profitability of selfish mining under the volatile model is that the attack increases the average generation time of adversarial blocks before the first DAM. While this does not affect the protocol reward per block, it boosts the average transaction fee reward per adversarial block. 

\begin{figure}[t]
    \centering
    \includegraphics[height=1.8in]{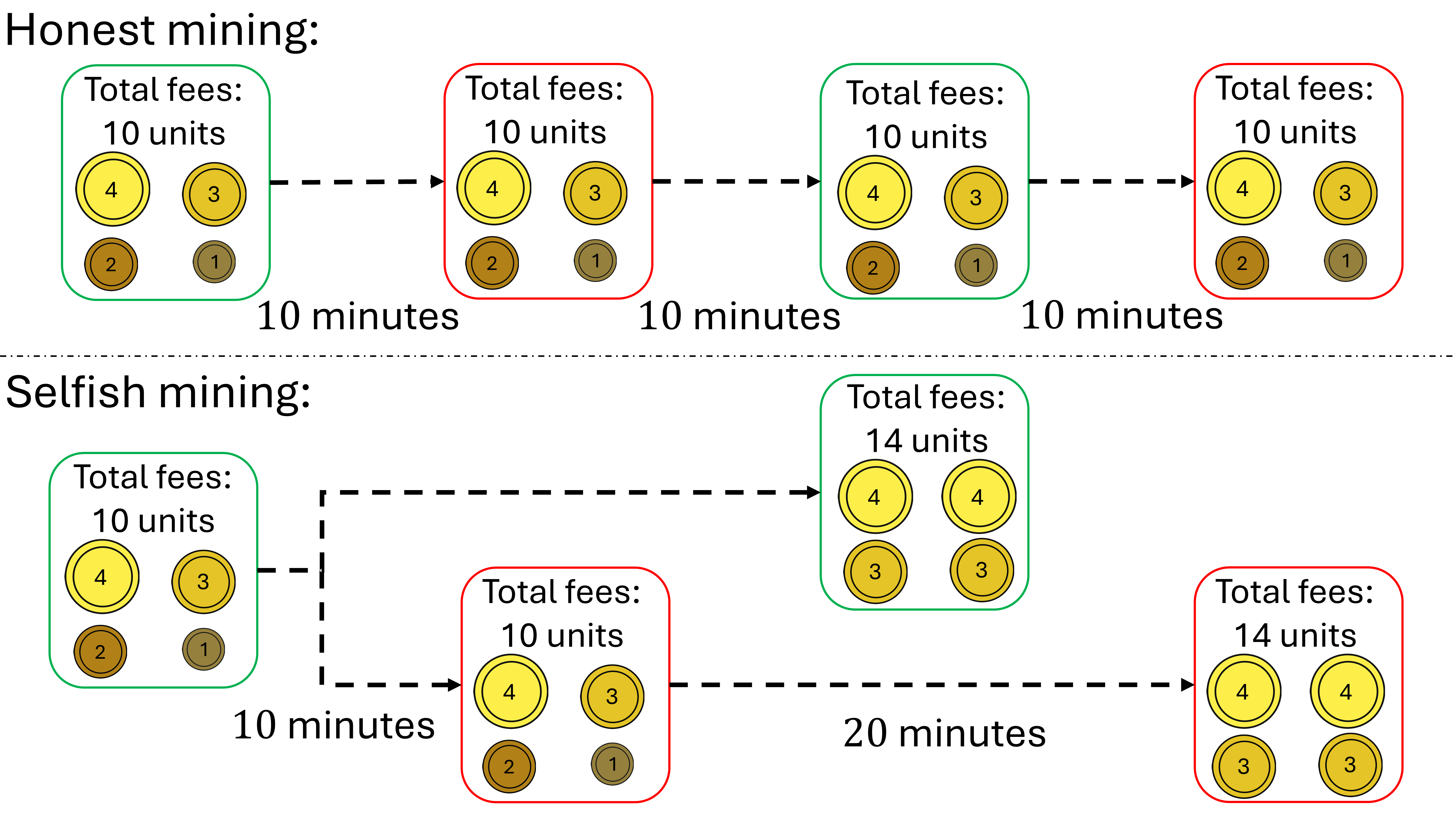}
    \caption{Selfish mining under the volatile reward model. Assume that transaction fees can take values $i \in \{1,2,3,4\}$. The rate of transaction arrival for transactions with a fee value $i \in \{1,2,3,4\}$ is one transaction per 10 minutes. Furthermore, each block can include up to 4 transactions. Adversarial (honest) blocks are depicted in red (green). Under honest mining, the block generation time for all blocks is 10 minutes, and each adversarial block receives 10 units of fee. In the case of selfish mining, the block generation time for the second adversarial block increases. If the attack successfully orphans the middle honest block, the second adversarial block can collect an increased fee of 14 units.}
    \label{fig:tnx_fee_effect_on_SM_before_DAM}
\end{figure}

Let $\lambda_1$\footnote{Note that $\lambda_1 = \lambda \alpha$, where $\lambda$ is the total block generation rate.} and $R_1$ denote the adversarial canonical block mining rate and the average block reward under the volatile reward model when all the miners, including the adversary, mine honestly. Also, let $\lambda_2$ and $R_2$ denote the adversarial canonical block mining rate and the average block reward under the volatile reward model during the first epoch of selfish mining. As discussed in Section~\ref{sec:non_profit_before_DAM}, the adversary's time-averaged profit during the honest mining and the first epoch of selfish mining are equal to $\lambda_1 R_1$ and $\lambda_2 R_2$, respectively. By conducting a selfish mining attack, the adversarial canonical block generation rate decreases in the first epoch of the attack as some of the adversarial blocks may become orphaned, implying that $\lambda_2 \le \lambda_1$. However, unlike the fixed block reward model, $R_2$ is not necessarily equal to $R_1$. In the volatile model, as a result of the increase in the block generation time\footnote{The network block generation time is inversely proportional to the total block generation rate.} during the first epoch of the attack, the average block reward may increase, implying that  $R_2 \ge R_1$. The reason behind this increase in the block reward is intuitively illustrated in Figure~\ref{fig:tnx_fee_effect_on_SM_before_DAM}. Therefore, one of the time-averaged profits $\lambda_i R_i$ for $i \in \{1,2\}$ may be greater than the other depending on the adversary's mining parameters and the mempool statistics. This shows that under the volatile reward block model, selfish mining can be profitable even in the first epoch of the attack.

\subsection{Pre-DAM Selfish Mining in the Simplified Environment}\label{sec:result_simplified_model_before_DAM}
In this section, we analyze the profitability of selfish mining before the difficulty adjustment under the simplified model introduced in Section~\ref{sec:simplified_volatile_reward_model} using our MDP-based tool. Let $\mathbf{t}^B_\texttt{ideal}$ denote Bitcoin's ideal average block generation time, and $R_\mathcal{A}(t; \pi)$ represent the total reward included in adversarial blocks added to the canonical chain at time step $t$ under strategy $\pi$. Under pre-DAM selfish mining, the average block generation time (considering both orphan and canonical blocks) remains $\mathbf{t}^B_\texttt{ideal}$. Since a new block is mined at each step in the environment, the expected elapsed time up to step $T$ is $T\mathbf{t}^B_\texttt{ideal}$. By normalizing $\mathbf{t}^B_\texttt{ideal}$ to 1 and ignoring mining costs, the time-averaged profit of selfish mining before difficulty adjustment, i.e., the objective function, can be expressed as:
\begin{equation} 
    \texttt{Profit}^\mathcal{A}(\pi) = \lim_{T\to\infty} {\frac{\sum_{t=1}^{T}{R_\mathcal{A}(t; \pi)}}{T}} \enspace.
\end{equation}
Since, before DAM, the difficulty is not adjusted based on the adversary's strategy, the generation time of each block follows an exponential distribution with rate $\lambda$. Therefore, at each step of mining a new block $B$, we sample a random number from an exponential distribution with rate $\lambda$ to represent its generation time.

\noindent\textbf{Security threshold (upcoming years).} To analyze the security threshold of Bitcoin in upcoming years, we need an estimation of the time-reward equation of Bitcoin blocks. The reward-fee equation takes the form $\texttt{Reward}(t) = R + \texttt{fee}_0 + r_{\texttt{fee}} t,$ where $R$ is the protocol reward, $t$ is the block generation time in minutes, $\texttt{fee}_0$ is the base fee, and $r_{\texttt{fee}}$ is the fee increase rate. We extracted the Bitcoin block transaction fee for all blocks mined during the period from January 2023 to December 2024. Then, for each day in this period, we used regression techniques to obtain the reward-fee equation based on different values of the protocol reward $R$. For each protocol reward $R$, we calculated the average value of the \emph{block reward increase ratio} defined as $\frac{r_{\texttt{fee}}}{R + \texttt{fee}_0}$ among those promising days whose block reward increase ratio is in the top 10 percent. According to our analysis, the block reward increase ratio during promising days under the current protocol reward of 3.125 BTC, after 1, 2, 3, 4 protocol reward halvings, and under the eventual protocol reward of 0, can be estimated as $0.0091$, $0.0150$, $0.0230$, $0.0336$, $0.0459$, and $0.2057$, respectively.
Figure~\ref{fig:security_threshold_before_DAM} illustrates the security threshold for Bitcoin mining before difficulty adjustment based on the extracted block reward increase ratios, considering three different adversarial communication capabilities of $0.5$, $0.75$, and $0.9$. Note that for the same block reward increase ratio, the security threshold for pre-DAM selfish mining is higher than that for the difficulty-adjusted case. For miners whose mining share exceeds the former threshold, selfish mining does not incur an initial loss period and becomes profitable immediately. However, for miners whose mining share falls between the two thresholds, selfish mining results in losses initially and becomes profitable only in the long term. We can conclude that in the coming years, as transaction fees gradually become the primary source of rewards, the threat of selfish mining becomes more significant. This is because not only does the minimum mining power required for a successful attack decrease, but the attacker would also gain immediate profits upon executing the attack.

\begin{figure}[t]
    \centering
    \includegraphics[height=2.3in]{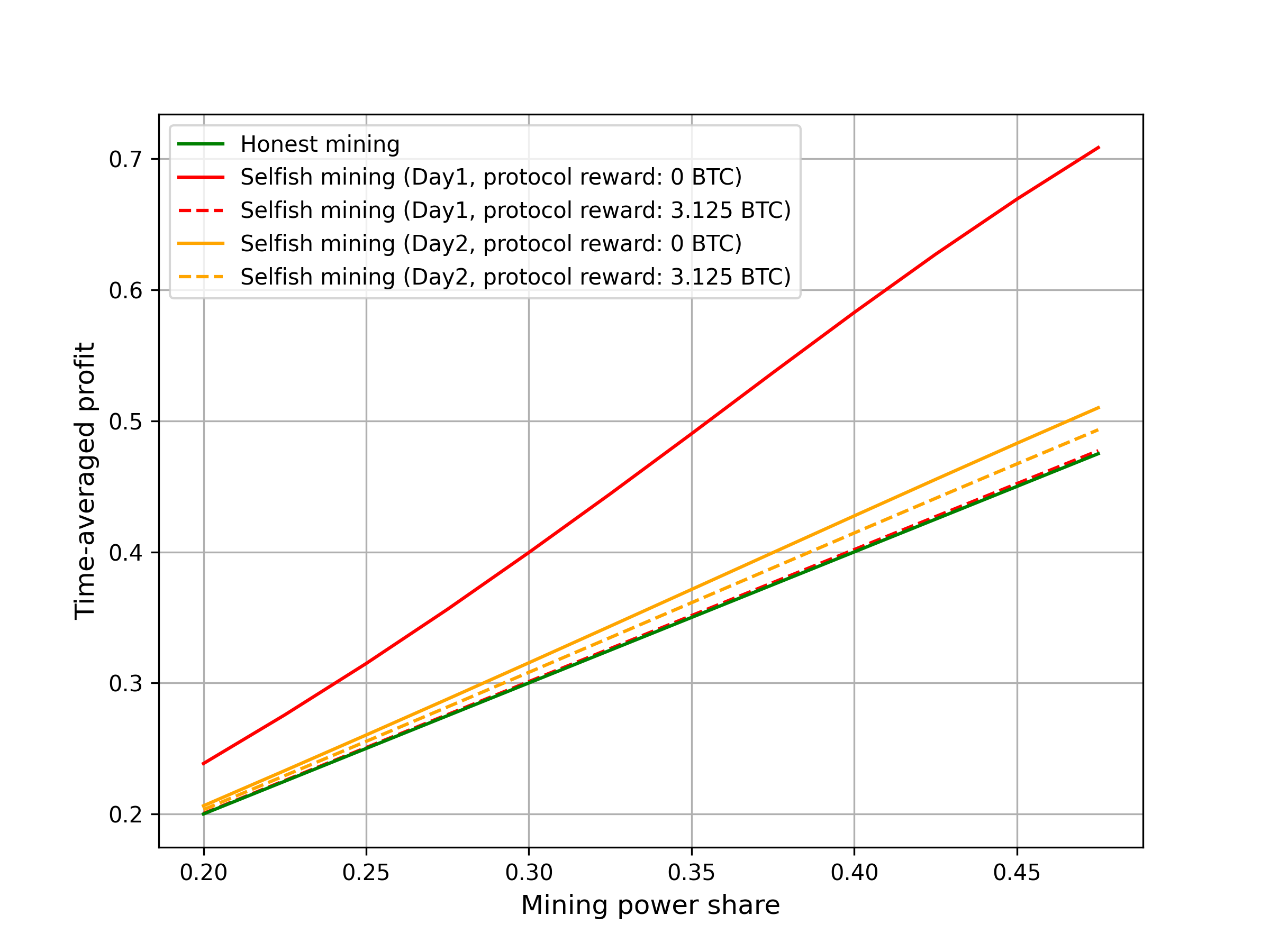}
    \caption{Time-averaged profit before difficulty adjustment.}
    \label{fig:minimum_reward_share}
\end{figure}

\noindent\textbf{Days susceptible to selfish mining.}
To identify the days most prone to selfish mining attacks, we compare the profits from selfish mining before difficulty adjustment on $\texttt{Day}_1$: October 8, 2023, with the time-fee equation $\texttt{fee}(t) = 0.0069 t + 0.0359$, and on $\texttt{Day}_2$: December 17, 2023, with the time-fee equation $\texttt{fee}(t) = 0.0659 t + 3.0457$, under two conditions: when the protocol reward is still substantial and after it has converged to zero. In Figure~\ref{fig:minimum_reward_share}, we present the time-averaged profit a selfish miner can achieve before the difficulty adjustment on $\texttt{Day}_1$ and $\texttt{Day}2$, under both protocol rewards of 0 and 3.125 BTC, as a function of the selfish miner's mining share. This graph assumes a communication capability of 1 for the selfish miner and normalizes the time-averaged profit of the entire network under honest mining to 1. Selfish mining becomes profitable on days when the block reward increase ratio $\frac{r_{\texttt{fee}}}{R + \texttt{fee}_0}$ is significant. To meet this condition with a substantial protocol reward, transaction fees must also be high and comparable to the protocol reward ($\texttt{Day}_2$). However, if the protocol reward becomes negligible, the Bitcoin network becomes more vulnerable to selfish mining on days when the rate of increase in transaction fees is high, even if those days do not coincide with high transaction fees per block ($\texttt{Day}_2$).

\section{Enhanced Volatile Reward Model}
\label{sec:Enhanced_volatile_reward}
Although the simplified volatile reward model introduced in Section~\ref{sec:simplified_volatile_reward_model} improves upon existing models by accounting for multiple fee levels, it still has inherent limitations that hinder precise analysis of Bitcoin mempool patterns. These limitations include:
\begin{itemize}[leftmargin=*]
    \item \textbf{Linear assumption on the time-fee relationship.} The fee accumulated over time by a block does not necessarily increase linearly. In practice, the time-fee relationship is often concave, suggesting that the rate of increase in block reward diminishes as generation time increases.
    
    \item \textbf{Block reward estimation based on a single dependent variable, "block generation time".} This assumption disregards the correlation between a block’s transaction fee reward and the generation times of preceding blocks. In practice, the prolonged generation time of a block not only increases its own reward but also affects the rewards of subsequent blocks.
    
    \item \textbf{Imprecision in time-fee estimation due to sparsity of block data.} Using sparse block data for time-fee regression (on average, one data point every 10 minutes) can lead to reduced accuracy in modeling mempool behavior for a specified period, causing deviations between real-time patterns and the model’s outcomes.
\end{itemize}
We elaborate further on the limitations of the simplified model in the Appendix~\ref{appendix: limitations_simplified_env}.

To address the limitations of the simplified model, this section introduces our Asynchronous Advantage Actor-Critic (A3C)-based implementation, designed to derive a near-optimal mining policy while accounting for the complexities of a real-world Bitcoin mempool. While implementing the Bitcoin environment, we observed that accurately modeling the mempool requires a significantly large state space. This large state space is needed to store fine-grained information about the mempool state, particularly with respect to any blocks involved in a fork race. This expanded state space makes traditional MDP tools impractical. Consequently, we adopt A3C---a reinforcement learning approach well-suited for high-dimensional environments. The A3C algorithm~\cite{mnih2016asynchronous}, introduced by DeepMind, is an actor–critic method~\cite{konda1999actor} that leverages asynchronous parallelism. In A3C, multiple agents interact independently with separate instances of the environment and asynchronously update a shared global network. This parallelism not only accelerates exploration and learning but also enhances convergence stability in complex environments with high-dimensional state spaces, making A3C a superior option compared to traditional reinforcement learning methods.

A3C enables us to implement an \emph{enhanced mempool}, which allows for a more accurate analysis of mining strategies in the transaction-fee era by overcoming the oversimplifications of earlier models. Specifically, the enhanced mempool incorporates polynomial and logarithmic regression techniques to better estimate the time-fee relationship. The mempool state is dynamically updated based on the passage of time and block generation events, capturing the impact of historical activity on the current state. Additionally, it leverages fine-grained mempool data to model the time-fee pattern more precisely, enabling shorter and more stable regression periods that mitigate sample deviation issues.

Our A3C-based implementation codes are provided in~\cite{our_implementation}. Additionally, a discussion on the implementation of the A3C-based tool is provided in Appendix~\ref{appendix:A3C_implementation}. This discussion includes a comprehensive description of the objective function (Appendix~\ref{appendix:objective_function}) and the A3C implementation details (Appendices~\ref{appendix: implementation} and~\ref{appendix:undercut_implementation}).



\subsection{Enhanced Mempool Implementation}
In the simplified model, we estimate the time-fee equation using only data on block transaction fees and generation times. To better estimate how transaction fees fluctuate over time, we can improve the model by incorporating the behavior of the Bitcoin mempool. The mempool is a storage space where unconfirmed transactions are stored before they are included in a block. To increase the chance of inclusion in a block, each Bitcoin transaction pays a fee to the block miner. The total fee that a transaction pays is calculated based on the transaction's weight, which is measured in virtual bytes (vBytes). Virtual bytes are a weight unit introduced by the SegWit upgrade. The fee is typically expressed in satoshis per virtual byte (sat/vByte), where each satoshi is $10^{-8}$ BTC. Each Bitcoin block has a size limit of 1 virtual megabyte (vMB)~\cite{Bitcoin_wiki}. Miners prioritize filling the block with the most valuable transactions from the mempool, where the term \emph{valuable} refers to transactions that offer the highest fees in satoshis per virtual byte. In the context of the enhanced volatile reward model, we define the \emph{base fee} for a given mempool period as the fee for which there is at least 1 virtual megabyte of transactions offering that fee or higher at all times during that period.

Our implementation categorizes the transactions available in the mempool into $N^\text{range}$ sat/vByte ranges, denoted as 
$\{\texttt{fee}_0, \texttt{fee}_1, \ldots,$ $ \texttt{fee}_{N^\text{range}-1}\} \enspace,$
where the fee levels are ordered in ascending sequence. 
A transaction paying $f$ sat/vByte, where $\texttt{fee}_j \leq f < \texttt{fee}_{j+1}$, is assigned to the range corresponding to $\texttt{fee}_j$. $\texttt{fee}_0$ denotes the base fee, implying that all transactions available in the mempool pay at least $\texttt{fee}_0$ sat/vByte. In our implementation, we assume an unlimited number of transactions paying the base fee. 
The weight of transactions in the range $\texttt{fee}_j$ increases over time according to a specific equation, called the \emph{weight-time} equation, which is estimated based on the transaction weight growth in the range $\texttt{fee}_j$.
Whenever a new block is mined, a total weight equivalent to 1 virtual megabyte is deducted from the mempool, starting from the highest sat/vByte ranges. To estimate the weight increase function for each sat/vByte range over a given time period, we analyzed mempool statistics during that period using the information presented in~\cite{mempool}. Then, we applied regression techniques to derive the weight-time equation for each sat/vByte range.

\begin{figure}[t]
    \centering
    \includegraphics[height=2.2in]{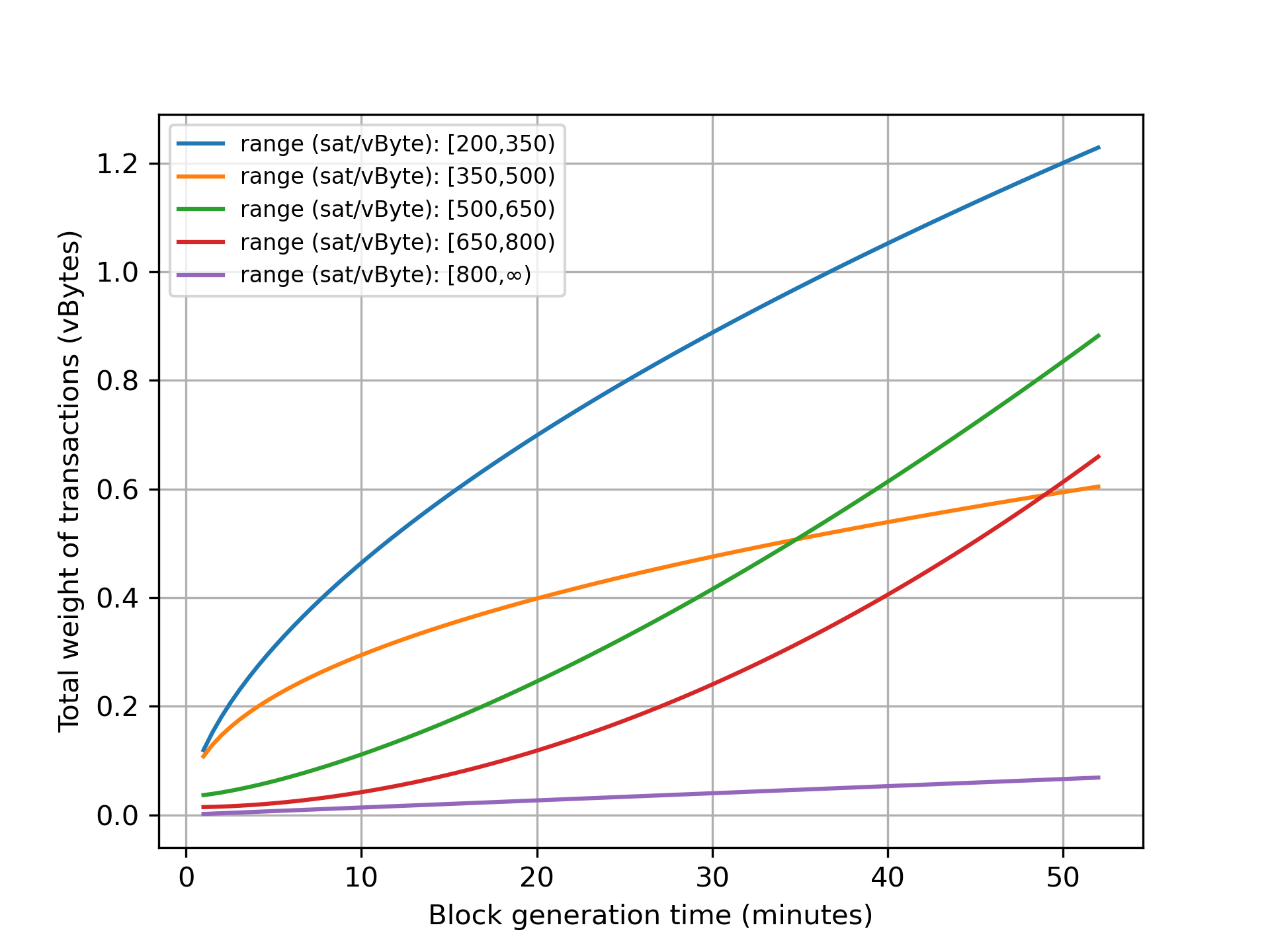}
    \caption{Total transaction weight for different ranges of sat/vByte as a function of block generation time (December 17, 2023).}
    \label{fig:weight_time_curves}
\end{figure}

The estimated weight-time curves based on mempool statistics from December 17, 2023, are shown in Figure~\ref{fig:weight_time_curves}. To depict Figure~\ref{fig:weight_time_curves}, the transactions in mempool are categorized into five ranges of sat/vByte, with the base fee equal to $200$ sat/vByte. This base fee is chosen because, at all times on December 17, 2023, there was at least 1 virtual MB of transactions in the mempool paying a fee higher than $200$ sat/vByte. This indicates that, regardless of the block generation time, a block mined on this day could collect at least 200 million satoshis (2 BTC). An interesting observation from Figure~\ref{fig:weight_time_curves} is that, although the weights of transactions in all sat/vByte ranges are increasing over time, the rate of increase for weights in the lower sat/vByte ranges decreases over time (indicating a concave weight-time equation). In contrast, the rate of increase for weights in the higher sat/vByte ranges accelerates (indicating a convex weight-time equation). This observation aligns with the competition that exists in practice among transactions for inclusion in blocks. As the generation time for the next potential block increases, this competition intensifies, where newer transactions need to submit higher fees to maintain their chances of being included in the next block.

\subsection{A3C Actions and State Representation}
\label{sec:actions_state}
In our A3C-based implementation, we assume that adversary $\mathcal{A}$ mines on top of its secret chain, while the honest miners $\mathcal{H}$ mine on top of the single public chain. These two chains share a common subchain, after which they diverge from each other. The common subchain is referred to as the canonical chain, whose included blocks cannot later be orphaned by the adversarial chain. The adversary’s chain (honest chain) proposed on top of the canonical chain tip is referred to as the adversarial fork (honest fork). Let $l_\mathcal{A}$ ($l_\mathcal{H}$) denote the adversarial fork length (the honest fork length). We first explain the set of possible actions that the adversary can take.

\texttt{override}: The action represents that the adversary publishes a sub-fork of its secret fork whose length is one block longer than the honest fork. As a result of \texttt{override} action, the honest miners abandon their current fork and start mining on top of the longer, adversarial fork. This action is feasible if $l_\mathcal{A} > l_\mathcal{H}$.

\texttt{match}: The action represents that the adversary publishes a sub-fork of its secret fork whose length is equal to the honest fork. Action \texttt{match} results in a race between $2$ same-height forks. To encourage altruistic miners to mine on top of the adversarial fork, the adversary should immediately publish its fork after an honest block is released. By doing so, the adversarial fork can be delivered earlier to some of the altruistic nodes, where the ratio of these nodes is determined based on the adversarial communication capability factor. However, petty-compliant miners do not necessarily choose a fork based solely on its delivery time. To incentivize these miners, the adversarial fork must leave transactions with higher total fees in the mempool compared to the honest fork. This means that the adversarial fork should have a lower total transaction fee than the honest fork to improve its chances of winning the same-height fork race in the presence of petty-compliant miners. Typically, if the adversary has mined the block of height $l$ in its fork earlier than the block of height $l$ in the honest fork, the adversarial fork up to and including length $l$ contains a lower total transaction fee compared to the honest fork up to and including length $l$. This is because, during the time gap between mining the $l^\text{th}$ adversarial block and the $l^\text{th}$ honest block, several fee-rich transactions may arrive that honest miners can include in their blocks, which are absent from the adversarial fork up to and including block $l$. This action is feasible if $l_\mathcal{A} \ge l_\mathcal{H}$.

\texttt{wait}: This action represents that the adversary continues mining new blocks on top of its secret fork. This action is always feasible.

$\texttt{adopt}_i$: The action $\texttt{adopt}_i$, $i \in \{0,1,\ldots,k_1\}$ represents that the adversary abandons its adversarial fork and starts mining a new fork on top the longest public chain whose latest $i$ blocks are pruned. In the fixed block reward model, considering only the action $\texttt{adopt}_0$---namely adopting the longest chain---is enough, as $\texttt{adopt}_i$ for $i>0$ cannot lead to a more profitable strategy. However, in the volatile block reward model, a deep block in the honest fork may have an extraordinary amount of transaction fees, incentivizing the adversary to orphan the block and steal its transaction fees. A block with an extraordinary fee can result from including a whale transaction, as considered in~\cite{bar2022werlman}, or from having a long generation time. Action $\texttt{adopt}_i$ is feasible if $i \le l_\mathcal{H}$.

$\texttt{undercut}_\texttt{block}$: This action represents that the adversary tries to undercut the tip block of the longest public chain denoted by $B^\text{tip}_\mathcal{H}$. To perform block undercutting, the adversary: i) abandons its fork, ii) adopts $l_\mathcal{H}-1$ blocks in the honest fork, namely the entire honest fork except for the last block $B^\text{tip}_\mathcal{H}$, and iii) adjusts the total transaction fee in its potential next block to be less than the total transaction fee included in $B^\text{tip}_\mathcal{H}$. 
If, after taking the action $\texttt{undercut}_\texttt{block}$, the next block is mined by the adversary, the adversary's subsequent action is $\texttt{match}$. Choosing action $\texttt{match}$ will result in a race between two forks of the same height, differing only in their last block. In this scenario, the altruistic miners will continue mining on top of the honest fork since they received it earlier. However, the petty-compliant miners will switch to mining on top of the adversarial fork as it contains a lower amount of transaction fees. Our implementation considers the $\delta$-petty-compliant miners with an adjustable $\delta$, where the petty-compliant miners deviate from the honest strategy if doing so allows them to earn at least $\delta$ BTC more. If the amount of remaining transaction fee in the transaction pool is relatively low and the tip block $B^\text{tip}_\mathcal{H}$ includes a relatively large amount of transaction fee, undercutting block $B^\text{tip}_\mathcal{H}$ may be a logical action for the adversary as it enables the adversary to fill its block with transactions with higher fees. This action is always feasible.

$\texttt{undercut}_\texttt{fork}$: This action represents that the adversary tries to undercut the entire honest fork using its fork whose length is $1$ less than the honest fork length. To perform fork undercutting, the adversary adjusts the transaction fee in its potential next block in a way that the total transaction fee in the adversarial fork becomes less than that in the honest fork. For $l_\mathcal{H}=1$, this action is the same as $\texttt{undercut}_\texttt{block}$. Similar to the scenario explained for action $\texttt{undercut}_\texttt{block}$, if the next block is mined and published by the adversary, the petty-compliant miners are incentivized to mine on top of the adversarial fork. Under the presence of petty-compliant miners, action $\texttt{undercut}_\texttt{fork}$ increases the probability of orphaning the honest fork when the adversary's fork is lagging behind in the fork race. This action is feasible if $l_\mathcal{A} = l_\mathcal{H}-1$.

A state at step $t$ in our implementation represents a race between the honest and adversarial forks and has the following format: 
\begin{equation}
    \begin{split}
    s_t = &  \Big(l_\mathcal{A}, l_\mathcal{H}, \texttt{Latest}, \texttt{Match}, \texttt{Undercut}, \\
    & BR_\mathcal{A} = \{BR_\mathcal{A}^1, BR_\mathcal{A}^2, \ldots, BR_\mathcal{A}^\texttt{maxForkLen}\}, \\
    & \texttt{Pool}_\mathcal{A} = \{\texttt{Pool}_\mathcal{A}^1, \texttt{Pool}_\mathcal{A}^2, \ldots, \texttt{Pool}_\mathcal{A}^\texttt{maxForkLen}\}, \\
    & \texttt{Pool}_\mathcal{H} = \{\texttt{Pool}_\mathcal{H}^1, \texttt{Pool}_\mathcal{H}^2, \ldots, \texttt{Pool}_\mathcal{H}^\texttt{maxForkLen}\}, \\
    & \texttt{Pool}_\mathcal{C}
    \Big) \enspace.
    \end{split}
\end{equation}
In the state representation above, $l_\mathcal{A}$ ($l_\mathcal{H}$) denotes the adversarial (honest) fork length.
$\texttt{Latest}$ is a boolean variable that represents whether the latest mined block is honest or adversarial. If the last block is honest and $l_\mathcal{A} \ge l_\mathcal{H}$, the adversary can take action $\texttt{match}$.
$\texttt{Match}$ is a boolean variable that represents whether the same-height fork race resulting from the action \texttt{match} is resolved or not. If, after taking action $\texttt{match}$, the next block is honest, the fork race is resolved. However, if the next block is adversarial, the fork race persists.
$\texttt{Undercut}$ is a boolean variable that represents whether the next block mined after taking actions $\texttt{undercut}_\texttt{block}$ or $\texttt{undercut}_\texttt{fork}$ is adversarial or not. If the adversary manages to mine the next block, it should then take action $\texttt{match}$ to publish the undercutting block or fork to the rational miners.
$BR_\mathcal{A}^i$ for $i \in \{1,2,\ldots, \texttt{maxForkLen}\}$ denotes the block reward of the $i^\text{th}$ block in the adversarial fork. This block reward includes both the protocol reward and the transaction fees collected from the included transactions. Note that \texttt{maxForkLen} is a variable that determines the maximum length of a fork in our implementation.
$\texttt{Pool}_\mathcal{A}^i$ ($\texttt{Pool}_\mathcal{H}^i$) for $i \in \{1,2,\ldots, \texttt{maxForkLen}\}$ denotes the mempool statistics w.r.t. the chain whose tip block is the $i^\texttt{th}$ adversarial (honest) block. Once a new block is mined, it includes some of the transactions available in the mempool, resulting in the removal of those transactions from the mempool and consequently a modification in the mempool statistics. A state must store the mempool statistics w.r.t. the $i^\texttt{th}$ adversarial (honest) block as miners may need to abandon their current forks and continue mining on top of the $i^\texttt{th}$ adversarial (honest) block as a result of a specific action. Mempool statistics is an $N^{\text{range}}$-element array that contains the weight of transactions available in each of the satoshi-per-byte ranges $\{\texttt{fee}_0, \texttt{fee}_1, \ldots, \texttt{fee}_{N^{\text{range}}-1}\}$. 
The weight of transactions corresponding to each satoshi-per-byte range $\texttt{fee}_j$ increases over time according to the specific weight-time equation defined for the range $\texttt{fee}_j$. As mentioned earlier, once a new block is mined, a total weight of 1 virtual megabyte is deducted from the transactions in the highest satoshi-per-byte ranges. However, if the adversary takes one of the actions $\texttt{undercut}_\texttt{block}$ or $\texttt{undercut}_\texttt{fork}$, the block may be filled with less than 1 virtual megabyte of transactions to leave some valuable transactions in the mempool for petty-compliant miners. $\texttt{Pool}_\mathcal{A}^i$ ($\texttt{Pool}_\mathcal{H}^i$) is an array containing $N^\text{range}$ elements, where the $j^\text{th}$ element represents the weight of total transactions of the mempool that pay $\texttt{fee}_{j-1}$ satoshis per virtual byte.
Finally, $\texttt{Pool}_\mathcal{C}$ denotes the mempool statistics w.r.t. the canonical chain.

\subsection{Undercutting as a Continuous Action} \label{sec:undercut_countinuous}
The actions \texttt{override}, \texttt{adopt}, \texttt{match}, and \texttt{wait} can be considered discrete actions. This means that once a new block is mined in the system, the agent decides whether to take or not take these actions. However, the action \texttt{undercut} is more appropriately viewed as a continuous action. When dealing with continuous actions, the agent must decide on a value within a range that represents the intensity or duration of the action.
In the case of \texttt{undercut}, the adversary must determine how long to continue undercutting the tip of the canonical chain. After a certain point, the adversary should stop undercutting and resume mining on top of the tip block. The continuous nature of the \texttt{undercut} action arises because, depending on the available transaction fees in the mempool, undercutting may or may not be the best action to take.
As discussed in Section~\ref{appendix:mining attacks}, the \texttt{undercut} action is profitable when the mempool w.r.t. the tip of the chain is relatively empty, leading the undercutter to mine on top of the parent block of the tip to steal the transaction fees from the tip block. Over time, as the transaction fees available in the mempool increase, a point is reached where the fees w.r.t. the tip of the chain are large enough that continuing the \texttt{undercut} action becomes unprofitable. At this point, the agent should abandon undercutting and instead mine on top of the tip block.

In an A3C implementation, the neural network outputs a single scalar value representing the value of the current state, along with a probability distribution over the possible actions. For discrete actions, this probability distribution is a vector where each element represents the probability of selecting a specific action given the current state. For continuous actions, the network outputs the parameters of a probability distribution over the continuous action~\cite{jia2022policy}.
In our implementation, the neural network outputs a vector representing the probability distribution of the actions introduced in Section~\ref{sec:actions_state}. Additionally, it outputs two scalars representing the mean and variance of the duration of the undercutting action, which is modeled as a continuous action.

\subsection{Results and Implications: Mining Attacks under an Enhanced Volatile Model}
This section presents results from our A3C-based implementation~\cite{our_implementation}, analyzing mining attacks under the mempool patterns observed during different periods. Information regarding our A3C-based tool implementation is provided in~\ref{appendix: implementation}. To compare the time-averaged profits of a mining strategy $\pi$, we use the concept of the percentage increase in the miner's time-averaged profit, defined as follows:
\begin{equation}
     \texttt{PI}^\mathcal{A}(\pi) = \frac{ \texttt{Profit}^\mathcal{A}(\pi) -  \texttt{Profit}^\mathcal{A}(\pi^\mathcal{H})}{ \texttt{Profit}^\mathcal{A}(\pi^\mathcal{H})} * 100\% \enspace,
\end{equation}
where $\texttt{Profit}^\mathcal{A}(\pi^\mathcal{H})$ represents the time-averaged profit achieved by the adversary if all miners, including the adversary, follow the honest strategy.

In this section, we analyze the profitability of selfish mining before the difficulty adjustment, as well as the single-block undercutting attack. The results related to the profitability of selfish mining after the difficulty adjustment are presented in Appendix~\ref{appendix: selfish mining_after_DAM}.
\subsubsection{Mining Attacks Before a Difficulty Adjustment.}
\begin{figure}[t]
    \centering
    \includegraphics[height=2in]{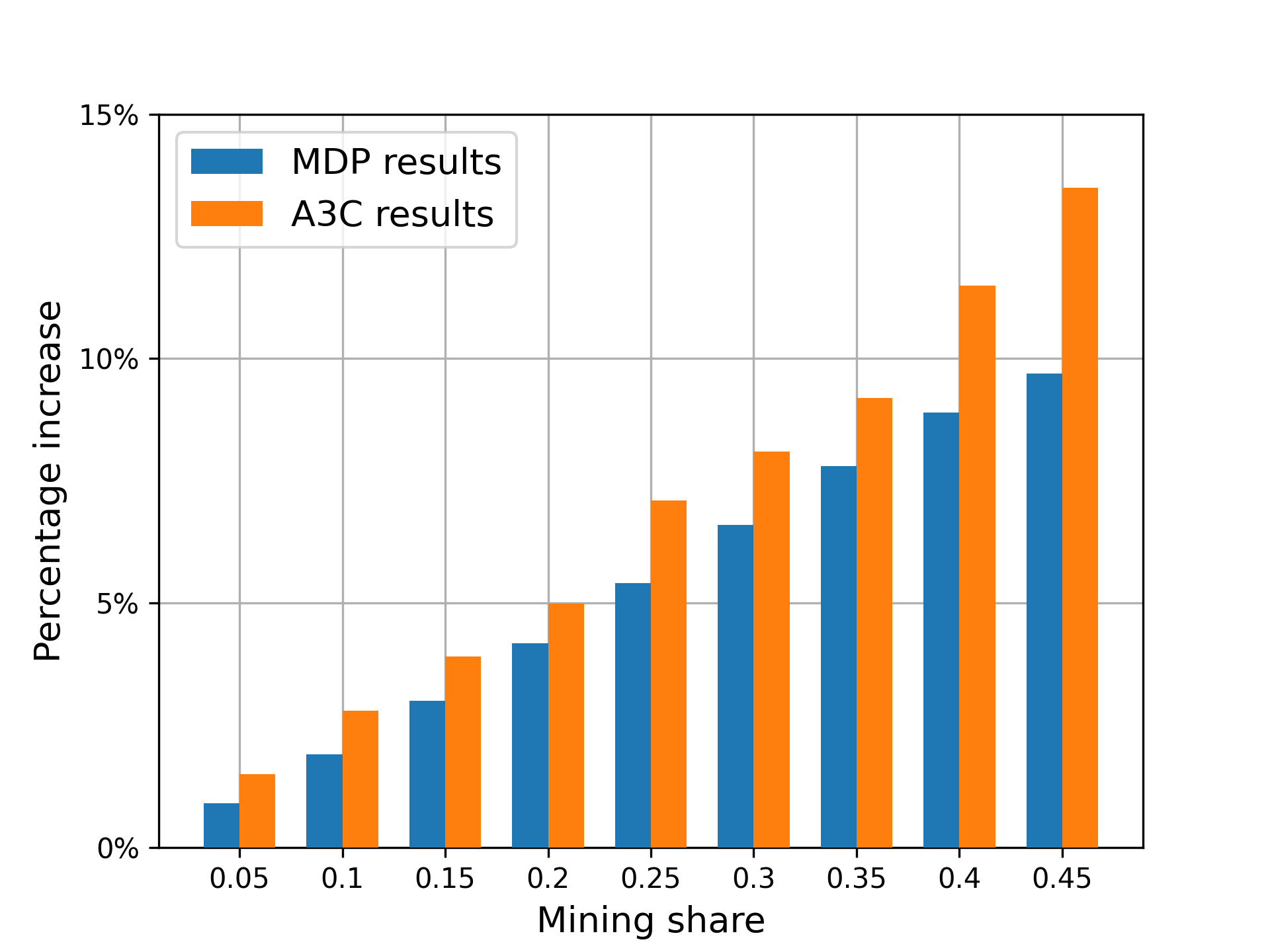}
    \caption{The percentage increase in time-averaged profit (December 18, 2023; protocol reward: $3.125$ BTC).}
    \label{fig:percentage_increase_reward_3_18_december}
\end{figure}

In Figure~\ref{fig:percentage_increase_reward_3_18_december}, the percentage increase in the adversary's time-averaged profit before a difficulty adjustment is shown as a function of the adversary's mining share. This figure is depicted based on the following configuration parameters:  
i) the protocol reward is set to $3.125$ BTC,  
ii) the mempool pattern observed on December 18, 2023, between 23:00 and 24:00 CET, has occurred,  
iii) the adversary's communication capability equals 1. 
Figure~\ref{fig:percentage_increase_reward_3_18_december} presents the results obtained from both our MDP implementation and A3C implementation. As shown in Figure~\ref{fig:percentage_increase_reward_3_18_december}, the strategy derived from the A3C implementation can outperform the strategy obtained from the MDP. Furthermore, as the adversary's mining share increases, the gap between the results obtained from these two approaches grows.
As can be seen in Figure~\ref{fig:percentage_increase_reward_3_18_december}, even with a protocol reward of $3.125$ BTC, on a day where the average transaction fee per block is comparable to the protocol reward (the average transaction fee per block on December 18, 2023, was $2.32$ BTC~\cite{explorer}), the selfish mining attack can lead to profitability immediately after the attack begins.

\begin{figure}[b]
    \centering
    \includegraphics[height=2in]{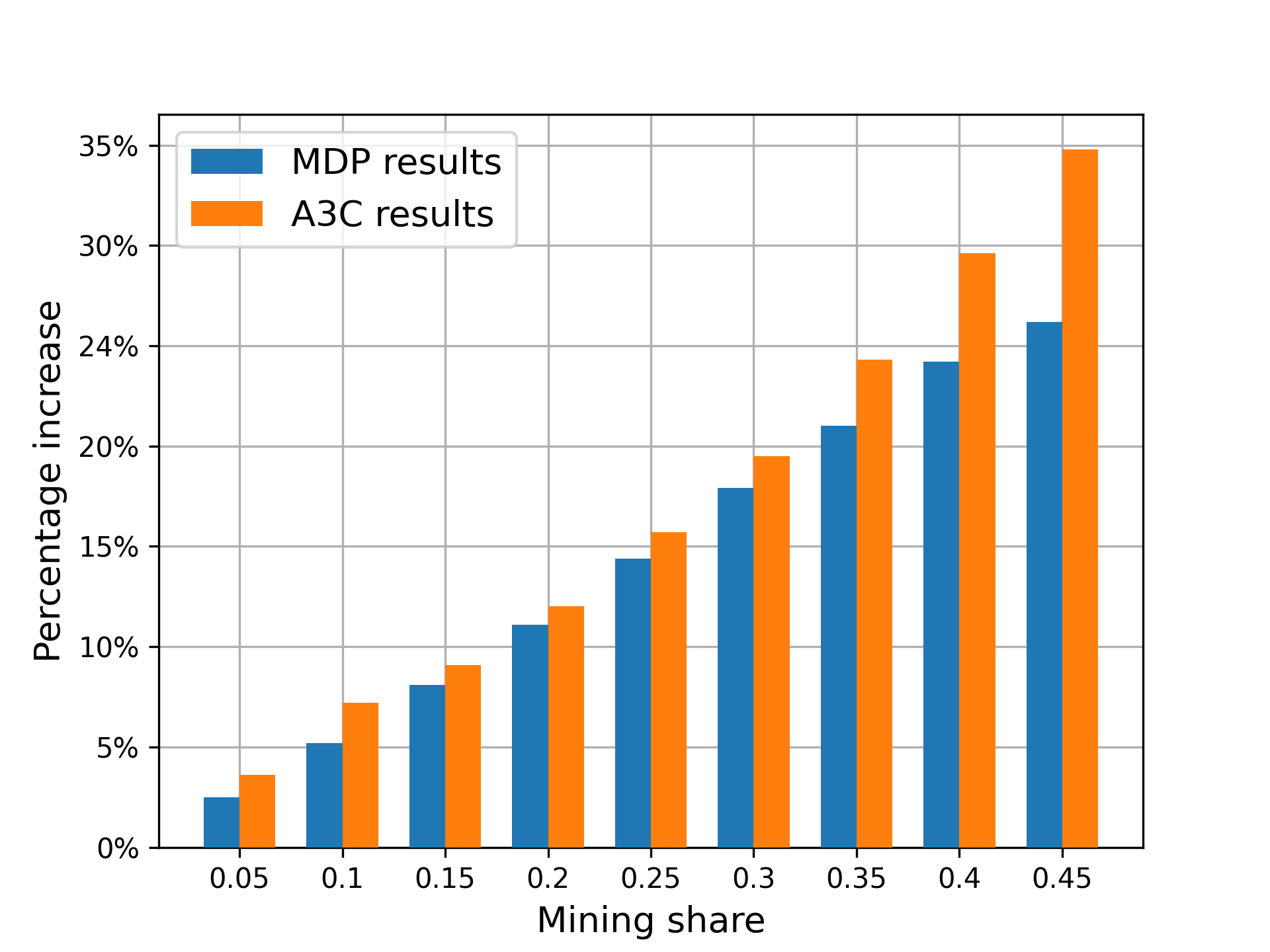}
    \caption{The percentage increase in time-averaged profit   (December 18, 2023; protocol reward: $0$ BTC).}
    \label{fig:percentage_increase_reward_0_18_december}
\end{figure}

Similarly, Figure~\ref{fig:percentage_increase_reward_0_18_december} illustrates the percentage increase in the adversary's time-averaged profit before a difficulty adjustment, using the same configuration parameters as in Figure~\ref{fig:percentage_increase_reward_3_18_december}, with the exception that the protocol reward is set to zero. By comparing Figures~\ref{fig:percentage_increase_reward_0_18_december} and~\ref{fig:percentage_increase_reward_3_18_december}, we observe that as the protocol reward decreases, deviations from the honest strategy become increasingly profitable, posing a potential threat to Bitcoin's progress in the near future.

\begin{figure}[b]
\centering
\includegraphics[height=2in]{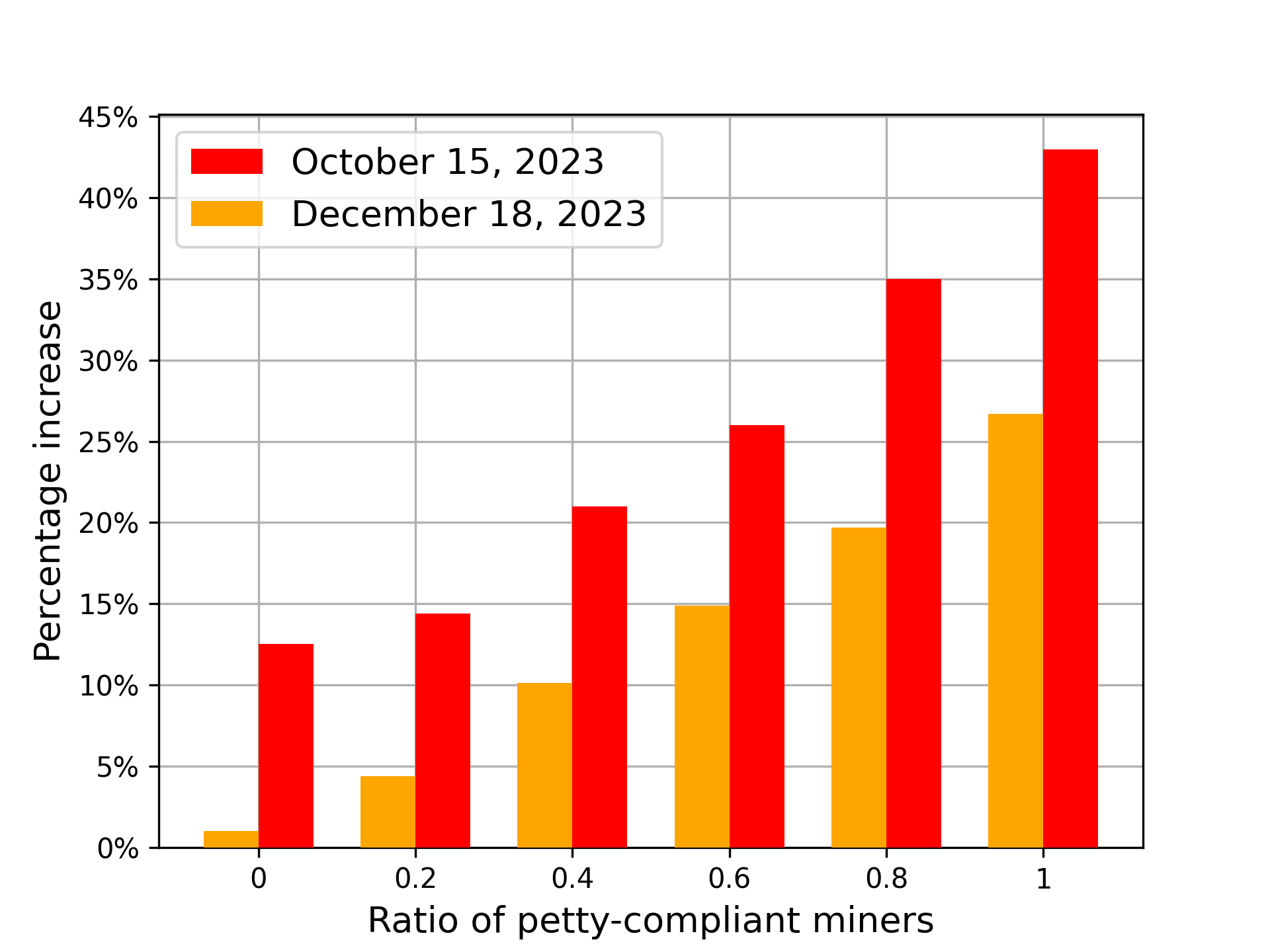}
\caption{The percentage increase in time-averaged profit (protocol reward: $0$ BTC).}
\label{fig:percentage_increase_reward_0_october}
\vspace{ - 15 pt}
\end{figure}

Although Figure~\ref{fig:percentage_increase_reward_0_18_december} shows that in the transaction-fee era, the same mempool pattern as occurred on December 18, 2023, can be concerning, there are some other days where the threat of adversarial attack is more alarming. Figure~\ref{fig:percentage_increase_reward_0_october} represents the percentage increase in profit before the difficulty adjustment from applying both selfish mining and undercutting attacks simultaneously under the mempool patterns of two different periods: December 18, 2023, between 23:00 and 24:00 CET (the same period as in Figures~\ref{fig:percentage_increase_reward_0_18_december} and~\ref{fig:percentage_increase_reward_3_18_december}), and October 15, 2023, between 8:00 and 9:00 CET. The figure is depicted as a function of different ratios of petty-compliant miners under the following configuration parameters: i) the protocol reward equals zero, ii) the adversarial mining share is $\frac{1}{3}$, and iii) the adversary has a normal communication capability of 50\%.
As can be seen in Figure~\ref{fig:percentage_increase_reward_0_october}, even if all the honest miners are altruistic, the deviation still yields a substantial immediate profit for the adversary under the mempool pattern of October 15, 2023. By increasing the ratio of petty-compliant miners in the system, the deviation can become even more profitable. The main reason for the higher profitability of mining attacks under the mempool pattern of October 15, 2023, compared to that of December 18, 2023, is the low level of base fee available in the mempool. Between 8:00 and 9:00 CET on October 15, 2023, the base fee was 1 sat/vByte~\cite{mempool}, implying that during this period, miners needed to fill at least some part of their blocks with transactions offering only 1 sat/vByte. In contrast, the base fee was 80 sat/vByte on December 18, 2023, between 23:00 and 24:00 CET~\cite{mempool}. This comparison shows that a higher base fee level for transactions available in the mempool can help mitigate mining attacks in the transaction-fee era.

In Appendix~\ref{appendix: selfish mining_before_DAM}, we present the selfish mining results obtained from our A3C implementation under the mempool behavior observed on two additional days.

\subsubsection{Single Block Undercutting Attack.}
As discussed in Section~\ref{sec:undercut_countinuous}, in our A3C-based implementation, we treat the undercut action as a continuous action, where the agent must not only decide whether to take this action but also determine the duration for which it should perform undercutting. Once the designated time has passed, the agent should stop undercutting the tip block and begin mining on top of it, as a sufficient number of transactions will have arrived in the system. To measure the profitability of the single-block undercutting attack, we have limited the adversary's possible action set, as outlined in Appendix~\ref{appendix:undercut_implementation}. Figure~\ref{fig:percentage_increase_undercut} shows the percentage increase in time-averaged profit achieved by performing the undercutting attack. The figure is depicted based on mempool patterns observed over different days between 8:00 and 9:00 CET, under the following configuration parameters:
i) the protocol reward is set to 0,
ii) the adversarial mining share and communication capability are set to $\frac{1}{3}$ and 50\%, respectively, and
iii) all non-adversarial miners are assumed to be petty-compliant.

\begin{figure}[t]
    \centering
    \includegraphics[height=2.7in]{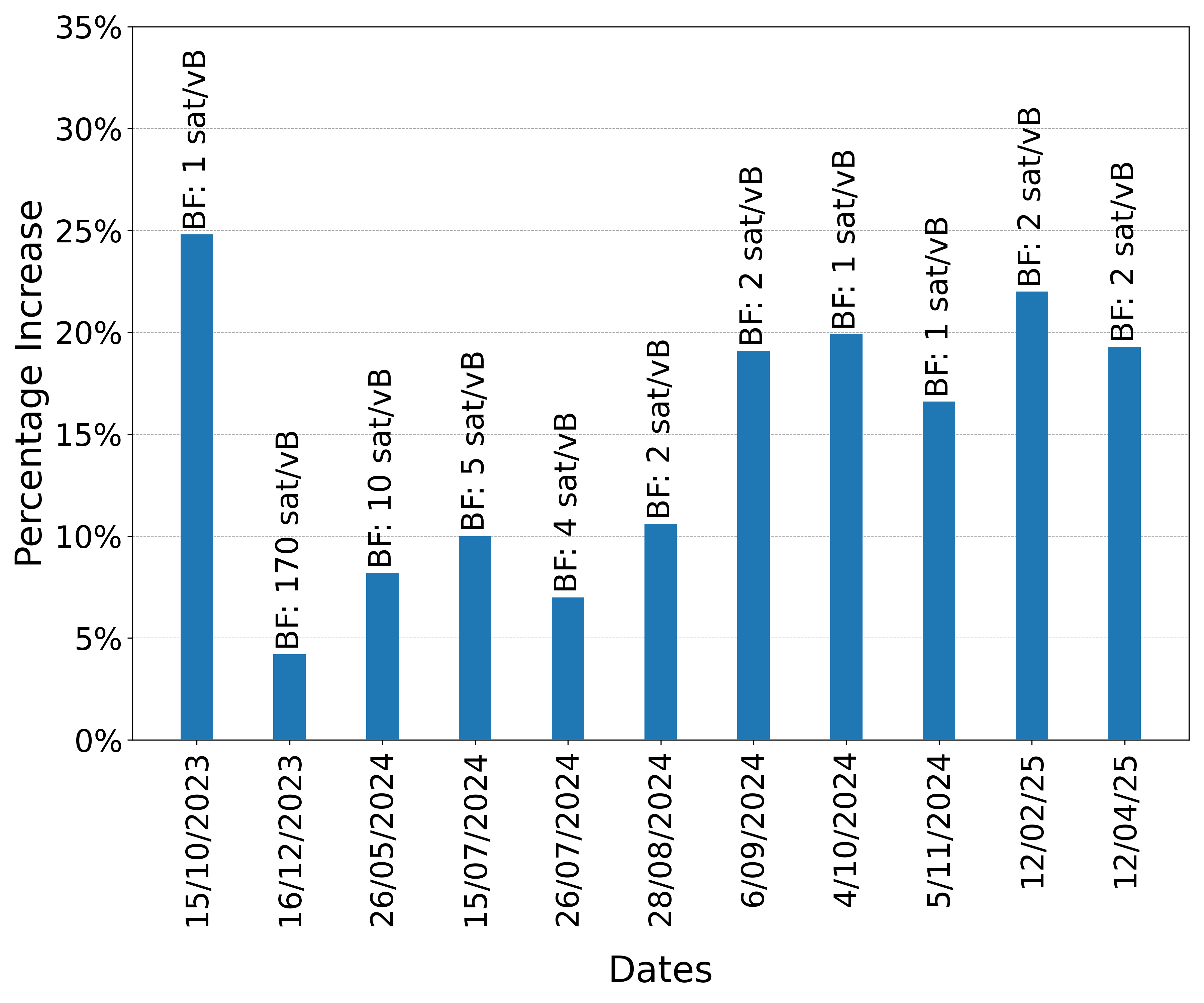}
    \caption{The profit percentage increase of undercutting. BF denotes the base fee.}
    \label{fig:percentage_increase_undercut}
\end{figure}

Lower base fees make the Bitcoin blockchain more vulnerable to undercutting. This is especially true when the tip block has a long generation time, as it stays exposed to attack longer while the mempool builds up enough fees.

\section{Conclusion}
\label{sec:conclusion}
In this paper, we analyzed the impact of volatility in block rewards on mining strategies in Bitcoin.
Under the fixed block reward model, which is primarily suitable for modeling Bitcoin's behavior during its first decade of existence, mining attacks such as selfish mining need to continue for at least one difficulty epoch (approximately two weeks) to become profitable. However, using historical Bitcoin statistics, we have demonstrated that under a volatile block reward model, when transaction fees become a major source of incentive, adversarial strategies can be profitable immediately, bypassing the initial loss period. The threat of deviant strategy profitability is concerning as the protocol reward diminishes to zero, and even under the current protocol reward, on days when average fees per block are comparable to the protocol reward.

Once assigning a transaction fee for their transactions, Bitcoin users should view their paid fee not only as a reward to compensate miners but also as a factor that can significantly impact the security of their valued blockchain. To keep Bitcoin secure, users should provide stable incentives for miners, giving them no, or at least minimal, motivation to deviate from the protocol.


\bibliographystyle{ACM-Reference-Format}
\balance
\bibliography{sample-base}


\begin{thebibliography}{28}


\ifx \showCODEN    \undefined \def \showCODEN     #1{\unskip}     \fi
\ifx \showDOI      \undefined \def \showDOI       #1{#1}\fi
\ifx \showISBNx    \undefined \def \showISBNx     #1{\unskip}     \fi
\ifx \showISBNxiii \undefined \def \showISBNxiii  #1{\unskip}     \fi
\ifx \showISSN     \undefined \def \showISSN      #1{\unskip}     \fi
\ifx \showLCCN     \undefined \def \showLCCN      #1{\unskip}     \fi
\ifx \shownote     \undefined \def \shownote      #1{#1}          \fi
\ifx \showarticletitle \undefined \def \showarticletitle #1{#1}   \fi
\ifx \showURL      \undefined \def \showURL       {\relax}        \fi
\providecommand\bibfield[2]{#2}
\providecommand\bibinfo[2]{#2}
\providecommand\natexlab[1]{#1}
\providecommand\showeprint[2][]{arXiv:#2}

\bibitem[Wer(2022)]%
        {Werlman_implementation}
 \bibinfo{year}{2022}\natexlab{}.
\newblock \bibinfo{title}{WeRLman Codes}.
\newblock
\newblock
\urldef\tempurl%
\url{https://github.com/roibarzur/pto-selfish-mining}
\showURL{%
\tempurl}


\bibitem[exp(2024)]%
        {explorer}
 \bibinfo{year}{2024}\natexlab{}.
\newblock \bibinfo{title}{Bitcoin Explore}.
\newblock
\newblock
\urldef\tempurl%
\url{https://btc.com/btc/blocks}
\showURL{%
\tempurl}


\bibitem[Bit(2024)]%
        {Bitcoin_wiki}
 \bibinfo{year}{2024}\natexlab{}.
\newblock \bibinfo{title}{Bitcoin Wiki}.
\newblock
\newblock
\urldef\tempurl%
\url{https://en.bitcoin.it/wiki/Bitcoin}
\showURL{%
\tempurl}


\bibitem[mar(2024)]%
        {market_cap}
 \bibinfo{year}{2024}\natexlab{}.
\newblock \bibinfo{title}{Coin Market Cap}.
\newblock
\newblock
\urldef\tempurl%
\url{https://coinmarketcap.com/}
\showURL{%
\tempurl}


\bibitem[gdp(2024)]%
        {gdp}
 \bibinfo{year}{2024}\natexlab{}.
\newblock \bibinfo{title}{GDP by Country}.
\newblock
\newblock
\urldef\tempurl%
\url{https://www.worldometers.info/gdp/gdp-by-country/}
\showURL{%
\tempurl}


\bibitem[Blo(2024)]%
        {Blockchair}
 \bibinfo{year}{2024}\natexlab{}.
\newblock \bibinfo{title}{Historical data on block transaction fees and generation times}.
\newblock
\newblock
\urldef\tempurl%
\url{https://blockchair.com/}
\showURL{%
\tempurl}


\bibitem[mem(2024)]%
        {mempool}
 \bibinfo{year}{2024}\natexlab{}.
\newblock \bibinfo{title}{Johoe's Bitcoin Mempool Statistics}.
\newblock
\newblock
\urldef\tempurl%
\url{https://jochen-hoenicke.de/queue/}
\showURL{%
\tempurl}


\bibitem[our(2025)]%
        {our_implementation}
 \bibinfo{year}{2025}\natexlab{}.
\newblock \bibinfo{title}{Implementation of our MDP-based and A3C-based tools to analyze Bitcoin mining under volatile block rewards}.
\newblock
\newblock
\urldef\tempurl%
\url{https://github.com/RoozbehSrnch/Bitcoin-Volatile-Reward}
\showURL{%
\tempurl}


\bibitem[Alzubaidi et~al\mbox{.}(2021)]%
        {alzubaidi2021review}
\bibfield{author}{\bibinfo{person}{Laith Alzubaidi}, \bibinfo{person}{Jinglan Zhang}, \bibinfo{person}{Amjad~J Humaidi}, \bibinfo{person}{Ayad Al-Dujaili}, \bibinfo{person}{Ye Duan}, \bibinfo{person}{Omran Al-Shamma}, \bibinfo{person}{Jos{\'e} Santamar{\'\i}a}, \bibinfo{person}{Mohammed~A Fadhel}, \bibinfo{person}{Muthana Al-Amidie}, {and} \bibinfo{person}{Laith Farhan}.} \bibinfo{year}{2021}\natexlab{}.
\newblock \showarticletitle{Review of deep learning: Concepts, CNN architectures, challenges, applications, future directions}.
\newblock \bibinfo{journal}{\emph{Journal of big Data}}  \bibinfo{volume}{8} (\bibinfo{year}{2021}), \bibinfo{pages}{1--74}.
\newblock


\bibitem[Bar-Zur et~al\mbox{.}(2023)]%
        {bar2022werlman}
\bibfield{author}{\bibinfo{person}{Roi Bar-Zur}, \bibinfo{person}{Ameer Abu-Hanna}, \bibinfo{person}{Ittay Eyal}, {and} \bibinfo{person}{Aviv Tamar}.} \bibinfo{year}{2023}\natexlab{}.
\newblock \showarticletitle{WeRLman: To Tackle Whale (Transactions), Go Deep (RL)}. In \bibinfo{booktitle}{\emph{2023 IEEE Symposium on Security and Privacy (SP)}}. IEEE, \bibinfo{pages}{93--110}.
\newblock


\bibitem[Brown-Cohen et~al\mbox{.}(2019)]%
        {brown2019formal}
\bibfield{author}{\bibinfo{person}{Jonah Brown-Cohen}, \bibinfo{person}{Arvind Narayanan}, \bibinfo{person}{Alexandros Psomas}, {and} \bibinfo{person}{S~Matthew Weinberg}.} \bibinfo{year}{2019}\natexlab{}.
\newblock \showarticletitle{Formal barriers to longest-chain proof-of-stake protocols}. In \bibinfo{booktitle}{\emph{Proceedings of the 2019 ACM Conference on Economics and Computation}}. \bibinfo{pages}{459--473}.
\newblock


\bibitem[Carlsten et~al\mbox{.}(2016)]%
        {carlsten2016instability}
\bibfield{author}{\bibinfo{person}{Miles Carlsten}, \bibinfo{person}{Harry Kalodner}, \bibinfo{person}{S~Matthew Weinberg}, {and} \bibinfo{person}{Arvind Narayanan}.} \bibinfo{year}{2016}\natexlab{}.
\newblock \showarticletitle{On the instability of bitcoin without the block reward}. In \bibinfo{booktitle}{\emph{Proceedings of the 2016 ACM SIGSAC conference on computer and communications security}}. \bibinfo{pages}{154--167}.
\newblock


\bibitem[Eyal and Sirer(2018)]%
        {eyal2018majority}
\bibfield{author}{\bibinfo{person}{Ittay Eyal} {and} \bibinfo{person}{Emin~G{\"u}n Sirer}.} \bibinfo{year}{2018}\natexlab{}.
\newblock \showarticletitle{Majority is not enough: Bitcoin mining is vulnerable}.
\newblock \bibinfo{journal}{\emph{Commun. ACM}} \bibinfo{volume}{61}, \bibinfo{number}{7} (\bibinfo{year}{2018}), \bibinfo{pages}{95--102}.
\newblock


\bibitem[Gong et~al\mbox{.}(2022)]%
        {gong2022towards}
\bibfield{author}{\bibinfo{person}{Tiantian Gong}, \bibinfo{person}{Mohsen Minaei}, \bibinfo{person}{Wenhai Sun}, {and} \bibinfo{person}{Aniket Kate}.} \bibinfo{year}{2022}\natexlab{}.
\newblock \showarticletitle{Towards overcoming the undercutting problem}. In \bibinfo{booktitle}{\emph{International Conference on Financial Cryptography and Data Security}}. Springer, \bibinfo{pages}{444--463}.
\newblock


\bibitem[Grunspan and P{\'e}rez-Marco(2018)]%
        {grunspan2018profitability}
\bibfield{author}{\bibinfo{person}{Cyril Grunspan} {and} \bibinfo{person}{Ricardo P{\'e}rez-Marco}.} \bibinfo{year}{2018}\natexlab{}.
\newblock \showarticletitle{On profitability of selfish mining}.
\newblock \bibinfo{journal}{\emph{arXiv preprint arXiv:1805.08281}} (\bibinfo{year}{2018}).
\newblock


\bibitem[Grunspan and P{\'e}rez-Marco(2023)]%
        {grunspan2023profit}
\bibfield{author}{\bibinfo{person}{Cyril Grunspan} {and} \bibinfo{person}{Ricardo P{\'e}rez-Marco}.} \bibinfo{year}{2023}\natexlab{}.
\newblock \showarticletitle{Profit lag and alternate network mining}. In \bibinfo{booktitle}{\emph{The International Conference on Mathematical Research for Blockchain Economy}}. Springer, \bibinfo{pages}{115--132}.
\newblock


\bibitem[Jia and Zhou(2022)]%
        {jia2022policy}
\bibfield{author}{\bibinfo{person}{Yanwei Jia} {and} \bibinfo{person}{Xun~Yu Zhou}.} \bibinfo{year}{2022}\natexlab{}.
\newblock \showarticletitle{Policy gradient and actor-critic learning in continuous time and space: Theory and algorithms}.
\newblock \bibinfo{journal}{\emph{Journal of Machine Learning Research}} \bibinfo{volume}{23}, \bibinfo{number}{275} (\bibinfo{year}{2022}), \bibinfo{pages}{1--50}.
\newblock


\bibitem[Konda and Tsitsiklis(1999)]%
        {konda1999actor}
\bibfield{author}{\bibinfo{person}{Vijay Konda} {and} \bibinfo{person}{John Tsitsiklis}.} \bibinfo{year}{1999}\natexlab{}.
\newblock \showarticletitle{Actor-critic algorithms}.
\newblock \bibinfo{journal}{\emph{Advances in neural information processing systems}}  \bibinfo{volume}{12} (\bibinfo{year}{1999}).
\newblock


\bibitem[Kostrikov(2018)]%
        {pytorchaaac}
\bibfield{author}{\bibinfo{person}{Ilya Kostrikov}.} \bibinfo{year}{2018}\natexlab{}.
\newblock \bibinfo{title}{PyTorch Implementations of Asynchronous Advantage Actor Critic}.
\newblock \bibinfo{howpublished}{\url{https://github.com/ikostrikov/pytorch-a3c}}.
\newblock


\bibitem[Mnih et~al\mbox{.}(2016)]%
        {mnih2016asynchronous}
\bibfield{author}{\bibinfo{person}{Volodymyr Mnih}, \bibinfo{person}{Adria~Puigdomenech Badia}, \bibinfo{person}{Mehdi Mirza}, \bibinfo{person}{Alex Graves}, \bibinfo{person}{Timothy Lillicrap}, \bibinfo{person}{Tim Harley}, \bibinfo{person}{David Silver}, {and} \bibinfo{person}{Koray Kavukcuoglu}.} \bibinfo{year}{2016}\natexlab{}.
\newblock \showarticletitle{Asynchronous methods for deep reinforcement learning}. In \bibinfo{booktitle}{\emph{International conference on machine learning}}. PMLR, \bibinfo{pages}{1928--1937}.
\newblock


\bibitem[Nakamoto(2008)]%
        {nakamoto2008bitcoin}
\bibfield{author}{\bibinfo{person}{Satoshi Nakamoto}.} \bibinfo{year}{2008}\natexlab{}.
\newblock \showarticletitle{Bitcoin: A peer-to-peer electronic cash system}.
\newblock \bibinfo{journal}{\emph{Satoshi Nakamoto}} (\bibinfo{year}{2008}).
\newblock


\bibitem[Negy et~al\mbox{.}(2020)]%
        {negy2020selfish}
\bibfield{author}{\bibinfo{person}{Kevin~Alarc{\'o}n Negy}, \bibinfo{person}{Peter~R Rizun}, {and} \bibinfo{person}{Emin~G{\"u}n Sirer}.} \bibinfo{year}{2020}\natexlab{}.
\newblock \showarticletitle{Selfish mining re-examined}. In \bibinfo{booktitle}{\emph{International Conference on Financial Cryptography and Data Security}}. Springer, \bibinfo{pages}{61--78}.
\newblock


\bibitem[Sapirshtein et~al\mbox{.}(2017)]%
        {sapirshtein2017optimal}
\bibfield{author}{\bibinfo{person}{Ayelet Sapirshtein}, \bibinfo{person}{Yonatan Sompolinsky}, {and} \bibinfo{person}{Aviv Zohar}.} \bibinfo{year}{2017}\natexlab{}.
\newblock \showarticletitle{Optimal selfish mining strategies in bitcoin}. In \bibinfo{booktitle}{\emph{Financial Cryptography and Data Security: 20th International Conference, FC 2016, Christ Church, Barbados, February 22--26, 2016, Revised Selected Papers 20}}. Springer, \bibinfo{pages}{515--532}.
\newblock


\bibitem[Sarenche et~al\mbox{.}(2024a)]%
        {sarenche2024deep}
\bibfield{author}{\bibinfo{person}{Roozbeh Sarenche}, \bibinfo{person}{Svetla Nikova}, {and} \bibinfo{person}{Bart Preneel}.} \bibinfo{year}{2024}\natexlab{a}.
\newblock \showarticletitle{Deep selfish proposing in longest-chain proof-of-stake protocols}. In \bibinfo{booktitle}{\emph{International Conference on Financial Cryptography and Data Security}}. Springer, \bibinfo{pages}{24--40}.
\newblock


\bibitem[Sarenche et~al\mbox{.}(2024b)]%
        {sarenche2024selfish}
\bibfield{author}{\bibinfo{person}{Roozbeh Sarenche}, \bibinfo{person}{Ren Zhang}, \bibinfo{person}{Svetla Nikova}, {and} \bibinfo{person}{Bart Preneel}.} \bibinfo{year}{2024}\natexlab{b}.
\newblock \showarticletitle{Selfish Mining Time-Averaged Analysis in Bitcoin: Is Orphan Reporting an Effective Countermeasure?}
\newblock \bibinfo{journal}{\emph{IEEE Transactions on Information Forensics and Security}} (\bibinfo{year}{2024}).
\newblock


\bibitem[Silver et~al\mbox{.}(2017)]%
        {silver2017mastering}
\bibfield{author}{\bibinfo{person}{David Silver}, \bibinfo{person}{Julian Schrittwieser}, \bibinfo{person}{Karen Simonyan}, \bibinfo{person}{Ioannis Antonoglou}, \bibinfo{person}{Aja Huang}, \bibinfo{person}{Arthur Guez}, \bibinfo{person}{Thomas Hubert}, \bibinfo{person}{Lucas Baker}, \bibinfo{person}{Matthew Lai}, \bibinfo{person}{Adrian Bolton}, {et~al\mbox{.}}} \bibinfo{year}{2017}\natexlab{}.
\newblock \showarticletitle{Mastering the game of go without human knowledge}.
\newblock \bibinfo{journal}{\emph{nature}} \bibinfo{volume}{550}, \bibinfo{number}{7676} (\bibinfo{year}{2017}), \bibinfo{pages}{354--359}.
\newblock


\bibitem[Tsabary and Eyal(2018)]%
        {tsabary2018gap}
\bibfield{author}{\bibinfo{person}{Itay Tsabary} {and} \bibinfo{person}{Ittay Eyal}.} \bibinfo{year}{2018}\natexlab{}.
\newblock \showarticletitle{The gap game}. In \bibinfo{booktitle}{\emph{Proceedings of the 2018 ACM SIGSAC conference on Computer and Communications Security}}. \bibinfo{pages}{713--728}.
\newblock


\bibitem[Zur et~al\mbox{.}(2020)]%
        {zur2020efficient}
\bibfield{author}{\bibinfo{person}{Roi~Bar Zur}, \bibinfo{person}{Ittay Eyal}, {and} \bibinfo{person}{Aviv Tamar}.} \bibinfo{year}{2020}\natexlab{}.
\newblock \showarticletitle{Efficient MDP analysis for selfish-mining in blockchains}. In \bibinfo{booktitle}{\emph{Proceedings of the 2nd ACM Conference on Advances in Financial Technologies}}. \bibinfo{pages}{113--131}.
\newblock


\end{thebibliography}

\clearpage
\appendix

\section{Mining Attacks} \label{appendix:mining attacks}
\subsection{Selfish Mining} \label{appendix:selfish mining}
In the selfish mining attack, introduced by Eyal et al. in~\cite{eyal2018majority}, a selfish miner withholds a newly mined block \( B_1 \) instead of immediately publishing it to the network. By keeping \( B_1 \) secret, the selfish miner creates a private fork one block ahead of the public chain. While other miners continue mining on the public chain, the selfish miners work to extend their private forks.
If a selfish miner mines a second block \( B_2 \) on top of its private fork, its chain gains a two-block lead over the public chain. When an honest miner mines a new block \( B_3 \) on top of the public chain, the selfish miner can reveal its secret fork. Since the selfish miner's chain, extended by blocks \( B_1 \) and \( B_2 \), is longer compared to the public chain, extended by block \( B_3 \), the selfish miner’s chain becomes the canonical chain. Consequently, block \( B_3 \), mined by honest miners, becomes orphaned and is excluded from the canonical chain. 
However, it is not necessarily the case that the selfish miner can always achieve a two-block lead. If the selfish miner keeps its block secret and honest miners mine the next block, a fork race occurs between two forks of the same height, with only one fork becoming the canonical chain. The probability of winning this race depends on how quickly the selfish miner can propagate its blocks through the network, encouraging other miners to build on top of the selfish miner's fork. This attack can increase the selfish miner's block reward relative to other miners. 
In~\cite{sapirshtein2017optimal}, the authors implemented an MDP-based tool to determine the optimal selfish mining strategy that a selfish miner can follow in a setting where the reward for all blocks is the same, referred to as the fixed block reward model. In~\cite{grunspan2018profitability}, the authors argued that the block ratio is not a good benchmark for analyzing selfish mining profitability and that one should use the time-averaged profit definition for this analysis. They demonstrated that selfish mining during the first difficulty epoch cannot be more profitable than honest mining, resulting in an initial loss period for selfish mining. However, from the subsequent epoch onward, the attack becomes profitable as the difficulty decreases due to the selfish mining attack. To reduce the initial loss period, the authors in~\cite{negy2020selfish} introduced intermittent selfish mining, where the selfish miner alternates between one epoch of selfish mining and one epoch of honest mining. The authors in~\cite{grunspan2023profit} introduced the term \textit{profit lag} to describe the point at which the attack consistently becomes more profitable than honest mining in order to analyze the initial period of loss. They argued that although intermittent selfish mining's profitability may exceed that of honest mining at certain points in the early cycles of the attack, its profit lag is actually longer than the profit lag of standard selfish mining. In~\cite {sarenche2024selfish}, the authors introduced smart intermittent selfish mining, where the selfish miner performs selfish mining for half of each block. Smart intermittent selfish mining can outperform the intermittent selfish mining introduced in~\cite{negy2020selfish} and reduces the profit lag. The selfish mining attack is not specific to the Bitcoin network and can be applied to any longest-chain-based blockchain. As analyzed in~\cite{brown2019formal, sarenche2024deep}, selfish mining can be even more destructive in the context of longest-chain Proof-of-Stake protocols due to factors such as proposer predictability and the nothing-at-stake phenomenon.

\subsection{Undercutting} \label{appendix:undercutting_definition}
The undercutting attack, introduced by Carlsten et al. in~\cite{carlsten2016instability}, can be profitable in scenarios where a significant portion of the mining power is petty-compliant. Petty-compliant miners are those who may deviate from the honest mining strategy to earn higher profits. As the protocol reward approaches zero, the undercutting attack poses an increasingly serious threat to the stability of the Bitcoin network.
In this attack, when an undercutter becomes aware of a new block $B_1$ that extends the longest chain, it may choose not to accept it. If $B_1$ contains the majority of the transaction fees available in the mempool, the undercutter might be disincentivized from mining a relatively empty block on top of it. Instead, the undercutter may opt to mine on the parent of $B_1$ to create a competing block. The key aspect of the undercutting attack is that the undercutter should include fewer transaction fees in the competing block compared to $B_1$, leaving a substantial portion of the fees in the mempool for the future block. This strategy of generously leaving some transactions is crucial for the attack's success because if the undercutter successfully mines the competing block $B_2$, a fork will occur in the network between blocks $B_1$ and $B_2$. In the event of such a fork, petty-compliant miners would prefer to mine on top of the block that offers the most available transaction fees, rather than the oldest block—--in this case, the competing block $B_2$. Attracting the mining power of rational miners to $B_2$ increases the undercutter's chances of winning the fork. If the next block is mined on top of the competing block $B_2$, the undercutter has successfully undercut block $B_1$. Carlsten et al. find that undercutting can become the equilibrium strategy for miners, leading to instability as miners undercut each other. However, this result is based on a model disregarding the block size limit. In paper~\cite{gong2022towards}, the authors have analyzed the undercutting attack with consideration of the block size limit and derived closed-form conditions on the pending transaction set in the mempool that make undercutting profitable. They have also proposed an alternative transaction selection rule to counter undercutting; instead of fitting all available transactions into a block, miners include only a portion of the pending transaction fees to mitigate the risk of being undercut.

\section{Non-Predictable WeRLman}\label{appendix:non_predictable_werlman}
In order to modify the original WeRLman environment to a non-predictable version, we updated the function \\$\texttt{tryGetWithAndWithoutNewFee}$ in the file \\$\texttt{bitcoin\_simplified\_fee\_model.py}$, which is available in the repository referenced in~\cite{Werlman_implementation}, as presented in Algorithm~\ref{alg:tryGetWithAndWithoutNewFee}.

\begin{algorithm}[ht]
\caption{Updated function \texttt{tryGetWithAndWithoutNewFee}}
\label{alg:tryGetWithAndWithoutNewFee}
\begin{algorithmic}[1]
\State \textbf{Input:} Current state $CS$ (self), and previous state $PS$.
\State \textbf{Output:} Updated states $new\_with\_fee$, $new\_without\_fee$.
\State $current\_max\_fork \gets \max(CS.h, CS.a)$
\State $prev\_max\_fork \gets \max(PS.h, PS.a)$
\State Create a copy of the current state: $new\_with\_fee \gets \text{copy}(CS)$
\State Create another copy of the current state: $new\_without\_fee \gets \text{copy}(CS)$
\If{$prev\_max\_fork < current\_max\_fork$ and $CS.pool < max\_pool$}
    \State $new\_with\_fee.pool \gets new\_with\_fee.pool + 1$
    \State $new\_honest \gets \text{True}$ if $(CS.h - PS.h) > 0$ else $\text{False}$
    \If{$new\_honest$}
        \State $with\_fee \gets 1$ if $CS.T\_h = PS.T\_h$ else $0$
        \State $new\_with\_fee.T\_h \gets new\_with\_fee.T\_h + with\_fee$
    \Else
        \If{$a \leq h$}
            \State $with\_fee \gets 1$ if $CS.T\_a = PS.T\_a$ else $0$
            \State $new\_with\_fee.T\_a \gets new\_with\_fee.T\_a + with\_fee$
        \Else
            \State $new\_with\_fee.L[-1] \gets 1$
        \EndIf
    \EndIf
\EndIf
\State \textbf{Return} $new\_with\_fee$, $new\_without\_fee$
\end{algorithmic}
\end{algorithm}



\section{Adversarial Strategies in the Original and Non-predictable WeRLman Environments} \label{appendix:adversarial_strategies}
Let $\alpha$, $g$, $p$, and $F$ denote the adversarial mining share, the adversarial communication capability, the whale transaction probability, and the extra fee of a whale transaction, respectively. We denote by $N_\mathcal{A}$, $N_\mathcal{H}$, and $R_\mathcal{A}$ the normalized average number of canonical adversarial blocks, the normalized average number of canonical honest blocks, and the normalized average adversarial block reward. The time-averaged profit of strategy $\pi$ can be calculated as $\texttt{Profit}(\pi) = \frac{R_\mathcal{A}{(\pi)}}{N_\mathcal{A}{(\pi)}+N_\mathcal{H}{(\pi)}}$. Note that the time-averaged profit of the adversary if mining honestly is equal to $\texttt{Profit}(\pi^\mathcal{H}) = \alpha (1+Fp)$.

We introduce three strategies: $\pi_1^{\text{WeRLman}}$, $\pi_1^{\text{Non-predictable}}$, and $\pi_2^{\text{Non-predictable}}$. Each strategy specifies the adversary's action in each state of the fork race. The adversary takes these actions upon the event of block mining. A newly mined block is adversarial with probability $\alpha$ and honest with probability $1 - \alpha$. 
At each event of mining a block at a \emph{new} height, a transaction is sampled: it is a whale transaction with probability $p$ and a normal transaction with probability $1 - p$. Note that in the original WeRLman environment (used by strategy $\pi_1^{\text{WeRLman}}$), the sampled transaction is added to the mempool and included in the next block. In contrast, in the non-predictable WeRLman environment (used by strategies $\pi_1^{\text{Non-predictable}}$ and $\pi_2^{\text{Non-predictable}}$), the sampled transaction is assumed to have already been included in the newly mined block. In a same-height fork race between a published adversarial fork and an honest fork, in addition to the adversarial mining power, a fraction $g$ of honest miners will also mine on top of the adversarial fork.

We use four actions to describe the strategies: \texttt{adopt}, denoting acceptance of the honest fork as the canonical chain; \texttt{wait}, denoting continued mining on top of the adversarial fork; \texttt{publish}, denoting the release of a longer adversarial fork; and \texttt{match}, denoting the release of an adversarial fork at the same height as the honest fork immediately after the publication of a new honest block.

\subsection{Strategy $\pi_1^\text{WeRLman}$} \label{appendix:strategy_werlman_1}
We use a tuple $(n_\mathcal{A}, n_\mathcal{H}, \texttt{whale}^\texttt{pool}, \texttt{whale}^\texttt{block})$ to represent a state during a fork race. Here, $n_\mathcal{A}$ and $n_\mathcal{H}$ denote the number of blocks in the adversarial and honest forks, respectively. The variable $\texttt{whale}^\texttt{pool}$ indicates the type of transaction currently in the pool w.r.t. the adversarial fork, which can be included in the next block: $\texttt{whale}^\texttt{pool} = 1$ if there is a whale transaction, and $\texttt{whale}^\texttt{pool} = 0$ if there is a normal transaction. $\texttt{whale}^\texttt{block}$ indicates the type of transaction included in the adversarial block: $\texttt{whale}^\texttt{block} = 1$ if the block includes a whale transaction, and $\texttt{whale}^\texttt{block} = 0$ if it includes a normal transaction.

We define four states as follows:
\begin{itemize}[leftmargin = *]
    \item State $S_{0,0}: (n_\mathcal{A}=0, n_\mathcal{H}=0, \texttt{whale}^\texttt{pool} = 0, \texttt{whale}^\texttt{block} = \bot)$. This state denotes that both the adversary and honest miners are mining on top of the same canonical chain, and the existing transaction in the mempool is normal.
    
    \item State $S_{0,0}': (n_\mathcal{A}=0, n_\mathcal{H}=0, \texttt{whale}^\texttt{pool} = 1, \texttt{whale}^\texttt{block} = \bot)$. This state denotes that both the adversary and honest miners are mining on top of the same canonical chain, but the existing transaction in the mempool is a whale transaction.
    
    \item State $S_{1,0}: (n_\mathcal{A}=1, n_\mathcal{H}=0, \texttt{whale}^\texttt{pool} = 1, \texttt{whale}^\texttt{block} = 0)$. This state denotes that the secret adversarial fork contains a single block with a normal transaction, while a whale transaction remains in the mempool to be included in the next block.
    
    \item State $S_{1,1}: (n_\mathcal{A}=1, n_\mathcal{H}=1, \texttt{whale}^\texttt{pool} = 1, \texttt{whale}^\texttt{block} = 0)$. This state denotes that there is a fork race between the adversarial and honest forks, both with a height of one block. The adversarial block includes a normal transaction, while the honest block includes a whale transaction. The next block in the adversarial fork can steal the whale transaction from the honest block. Note that upon transferring to this state, a new transaction is not sampled, as the same height block already exists.
\end{itemize}

Strategy $\pi_1^\text{WeRLman}$ is introduced in Table~\ref{tab:strategy_werlman_1}.

\begin{table*}[th]
\caption{Strategy $\pi_1^\text{WeRLman}$.}
\centering
\renewcommand{\arraystretch}{1.2}
\begin{tabular}{|c|c|c|c|c|}
\hline
\textbf{Initial State} & \textbf{Action} & \textbf{Probability} & \textbf{Final State} & \textbf{Reward} \\
\hline

\multirow{4}{*}{$S_{0,0}$} 
& \multirow{4}{*}{%
\begin{tabular}[c]{@{}l@{}}\texttt{wait}. \\ If the next block is honest, then switch to \texttt{adopt}.\\ If the next block is adversarial and the next transaction \\ is normal,  then switch to \texttt{publish}.\end{tabular}
}
& $\alpha p$ & $S_{1,0}$ & 0 \\
\cline{3-5}
& & $\alpha (1-p)$ & $S_{0,0}$ & 1 \\
\cline{3-5}
& & $(1-\alpha)p$ & $S_{0,0}'$ & 0 \\
\cline{3-5}
& & $(1-\alpha)(1-p)$ & $S_{0,0}$ & 0 \\
\hline

\multirow{4}{*}{$S_{0,0}'$} 
& \multirow{4}{*}{%
\begin{tabular}[c]{@{}l@{}}\texttt{wait}. \\ If next block is honest, then switch to \texttt{adopt}.\\ If the next block is adversarial, then switch to \texttt{publish}.\end{tabular}
}
& $\alpha p$ & $S_{0,0}'$ & $1+F$ \\
\cline{3-5}
& & $\alpha (1-p)$ & $S_{0,0}$ & $1+F$ \\
\cline{3-5}
& & $(1-\alpha)p$ & $S_{0,0}'$ & 0 \\
\cline{3-5}
& & $(1-\alpha)(1-p)$ & $S_{0,0}$ & 0 \\
\hline

\multirow{3}{*}{$S_{1,0}$} 
& \multirow{3}{*}{%
\begin{tabular}[c]{@{}l@{}}\texttt{wait}. \\ If the next block is adversarial, \texttt{wait} until $n_\mathcal{A} - n_\mathcal{H} = 1$, \\ then switch to \texttt{publish}.\end{tabular}
}
& $\alpha p$ & $S_{0,0}'$ & $2+F+(1+pF)\frac{\alpha}{1-2\alpha}$ \\
\cline{3-5}
& & $\alpha (1-p)$ & $S_{0,0}'$ & $2+F+(1+pF)\frac{\alpha}{1-2\alpha}$ \\
\cline{3-5}
& & $(1-\alpha)$ & $S_{1,1}$ & 0 \\
\hline

\multirow{6}{*}{$S_{1,1}$} 
& \multirow{6}{*}{\texttt{match}}
& $\alpha p$ & $S_{0,0}'$ & $2+F$ \\
\cline{3-5}
& & $\alpha (1-p)$ & $S_{0,0}$ & $2+F$ \\
\cline{3-5}
& & $(1-\alpha)g p$ & $S_{0,0}'$ & 1 \\
\cline{3-5}
& & $(1-\alpha)g(1-p)$ & $S_{0,0}$ & 1 \\
\cline{3-5}
& & $(1-\alpha)(1-g)p$ & $S_{0,0}'$ & 0 \\
\cline{3-5}
& & $(1-\alpha)(1-g)(1-p)$ & $S_{0,0}$ & 0 \\
\hline

\end{tabular}
\label{tab:strategy_werlman_1}
\end{table*}

Let $P_{0,0}$, $P_{0,0}'$, $P_{1,0}$, $P_{1,1}$ denote the stationary probability of being in states $S_{0,0}$, $S_{0,0}'$, $S_{1,0}$, $S_{1,1}$, respectively. We have:
\begin{equation}
    \begin{split}
        & P_{0,0} = \frac{1-p}{1+p(1-p)\alpha(1-\alpha)}, \quad 
        P_{1,0} = \frac{\alpha p(1-p)}{1+p(1-p)\alpha(1-\alpha)}, \quad \\
        & P_{1,1} = \frac{p(1-p)\alpha(1-\alpha)}{1+p(1-p)\alpha(1-\alpha)}, \quad
        P_{0,0}' = 1-P_{0,0}-P_{1,0}-P_{1,1}.
    \end{split}
\end{equation}

The normalized numbers of canonical honest blocks, canonical adversarial blocks, and the normalized adversarial block reward achieved under strategy $\pi_1^\text{WeRLman}$ can be obtained as follows:

\begin{equation}
    \begin{split}
        N_\mathcal{H} =  & P_{0,0}\left(1-\alpha\right) + 2P_{1,1}\left(1-\alpha\right)\left(1-g\right) \\
        & + P_{1,1}\left(1-\alpha\right)g + P_{0,0}'\left(1-\alpha\right), \\
        N_\mathcal{A} =  & P_{0,0} \alpha \left(1-p\right) + \left(2+\frac{\alpha}{1-2\alpha}\right) P_{1,0} \alpha + 2P_{1,1} \alpha \\ 
        & + P_{1,1} \left(1-\alpha\right)g + P_{0,0}' \alpha, \\
        R_\mathcal{A} =  & P_{0,0} \alpha \left(1-p\right) + P_{0,0}' \alpha \left(1+F\right) + P_{1,1} \alpha \left(2+F\right) + P_{1,1} \left(1-\alpha\right)g \\
        &+ P_{1,0} \alpha \left(2+F\right) + P_{1,0} \alpha \frac{\alpha}{1-2\alpha}\left(1+pF\right) \enspace.
    \end{split}
\end{equation}

\subsection{Strategy $\pi_1^\text{Non-predictable}$}
We represent a state during a fork race using the following tuple: $(n_\mathcal{A}, n_\mathcal{H}, \texttt{whale}^\texttt{block})$. The descriptions of these parameters are the same as those provided in Appendix~\ref{appendix:strategy_werlman_1}. We define three states as follows:
\begin{itemize}[leftmargin = *]
    \item State $S_{0,0}: (n_\mathcal{A}=0, n_\mathcal{H}=0, \texttt{whale}^\texttt{block} = \bot)$. This state denotes that both the adversary and honest miners are mining on top of the same canonical chain.
    
    \item State $S_{1,0}: (n_\mathcal{A}=1, n_\mathcal{H}=0, \texttt{whale}^\texttt{block} = 0)$. This state denotes that the secret adversarial fork contains a single block with a normal transaction.
    
    \item State $S_{1,1}: (n_\mathcal{A}=1, n_\mathcal{H}=1, \texttt{whale}^\texttt{block} = 0)$. This state denotes that there is a fork race between the adversarial and honest forks, both with a height of one block. The adversarial block includes a normal transaction. Note that upon transferring to this state, a new transaction is not sampled, as the same height block already exists.
\end{itemize}

Strategy $\pi_1^\text{Non-predictable}$ is presented in Table~\ref{tab:strategy_nonpredictable_1}.

\begin{table*}[t!]
\centering
\caption{Strategy $\pi_1^\text{Non-predictable}$.}
\renewcommand{\arraystretch}{1.2}
\begin{tabular}{|c|c|c|c|c|}
\hline
\textbf{Initial State} & \textbf{Action} & \textbf{Probability} & \textbf{Final State} & \textbf{Reward} \\
\hline

\multirow{3}{*}{$S_{0,0}$} 
& \multirow{3}{*}{%
\begin{tabular}[c]{@{}l@{}}\texttt{wait}. \\ If the next block is honest, then switch to \texttt{adopt}.\\ If the next block is adversarial with a whale, then \texttt{publish}. \end{tabular}
}
& $\alpha p$ & $S_{0,0}$ & $1+F$ \\
\cline{3-5}
& & $\alpha (1-p)$ & $S_{1,0}$ & 0 \\
\cline{3-5}
& & $(1-\alpha)$ & $S_{0,0}$ & 0 \\
& &  &  & \\
\hline

\multirow{3}{*}{$S_{1,0}$} 
& \multirow{3}{*}{%
\begin{tabular}[c]{@{}l@{}}\texttt{wait}. \\ If the next block is adversarial, \texttt{wait} until $n_\mathcal{A} - n_\mathcal{H} = 1$, \\ then  switch to \texttt{publish}. \end{tabular}
}
& $\alpha$ & $S_{0,0}$ & $1+(1+pF)\left(1+\frac{\alpha}{1-2\alpha}\right)$ \\
\cline{3-5}
& & $(1-\alpha)$ & $S_{1,1}$ & 0 \\
& &  &  & \\
\hline

\multirow{4}{*}{$S_{1,1}$} 
& \multirow{4}{*}{\texttt{match}}
& $\alpha p$ & $S_{0,0}$ & $2+F$ \\
\cline{3-5}
& & $\alpha(1-p)$ & $S_{0,0}$ & $2$ \\
\cline{3-5}
& & $(1-\alpha)g$ & $S_{0,0}$ & 1 \\
\cline{3-5}
& & $(1-\alpha)(1-g)$ & $S_{0,0}$ & 0 \\
\hline

\end{tabular}
\label{tab:strategy_nonpredictable_1}
\end{table*}

Let $P_{0,0}$, $P_{1,0}$, $P_{1,1}$ denote the stationary probability of being in states $S_{0,0}$, $S_{1,0}$, and $S_{1,1}$, respectively. We have:
\begin{equation}
    \begin{split}
        & P_{0,0} = \frac{1}{1 + \left(1-p\right)\alpha\left(2-\alpha\right)}, \quad 
        P_{1,0} = \alpha\left(1-p\right)P_{0,0}, \quad \\
        & P_{1,1} = \left(1-\alpha\right)P_{1,0}.
    \end{split}
\end{equation}

The normalized numbers of canonical honest blocks, canonical adversarial blocks, and the normalized adversarial block reward achieved under strategy $\pi_1^\text{Non-predictable}$ can be obtained as follows:
\begin{equation}
    \begin{split}
        N_\mathcal{H} = &P_{0,0}\left(1-\alpha\right) + P_{1,1}\left(1-\alpha\right)g + 2P_{1,1}\left(1-\alpha\right)\left(1-g\right), \\
        N_\mathcal{A} = & P_{0,0}\alpha p + P_{1,0}\alpha\left(2+\frac{\alpha}{1-2\alpha}\right) + 2P_{1,1}\alpha + P_{1,1}\left(1-\alpha\right)g, \\
        R_\mathcal{A} = & P_{0,0}\alpha p\left(1+F\right) + P_{1,0}\alpha\left(1+\left(1+pF\right)\left(1+\frac{\alpha}{1-2\alpha}\right)\right) \\ &+ P_{1,1}\left(1-\alpha\right)g + P_{1,1}\alpha\left(2+pF\right) \enspace .
    \end{split}
\end{equation}

\subsection{Strategy $\pi_2^\text{Non-predictable}$}
We represent a state during a fork race using the following tuple: $(n_\mathcal{A}, n_\mathcal{H}, \texttt{whale}^\texttt{block}_\mathcal{H})$. Here, $n_\mathcal{A}$ and $n_\mathcal{H}$ denote the number of blocks in the adversarial and honest forks, respectively. $\texttt{whale}^\texttt{block}_\mathcal{H}$ indicates the type of transaction included in the first block of the honest fork: $\texttt{whale}^\texttt{block}_\mathcal{H} = 1$ if the block includes a whale transaction, and $\texttt{whale}^\texttt{block}_\mathcal{H} = 0$ if it includes a normal transaction. We define the states as follows:

\begin{itemize}[leftmargin = *]
    \item State $S_{0,0}: (n_\mathcal{A}=0, n_\mathcal{H}=0, \texttt{whale}^\texttt{block}_\mathcal{H} = \bot)$. This state denotes that both the adversary and honest miners are mining on top of the same canonical chain.
    
    \item State $S_{0,1}: (n_\mathcal{A}=0, n_\mathcal{H}=1, \texttt{whale}^\texttt{block}_\mathcal{H} = 1)$. This state denotes that the honest fork contains a single block with a whale transaction.
    
    \item State $S_{i,i}: (n_\mathcal{A}=i, n_\mathcal{H}=i, \texttt{whale}^\texttt{block} = 1)$ for $i \ge 1$. This state denotes that there is a fork race between the adversarial and honest forks, both with a height of $i$ blocks. The first block in the honest fork includes a whale. The action \texttt{match} is not feasible in this state, as the $i^{\text{th}}$ adversarial block is mined after the $i^{\text{th}}$ honest block. Note that upon transferring to this state, a new transaction is not sampled, as the same height block already exists.

    \item State $S_{i,i+1}: (n_\mathcal{A}=i, n_\mathcal{H}=i+1, \texttt{whale}^\texttt{block} = 1)$ for $i \ge 1$. This state denotes that there is a fork race between the adversarial fork of height $i$ blocks with an honest fork of height $i+1$ blocks. The first block in the honest fork includes a whale. 
\end{itemize}

Strategy $\pi_2^\text{Non-predictable}$ is presented in Table~\ref{tab:strategy_nonpredictable_2}. Let $P_{0,0}$, $P_{0,1}$, $P_{i,i}$, $P_{i,i+1}$ denote the stationary probability of being in states $S_{0,0}$, $S_{0,1}$, $S_{i,i}$, $S_{i,i+1}$, respectively. We have:
\begin{equation}
    \begin{split}
        & P_{0,0} = \frac{1-(1-\alpha)(\alpha+p)}{1-(1-p)\alpha(1-\alpha)}, \quad
         P_{0,1} = \frac{(1-\alpha)p(1-\alpha(1-\alpha))}{1-(1-p)\alpha(1-\alpha)}, \quad \\
        & \sum_{i=1}^{\infty}{P_{i,i}} = P_{0,1} \frac{\alpha}{1-\alpha(1-\alpha)}, \quad
          \sum_{i=1}^{\infty}{P_{i,i+1}} = P_{0,1} \frac{\alpha(1-\alpha)}{1-\alpha(1-\alpha)}.
    \end{split}
\end{equation}

\begin{table*}[t!]
\centering
\caption{Strategy $\pi_2^\text{Non-predictable}$.}
\renewcommand{\arraystretch}{1.2}
\begin{tabular}{|c|c|c|c|c|}
\hline
\textbf{Initial State} & \textbf{Action} & \textbf{Probability} & \textbf{Final State} & \textbf{Reward} \\
\hline

\multirow{4}{*}{$S_{0,0}$} 
& \multirow{4}{*}{%
\begin{tabular}[c]{@{}l@{}}\texttt{wait}. \\If the next block is adversarial, then switch to \texttt{publish}.\\ If the next block is honest with a normal transaction, then \texttt{adopt}. \end{tabular}
}
& $\alpha p$ & $S_{0,0}$ & $1+F$ \\
\cline{3-5}
& & $\alpha(1-p)$ & $S_{0,0}$ & $1$ \\
\cline{3-5}
& & $(1-\alpha)p$ & $S_{0,1}$ & $0$ \\
\cline{3-5}
& & $(1-\alpha)(1-p)$ & $S_{0,0}$ & $0$ \\
\hline

\multirow{5}{*}{$S_{0,1}$} 
& \multirow{5}{*}{%
\begin{tabular}[c]{@{}l@{}}\texttt{wait}. \\ If the next block is honest with a normal transaction, then \texttt{adopt} \\ the whole honest fork.  \\ If the next block is honest with a whale transaction, then \texttt{adopt} \\ except the last honest block. \end{tabular}
}
& $\alpha$ & $S_{1,1}$ & $0$ \\
\cline{3-5}
& & $(1-\alpha)p$ & $S_{0,1}$ & $0$ \\
\cline{3-5}
& & $(1-\alpha)(1-p)$ & $S_{0,0}$ & $0$ \\
&& & & \\
&& & & \\
\hline

\multirow{2}{*}{$S_{i,i}$ $(i \geq 1)$} 
& \multirow{2}{*}{%
\begin{tabular}[c]{@{}l@{}}\texttt{wait}. \\ If the next block is adversarial, then \texttt{publish}.\end{tabular}
}
& $\alpha$ & $S_{0,0}$ & $1+F+i(1+pF)$ \\
\cline{3-5}
& & $(1-\alpha)$ & $S_{i,i+1}$ & $0$ \\
\hline

\multirow{5}{*}{$S_{i,i+1}$ $(i \geq 1)$} 
& \multirow{5}{*}{%
\begin{tabular}[c]{@{}l@{}}\texttt{wait}. \\ If the next block is honest with a normal transaction, then \texttt{adopt} \\ the whole honest fork. \\ If the next block is honest with a whale transaction, then \texttt{adopt} \\except the last honest block.\end{tabular}
}
& $\alpha$ & $S_{i+1,i+1}$ & $0$ \\
\cline{3-5}
& & $(1-\alpha)p$ & $S_{0,1}$ & $0$ \\
\cline{3-5}
& & $(1-\alpha)(1-p)$ & $S_{0,0}$ & $0$ \\
&& & & \\
&& & & \\
\hline

\end{tabular}
\label{tab:strategy_nonpredictable_2}
\end{table*}

The normalized numbers of canonical honest blocks, canonical adversarial blocks, and the normalized  adversarial block reward achieved under strategy $\pi_2^\text{Non-predictable}$ can be obtained as follows:

\begin{equation}
    \begin{split}
        N_{\mathcal{H}} &= P_{0,0} (1 - \alpha)(1 - p) +
           P_{0,1} (1 - \alpha) p +
           2 P_{0,1} (1 - \alpha)(1 - p) \\
           &\quad + \frac{P_{0,1} \alpha (1 - \alpha)^2}{(1 - \alpha (1 - \alpha))^2} +
           \frac{P_{0,1} (2 - p) \alpha (1 - \alpha)^2}{1 - \alpha (1 - \alpha)} \enspace, \\
        N_{\mathcal{A}} &= \left(\frac{P_{0,1} \alpha^2}{1 - \alpha (1 - \alpha)}\right)
            \left(
            \frac{1}{1 - \alpha (1 - \alpha)} + 1
            \right) + P_{0,0} \alpha \enspace , \\
        R_{\mathcal{A}} &= \left(\frac{P_{0,1} \alpha^2}{1 - \alpha (1 - \alpha)}\right)
            \left(\frac{(1 + pF)}{1 - \alpha (1 - \alpha)} +(1 + F) \right) 
            + P_{0,0} \alpha (1 + pF) \enspace .
\end{split}
\end{equation}

\section{Simplified Mempool Environment: Supplementary} \label{appendix:simplified_memppol}
Deriving a time-fee equation compatible with the Bitcoin environment is challenging, as transaction fees vary across different periods due to various reasons such as fluctuations in Bitcoin's supply and demand dynamics.
This variability indicates that the optimal mining strategy for an adversary can differ significantly across different periods. To maximize its profit, the adversary must accurately predict the time-fee equation for the targeted attack period.

\begin{figure}[b]
     \centering
     \begin{subfigure}{0.235\textwidth}
         \centering
         \includegraphics[width=\textwidth]{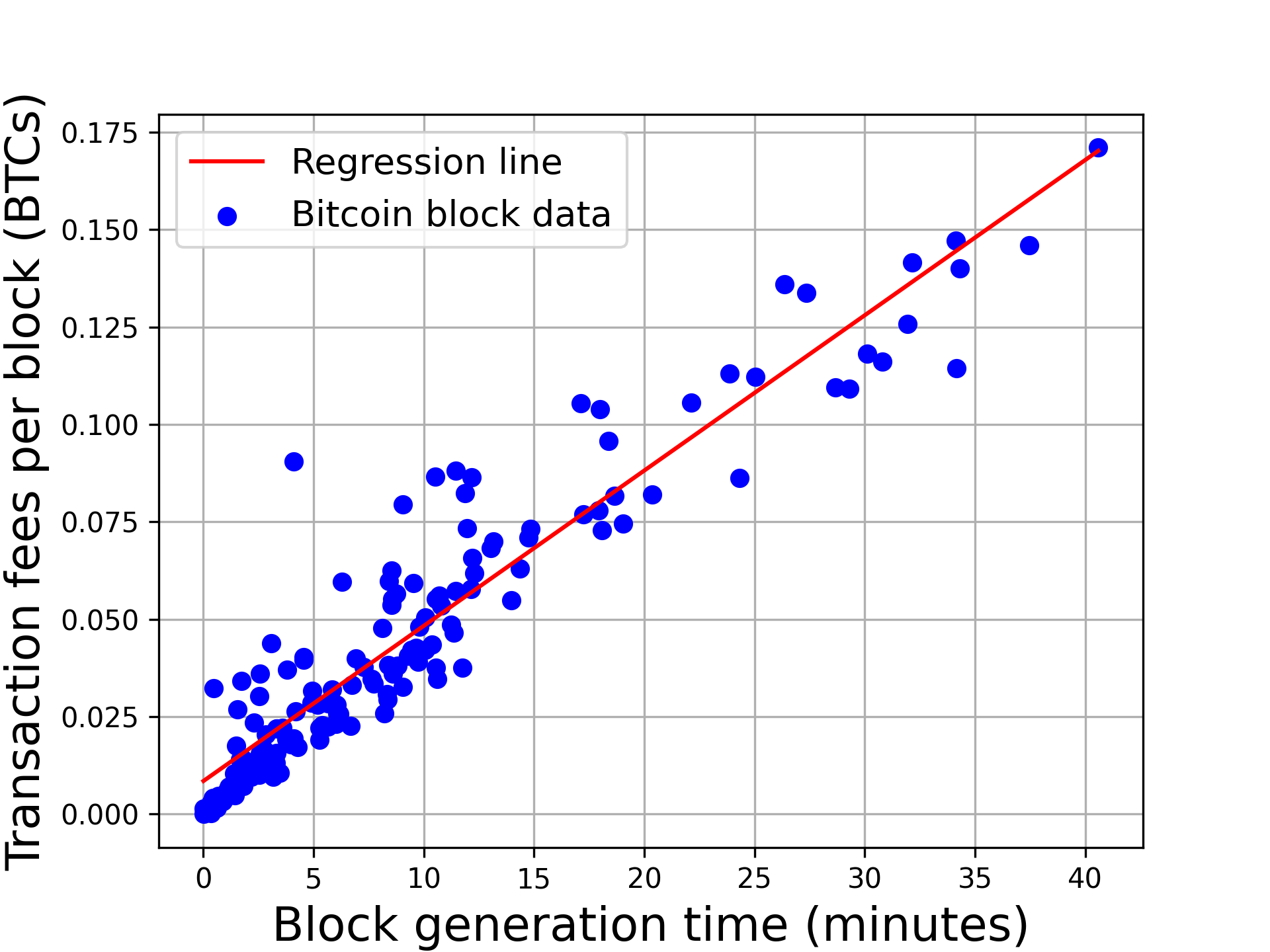}
         \caption{January 1, 2023}
         \label{fig:fee-time-one_dimension-linear-low_fee}
     \end{subfigure}
     \hfill
     \begin{subfigure}{0.235\textwidth}
         \centering
         \includegraphics[width=\textwidth]{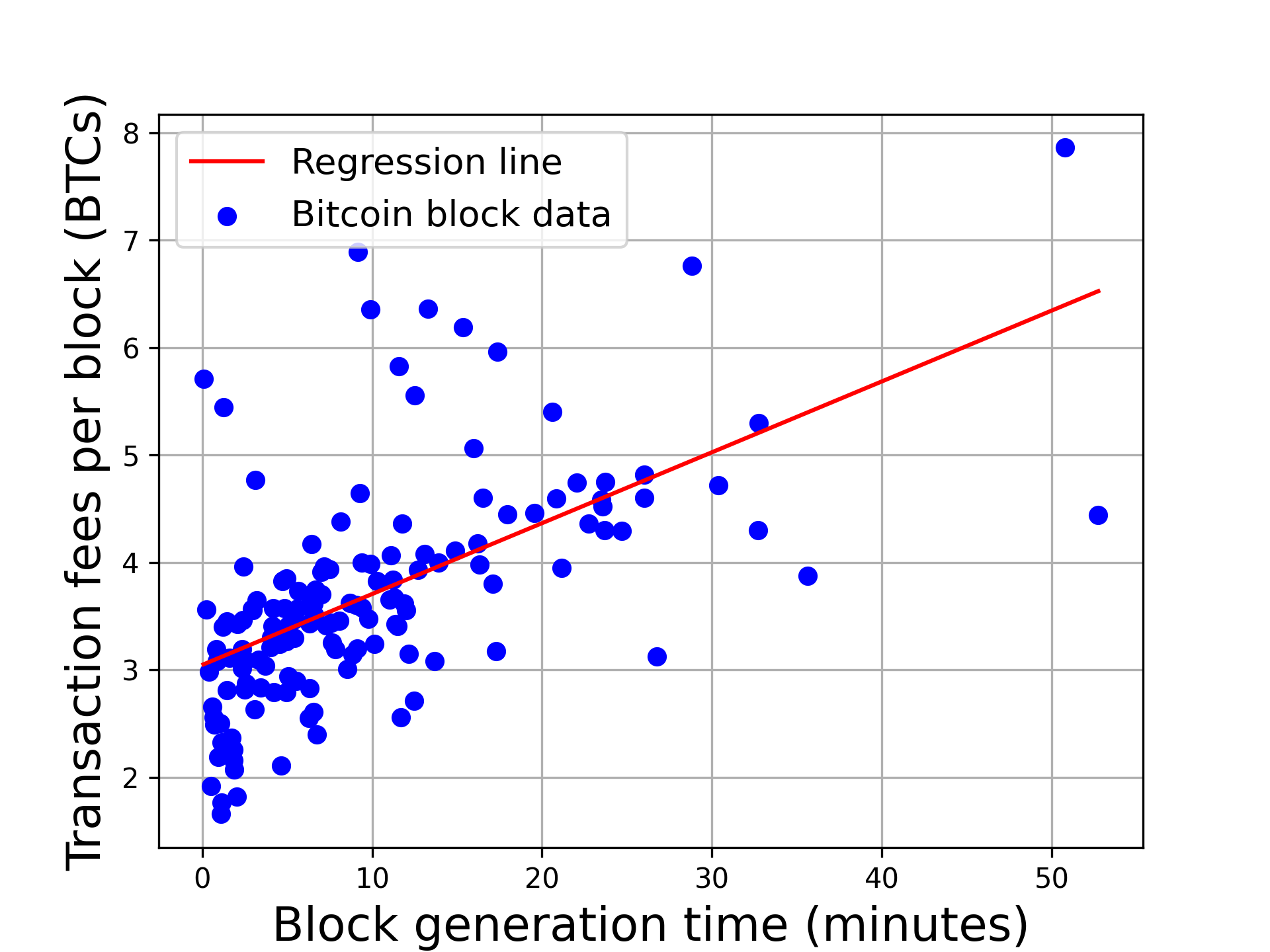}
         \caption{December 17, 2023}
         \label{fig:fee-time-one_dimension-linear-high_fee}
     \end{subfigure}
        \caption{Linear time-fee equation.}
        \label{fig:fee-time-one_dimension-linear}
\end{figure}

For the simplified mempool implementation, we extracted historical data~\cite{Blockchair} on total transaction fees and block generation times for Bitcoin blocks mined across different periods. Using regression techniques, we derived the time-fee equation for the target period. This time-fee equation enables us to estimate the reward function and analyze the profitability of mining strategies.
As an example, in Figure~\ref{fig:fee-time-one_dimension-linear}, the total transaction fee of a block in BTC, denoted by $\texttt{fee}$, is depicted as a function of its generation time in minutes, denoted by $t$, for two different periods. Each point represents a block, and the line represents the time-fee equation. Figure~\ref{fig:fee-time-one_dimension-linear-low_fee} corresponds to Bitcoin blocks mined on January 1, 2023, with relatively low transaction fees per block, while Figure~\ref{fig:fee-time-one_dimension-linear-high_fee} corresponds to blocks mined on December 17, 2023, during a period of high transaction fees.

As already mentioned in~\ref{sec:simplified_memppol}, a linear time-fee equation can be expressed as $\texttt{fee}(t) = \texttt{fee}_0 + r_{\texttt{fee}} t$, where $\texttt{fee}_0$ and $r_{\texttt{fee}}$ are base fee and fee increase rate, respectively. At $t=0$, the transaction reward of the next potential block is equal to the base fee. This reflects reality, as there are always some low-value transactions in the mempool that miners can use to fill their blocks. As block generation time increases, the collected transaction fee rises by the fee increase rate for each additional unit of time due to the arrival of new transactions.

We assume that the time-fee equation is a linear function, and time is divided into $M \ge 2$ discrete steps of the form $\{ t_i = i\Delta \mid i \in [M] \}$, $\Delta$ denotes the length of each time step. For a block with block generation time $t$, the transaction fee reward can be obtained as follows:
\begin{equation}
    \texttt{fee}(t)= 
    \begin{cases}
        \texttt{fee}(i\Delta), & \text{if } \enspace i\Delta \le t < (i+1)\Delta \enspace \text{for} \enspace i \in [M-1] \enspace\\
        \texttt{fee}((M-1)\Delta),              & \text{if } \enspace (M-1)\Delta \le t
    \end{cases}
\end{equation}
In our environment, block generation time follows the exponential distribution with rate $\lambda$.

\section{Limitations of the Simplified Environment}\label{appendix: limitations_simplified_env}
\begin{figure}[b]
     \centering
     \begin{subfigure}{0.23\textwidth}
         \centering
         \includegraphics[width=\textwidth]{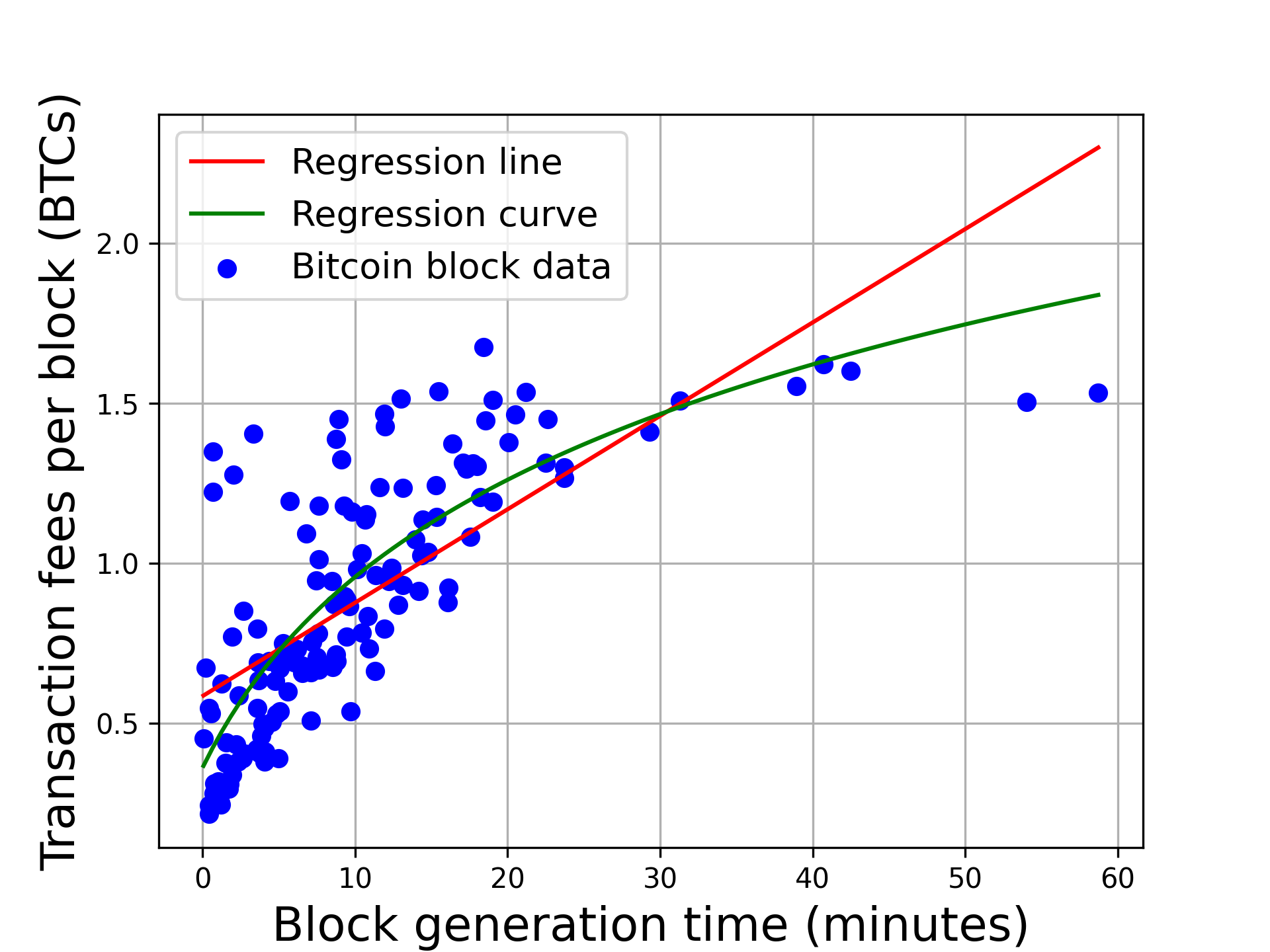}
         \caption{May 3, 2023}
         \label{fig:fee-time-one_dimension-polynomial}
     \end{subfigure}
     \hfill
     \begin{subfigure}{0.23\textwidth}
         \centering
         \includegraphics[width=\textwidth]{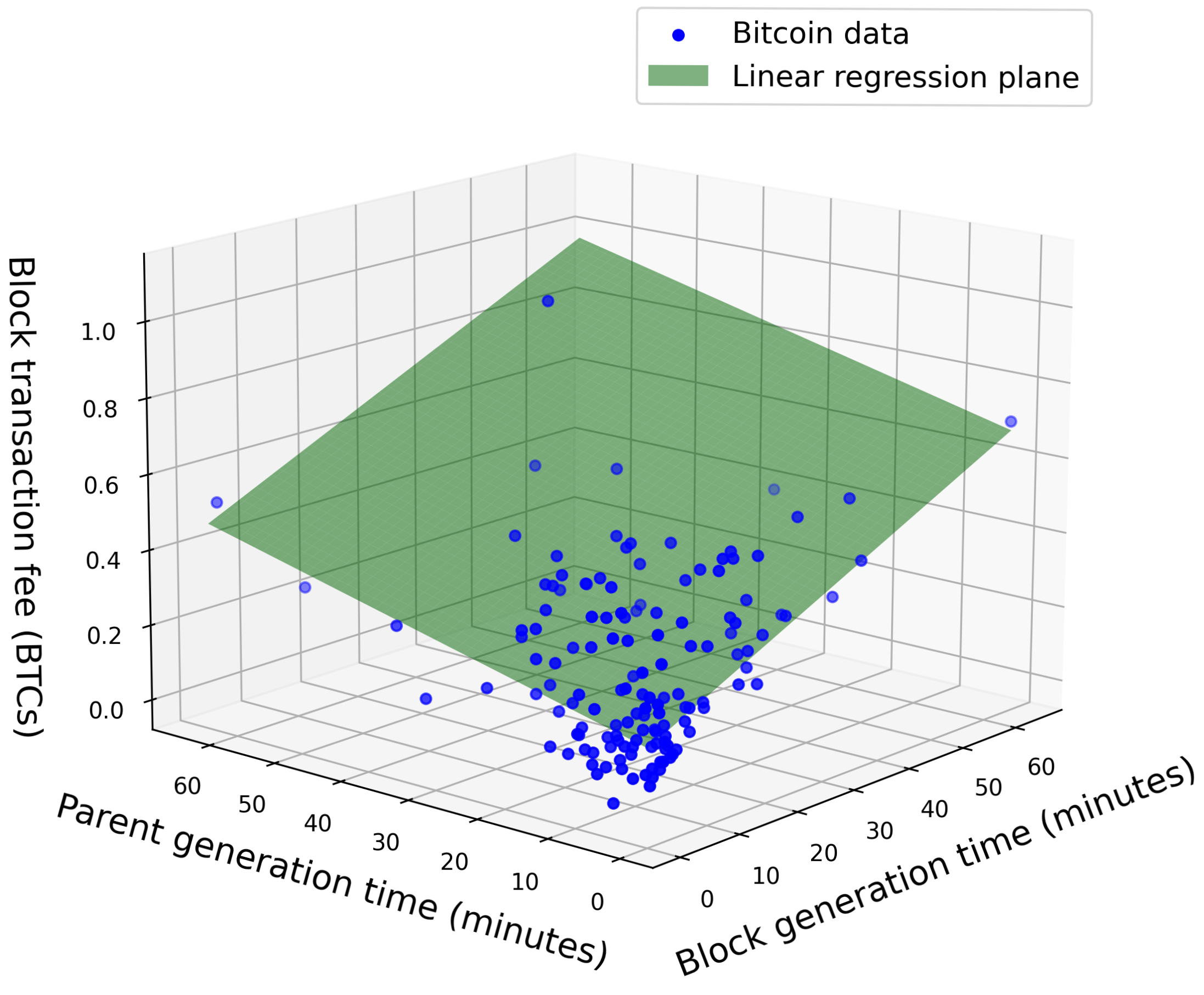}
         \caption{March 18, 2024}
         \label{fig:fee-time-two_dimension-linear}
     \end{subfigure}
        \caption{Time-fee equation.}
        \label{fig:fee-time-advanced}
\end{figure}

\noindent\textbf{Non-linearity of the time-fee equation:} In Figure~\ref{fig:fee-time-one_dimension-linear} presented in Appendix~\ref{appendix:simplified_memppol}, a linear regression model is used to represent the time-fee equation. However, the block transaction fee does not necessarily increase linearly with the block generation time. This occurs when the rate of fee increase is not constant and decreases as the block generation time increases. In such cases, using alternative regression techniques rather than a linear model can provide a time-fee equation that better fits the behavior observed in the Bitcoin mempool. 
In Figure~\ref{fig:fee-time-one_dimension-polynomial}, we compare linear regression with a curve regression model given by $f(t) = 0.6414 \; \texttt{ln}(t+6.5209)-0.8419$ to derive the time-fee equation for Bitcoin blocks mined on May 3, 2023. The coefficient of determination (R-squared) for the curve regression is $0.6243$, while for the linear regression it is $0.5180$. This indicates that the curve regression provides a better fit compared to the linear regression, suggesting that the fee-increase rate diminishes with longer block generation times.

\noindent\textbf{Dependence of fee rewards on the generation times of preceding blocks:} Another technique that can lead to a more realistic mempool implementation is using multivariate regression instead of univariate regression. In univariate regression, the block transaction fee is estimated based solely on a single predictor, namely the block generation time. In contrast, multivariate regression allows us to account for additional predictors, such as the generation time of the parent block, in determining the block transaction fee.
In Figure~\ref{fig:fee-time-two_dimension-linear}, we depict the regression plane for the block transaction fee on March 18, 2024, based on two predictors: the block generation time and the parent block generation time. The resulting time-fee equation is $f(t_1,t_2) = 0.1697 + 0.0079 * t_1 +  0.0046 * t_2$, where \(t_1\) and \(t_2\) represent the block generation time and the parent block generation time, respectively.
Note that if the parent block generation time is longer, more fee-valuable transactions will likely remain in the mempool after the parent block is mined, allowing the next block to collect these transactions. As the time-fee equation shows, the fee increase rate with respect to the block generation time is higher than the fee increase rate with respect to the parent block generation time. This implies that the block generation time is a more effective predictor of the transaction fee compared to the parent block generation time.

\section{A3C Implementation} \label{appendix:A3C_implementation}
\subsection{Objective Function} \label{appendix:objective_function}

The training process aims to enable agents to identify which action maximizes the objective function at a given state $s$. The objective function is defined in terms of the parameters that the agent, the adversary in our case, seeks to maximize. In our implementation, this objective is to maximize the mining time-averaged profit. We discuss how the objective function can be defined both before and after the difficulty adjustment. In our implementation, at each time step $t$, a block is generated either by the adversary or by the honest miners. However, depending on the adversarial strategy, the block generated at time step $t$ may not immediately be added to the canonical chain. This block could be added to the canonical chain or even be identified as orphaned at a future time step. Let $N_\mathcal{A}(t; \pi)$, $N_\mathcal{H}(t; \pi)$, and $R_\mathcal{A}(t; \pi)$ denote the number of adversarial blocks, the number of honest blocks, and the total reward included in the adversarial blocks, respectively, added to the canonical chain at time step $t$ under policy $\pi$. Also, let $N_\mathcal{T}(t; \pi)$ denote the total number of blocks both added to and removed (orphaned) from the canonical chain at time step $t$ under policy $\pi$. For each time step $t$, we have $N_\mathcal{T}(t; \pi) \ge N_\mathcal{A}(t; \pi) + N_\mathcal{H}(t; \pi)$. After the state transition at time step $t$, the environment returns the next state along with the following information to the agent: $\big(R_\mathcal{A}(t; \pi), N_\mathcal{A}(t; \pi) + N_\mathcal{H}(t; \pi), N_\mathcal{T}(t; \pi) \big)$.

We denote by $\mathbf{t}^B$ the average block generation time. We normalize the average block generation time achieved while all the miners including the adversary mine honestly to $\mathbf{t}^B_\texttt{ideal} = 1$. Let us ignore the mining cost since, as long as the adversary does not turn off some of its miners, its mining cost is not affected by the chosen strategy. Therefore, the time-averaged profit can be defined as follows:
\begin{equation} \label{eq: mining_time_averaged_profit}
    \texttt{Profit}^\mathcal{A}(\pi) = \lim_{T\to\infty} {\frac{\sum_{t=0}^{T}{R_\mathcal{A}(t; \pi)}}{\mathbf{t}^B \cdot T}} \enspace.
\end{equation}
Depending on the mining difficulty and the adversary's strategy, the average block generation time $\mathbf{t}^B$ may vary, leading to different objective functions before and after the difficulty adjustment.

\subsubsection{Before the difficulty adjustment} \label{sec: objective_function_before_DAM}
As discussed in Section~\ref{sec:non_profit_before_DAM}, before a difficulty adjustment, the adversary's mining strategy does not affect the average block generation time, and thus, $\mathbf{t}^B = 1$. Note that while the canonical block generation time may differ, the average duration of each time step in our environment before a difficulty adjustment is determined by the average of both canonical and non-canonical block generation times, which remains unchanged.
Therefore, the objective function before the difficulty adjustment can be defined as follows:
\begin{equation}
    O^{\mathcal{A}}(\pi) = \lim_{T\to\infty} {\frac{\sum_{t=0}^{T}{R_\mathcal{A}(t; \pi)}}{T}} \enspace.
\end{equation}
To find the policy that maximizes the optimization function above, the reward that agents receive after a transition at time step $t$, referred to as the \emph{time step reward} and denoted by $r_t$, is set to $R_\mathcal{A}(t; \pi)$. During our implementation, it was found that including the mining cost in the reward function (as a constant negative reward added after each time step to account for the passage of time) and normalizing the rewards helps the agent converge faster and in a more stable manner. Our implementation uses the following step reward function:
\begin{equation}
    r_t = \frac{R_\mathcal{A}(t; \pi)}{R^\texttt{norm}} - \texttt{cost} \enspace,
\end{equation}
where $R^\text{norm}$ denotes the total distributed reward on average per $\mathbf{t}^B$ units of time if all miners mine honestly, and $\texttt{cost}$ denotes the constant cost per time step.


\subsubsection{After the difficulty adjustment}
After the difficulty adjustment, the mining difficulty will be modified, resulting in a decrease in the block generation time. The average block generation time after difficulty adjustment and up to including time step $T$ can be obtained as follows:
\begin{equation} \label{eq: block generation time}
    \begin{split}
       & \mathbf{t}^B = \frac{\sum_{t=0}^{T}{N_\mathcal{A}(t; \pi) + N_\mathcal{H}(t; \pi)}}{\sum_{t=0}^{T}{ N_\mathcal{T}(t; \pi)}} \cdot \mathbf{t}^B_\texttt{ideal} = \\
       &\frac{\sum_{t=0}^{T}{N_\mathcal{A}(t; \pi) + N_\mathcal{H}(t; \pi)}}{T} \enspace.
    \end{split}
\end{equation}
The second equality in the equation above is achieved as $\mathbf{t}^B_\texttt{ideal} = 1$ and $\sum_{t=0}^{T}{ N_\mathcal{T}(t)} = T$. Based on the value of $\mathbf{t}^B$ and the time-averaged profit formula outlined in Equation~\ref{eq: mining_time_averaged_profit}, we define the objective function after difficulty adjustment as follows:
\begin{equation} \label{eq: mining_time_averaged_profit_after_adjustment}
    O^\mathcal{A}(\pi) = \lim_{T\to\infty} {\frac{\sum_{t=0}^{T}{R_\mathcal{A}(t; \pi)}}{\sum_{t=0}^{T}{N_\mathcal{A}(t; \pi) + N_\mathcal{H}(t; \pi)}}} \enspace.
\end{equation}
Since the objective function defined in equation~\ref{eq: mining_time_averaged_profit_after_adjustment} is non-linear, we cannot naively assign the time step reward as is done in Section~\ref{sec: objective_function_before_DAM}. To address this challenge, we adopt the approach introduced in~\cite{sapirshtein2017optimal}, which converts the non-linear optimization function into a linear one. We set the time step reward as follows: 
\begin{equation} \label{eq: time_step_reward_after_DAM}
    r_t(\rho) = R_\mathcal{A}(t; \pi) - \rho \big(N_\mathcal{A}(t; \pi) + N_\mathcal{H}(t; \pi)\big) \enspace,
\end{equation}
where $\rho$ represents the time-averaged profit the adversary earns by following the optimal strategy, i.e., $\rho = O^\mathcal{A}(\pi^*)$, where $\pi^*$ denotes the optimal strategy. At the start of training, the actual value of $\rho$ is unknown. Therefore, the training agents initialize with $\rho = O^\mathcal{A}(\pi^\mathcal{H})$, where $\pi^\mathcal{H}$ denotes the honest strategy. Certain test agents are tasked with measuring the time-averaged profit under the policy trained by the training agents at specific time steps during the training process. If this profit exceeds the highest time-averaged profit achieved so far, they prompt the training agents to update $\rho$ to the new highest time-averaged profit. For the detailed analysis of the non-linear objective function conversion, readers are referred to~\cite{sapirshtein2017optimal}. We briefly provide the intuition behind setting the time step reward as defined in equation~\ref{eq: time_step_reward_after_DAM} in the following.
The value function of the initializing state $s_0$ under the time step reward introduced in equation~\ref{eq: time_step_reward_after_DAM} can be obtained as follows:
\begin{equation}
    V^\mathcal{A}(s_0; \pi, \rho) = \sum_{t=0}^{\infty}{\gamma^t R_\mathcal{A}(t; \pi)} - \rho \sum_{t=0}^{\infty}{\gamma^t \big(N_\mathcal{A}(t; \pi) + N_\mathcal{H}(t; \pi)\big)} \enspace.
\end{equation}
Let $\rho_1$ be the time-averaged profit achieved under policy $\pi_1$, i.e., $\rho_1 = O^\mathcal{A}(\pi_1)$.
If, for a given policy $\pi_2$ and as $\gamma \rightarrow 1$, we have $V^\mathcal{A}(s_0; \pi_2, \rho_1) > 0$, then we can conclude $\pi_2$ dominates $\pi_1$. This is because $V^\mathcal{A}(s_0; \pi_2, \rho_1) > 0$ implies that:
$$\frac{\sum_{t=0}^{\infty}{\gamma^t R_\mathcal{A}(t; \pi_2)}}{\sum_{t=0}^{\infty}{\gamma^t \big(N_\mathcal{A}(t; \pi_2) + N_\mathcal{H}(t; \pi_2)\big)}} > \rho_1 \enspace .$$
According to equation~\ref{eq: mining_time_averaged_profit_after_adjustment} and considering $\gamma \rightarrow 1$, the inequality above results in $O^\mathcal{A}(\pi_2) > O^\mathcal{A}(\pi_1)$, which indicates that following strategy $\pi_2$ is more profitable than $\pi_1$. Therefore, if the training agents under step rewards \(r_t(\rho_1)\) are properly trained, the resulting policy will achieve a time-averaged profit \(\rho_2\), where \(\rho_2 \geq \rho_1\). If, in the next step, the agents are trained based on the step reward \(r_t(\rho_2)\), the time-averaged profit of the resulting policy will be \(\rho_3\) with \(\rho_3 \geq \rho_2\). By continuing this process, the trained policy will gradually converge to the optimal policy that yields the highest time-averaged profit.

Another critical point is that the average time step duration after difficulty adjustment needs recalibration in the environment. In our implementation, the duration of each time step follows an exponential distribution with rate $\frac{1}{t^B}$. The duration of each time step is of great importance, as it affects the amount by which transaction weight increases in the mempool.
Before the difficulty adjustment, $t^B$ is set to 10 minutes. However, after the difficulty adjustment, the value of block generation time $t^B$ needs to be adjusted according to the policy followed by the adversary. To calculate the block generation time after difficulty adjustment under policy $\pi$, we need to keep track of the ratio of the number of canonical blocks to the total number of blocks mined under policy $\pi$. To this end, the neural network outputs another value that aims to estimate the following sum:
\begin{equation}
    \sum_{t=0}^{T} \gamma^t \big(N_\mathcal{A}(t; \pi) + N_\mathcal{H}(t; \pi)\big) \enspace.
\end{equation}
The neural network is trained to estimate this sum based on the values of $N_\mathcal{A}(t; \pi)$ and $N_\mathcal{H}(t; \pi)$ at each time step. With access to the sum of the number of canonical blocks and equation~\ref{eq: block generation time}, the environment can calculate the block generation time. Note that the policy is only trained to maximize the optimization function defined in equation~\ref{eq: mining_time_averaged_profit_after_adjustment}, and this sum is not used in training the policy. 

\subsection{Implementation Details} \label{appendix: implementation}
Our A3C-based tool is adapted from the code provided in \cite{pytorchaaac}. Our implementation consists of two linear layers, each with 256 nodes, followed by an LSTM layer that processes the output from the last linear layer. The model uses a learning rate of $10^{-6}$, a discount factor of $0.99$, and an entropy coefficient of $0.01$. We employ 30 training agents to train the neural network and 2 testing agents to evaluate the results. During the training process, information related to the three best checkpoints is stored, and results are presented after running those checkpoints for $10^6$ steps.

\subsection{Implementation of the Undercutting Attack} \label{appendix:undercut_implementation}
To analyze the undercutting attack, the possible action set for the A3C agents is limited to the following actions:

\begin{itemize} [leftmargin=*]
    \item If the adversary mines a new block on top of the longest chain, it publishes the block for the other miners. 
    \item Whenever the tip of the longest chain, denoted by block $B_1$, is not an adversarial block, the adversary attempts to undercut block $B_1$. For a period of $t$ seconds, determined by the training process, the adversary mines on the parent of block $B_1$. If no block is mined during this time, the adversary abandons the undercutting attempt and resumes mining on top of block $B_1$. 
    \item If, during the undercutting of block $B_1$, the next block $B_2$ is mined by honest miners on top of $B_1$, the adversary stops undercutting block $B_1$ and begins undercutting block $B_2$. 
    \item If, while undercutting block $B_1$, the adversary successfully mines a competing block $B_\mathcal{A}$, it publishes block $B_\mathcal{A}$. If the next block $B_2$ is mined on top of $B_1$, the undercutting attempt fails, and the adversary tries to undercut block $B_2$. If block $B_2$ is mined on top of block $B_\mathcal{A}$, the attack succeeds. 
\end{itemize}
Note that this set of actions does not result in obtaining the optimal strategy and is only designed to measure the profitability of single-block undercutting under different mempool patterns.

\section{Selfish Mining After Difficulty Adjustment} \label{appendix: selfish mining_after_DAM}
\begin{figure}[b]
    \centering
    \includegraphics[height=2in]{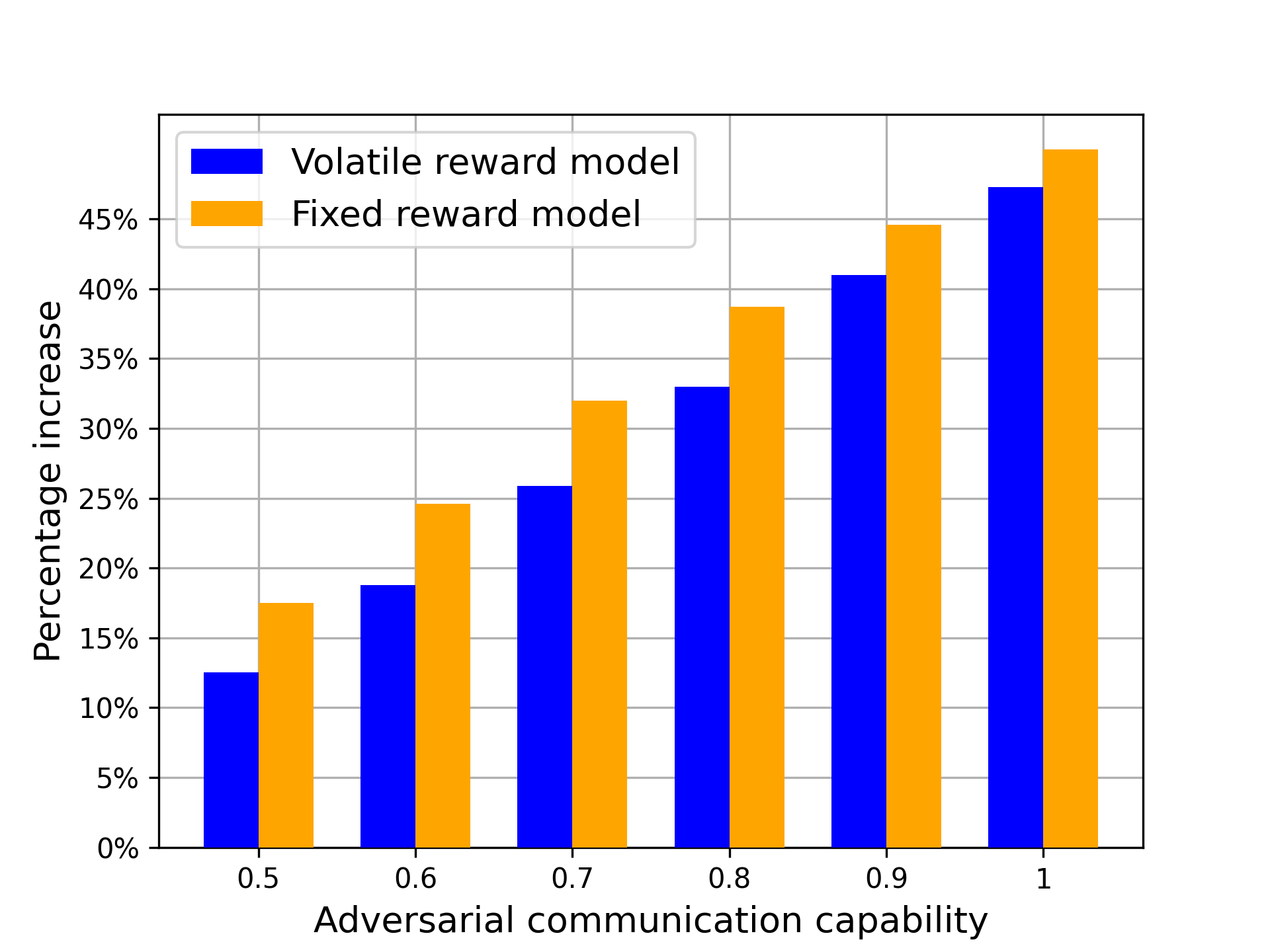}
    \caption{The profit percentage increase of selfish mining after difficulty adjustment.}
    \label{fig:percentage_increase_selfish_after_DAM}
\end{figure}

\begin{figure}[t]
    \centering
    \includegraphics[height=2in]{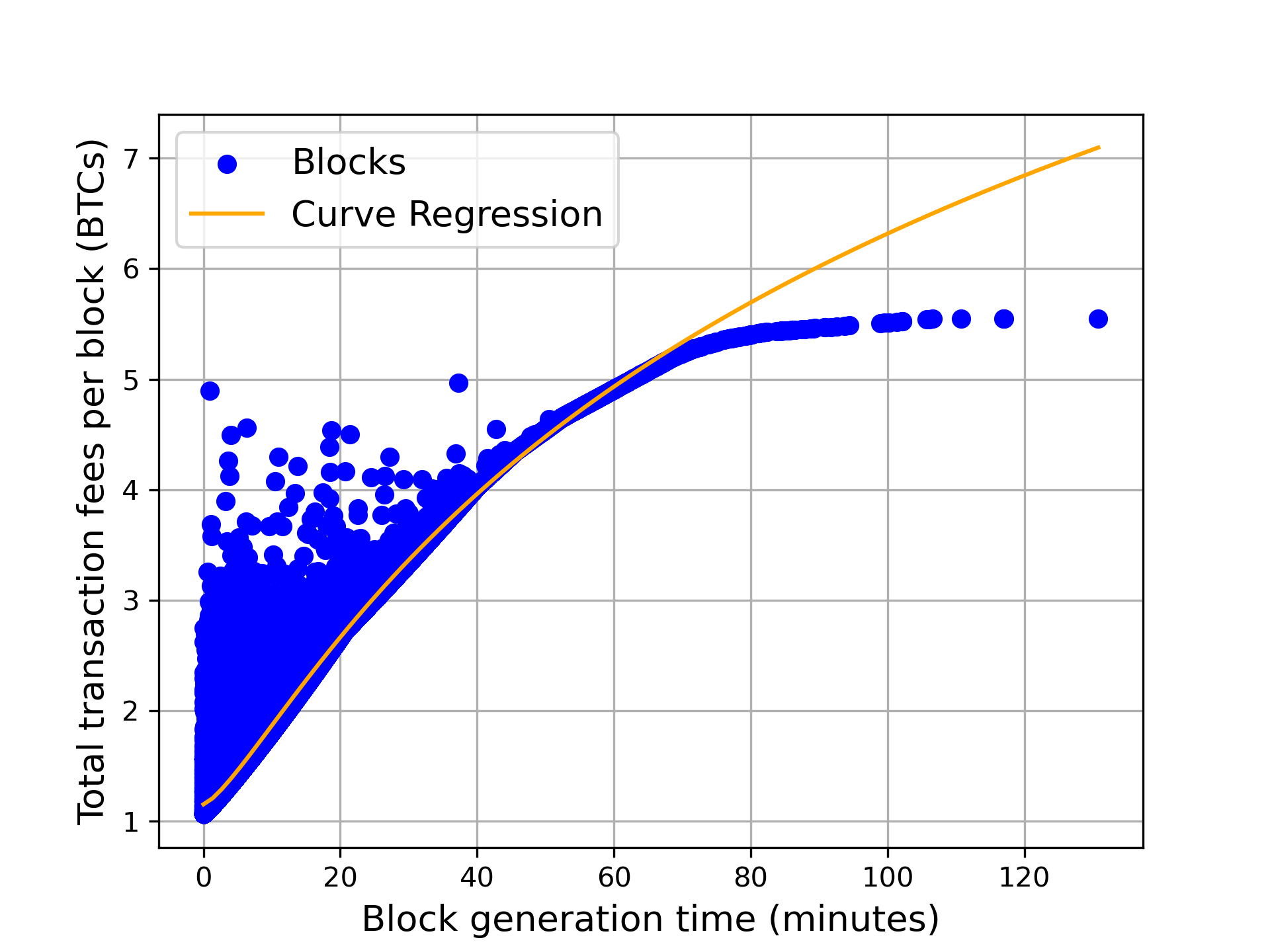}
    \caption{The concave time-fee function.}
    \label{fig:fee_time_concave}
\end{figure}

In Figure~\ref{fig:percentage_increase_selfish_after_DAM}, the percentage increase in time-averaged profit of selfish mining after a difficulty adjustment is depicted as a function of the adversarial communication capability. This figure is depicted based on the following assumptions:  
i) the protocol reward is set to $0$ BTC,  
ii) the mempool pattern observed on December 18, 2023, between 23:00 and 24:00 CET, has occurred,  
iii) the adversary's mining share is equal to $\frac{1}{3}$.  
As shown in Figure~\ref{fig:percentage_increase_selfish_after_DAM}, under the mempool pattern observed on December 18, 2023, difficulty-adjusted selfish mining is more profitable in the fixed reward model compared to the volatile reward model. This suggests that, under selfish mining, the rate of increase in collected transaction fees is lower than the rate of increase in the adversarial block ratio. One of the reasons contributing to the higher profitability of selfish mining in the fixed reward model is the concavity of the time-fee function during this period. Figure~\ref{fig:fee_time_concave} illustrates the time-fee function obtained under the mempool pattern from December 18, 2023, between 23:00 and 24:00 CET. To generate Figure~\ref{fig:fee_time_concave}, we simulated 100,000 blocks under this mempool pattern, collecting their generation times and fees. As depicted in Figure~\ref{fig:fee_time_concave}, the corresponding time-fee function for this period is concave. 

To better understand the effect of the concavity of the time-fee function on selfish mining profitability, we review the following theorem.

\begin{theorem}\label{theorem:selfish_after_DAM}
    If the time-fee function is strictly concave, the average fee of a canonical block under honest mining will be greater than the average fee of a canonical block under difficulty-adjusted selfish mining.
\end{theorem}

\begin{proof}
    Let \( f(t) \) denote the fee of a block with a block generation time of \( t \), where \( f(t) \) is strictly concave. Let \( \lambda \) and \( \lambda' \) denote the total block generation rates under honest mining and difficulty-adjusted selfish mining, respectively. Note that the block generation time follows an exponential distribution.
    The canonical block generation rate under honest mining is also equal to $\lambda$. Therefore, the average canonical block fee under honest mining, denoted by $r_\mathcal{H}$, can be obtained as follows: 
    \begin{equation}
        r_\mathcal{H} = \int_{0}^{\infty} f(t) \lambda e^{-\lambda t} \, dt \enspace .
    \end{equation}
     We divide canonical blocks mined under difficulty-adjusted selfish mining into groups $\{G_1, G_2, G_3, \cdots\}$. A canonical block $B$ belongs to group $G_i$ for $i \in \{1,2, 3, \cdots\}$ if the number of orphaned blocks mined between the parent of block $B$ and block $B$ itself is $i-1$. The average canonical block generation rate of blocks belonging to group $G_i$ is equal to $\frac{\lambda'}{i}$. Let $a_i$ denote the ratio of the number of blocks belonging to group $G_i$ to the total number of canonical blocks, where $\sum_{i=1}^{\infty}{a_i} = 1$. As the average canonical block generation rate under difficulty-adjusted selfish mining is also equal to $\lambda$, we have:
     \begin{equation} \label{eq:lambda2}
         \frac{1}{\lambda} = \sum_{i=1}^{\infty}{a_i \int_{0}^{\infty} t' \frac{\lambda'}{i} e^{-\frac{\lambda'}{i} t'} \, dt'}  \enspace .
     \end{equation}
     By calculating the result of integral, we have:
     \begin{equation} \label{eq:lambda_lambda'}
         \frac{1}{\lambda} = \frac{\sum_{i=1}^{\infty}{a_i i}}{\lambda'} \Longrightarrow \frac{\lambda'}{\lambda} = \sum_{i=1}^{\infty}{a_i i} \enspace .
     \end{equation}
     The average fee of a canonical block under difficulty-adjusted selfish mining, denoted by $r_\mathcal{S}$, can be obtained as follows:
     \begin{equation}
        r_\mathcal{S} = \sum_{i=1}^{\infty}{a_i \int_{0}^{\infty} f(t') \frac{\lambda'}{i} e^{-\frac{\lambda'}{i} t'} \, dt'} \enspace .
    \end{equation}
    By making a variable substitution of form $\lambda t = \frac{\lambda'}{i} t'$, we obtain:
    \begin{equation}
        r_\mathcal{S} = \sum_{i=1}^{\infty}{a_i \int_{0}^{\infty} f(\frac{\lambda}{\lambda'} it) \lambda e^{-\lambda t} \, dt} \enspace .
    \end{equation}
    As function $f(t)$ is strictly concave, we have:
    \begin{equation}
        r_\mathcal{S} < \int_{0}^{\infty} f(\frac{\lambda}{\lambda'}\big(\sum_{i=1}^{\infty}{a_i i}\big) t) \lambda e^{-\lambda t} \, dt\enspace .
    \end{equation}
    Using the result obtained in equation~\ref{eq:lambda_lambda'}, we can write the inequality above as follows:
    \begin{equation}
        r_\mathcal{S} < \int_{0}^{\infty} f(t) \lambda e^{-\lambda t} \, dt\enspace .
    \end{equation}
    The right hand side in inequality above is equal to $r_\mathcal{H}$. Therefore, we obtain $r_\mathcal{S} < r_\mathcal{H}$.
\end{proof}
According to Theorem~\ref{theorem:selfish_after_DAM}, although the average canonical block generation rate is the same under both honest mining and difficulty-adjusted selfish mining, the total transaction fees distributed among canonical blocks under difficulty-adjusted selfish mining decrease due to the concavity of the time-fee function. This results in a lower profit percentage increase in the volatile model compared to honest mining.

\section{Selfish Mining Before Difficulty Adjustment: Supplementary} \label{appendix: selfish mining_before_DAM}
Figure~\ref{fig:percentage_increase_reward_3_18_december} shows the percentage increase in the adversary's time-averaged profit before a difficulty adjustment resulting from conducting selfish mining. This figure is depicted based on different ratios of petty-compliant miners, over the mempool patterns observed on two different days: January 1, 2024, and March 1, 2024, both between 8:00 and 9:00 AM CET. The figure is generated using the following configuration parameters:
i) the protocol reward is set to $0$ BTC, and
ii) the adversary's mining share and communication capability are set to $\frac{1}{3}$ and $50\%$, respectively.
 
\begin{figure}[b]
    \centering
    \includegraphics[height=2.5in]{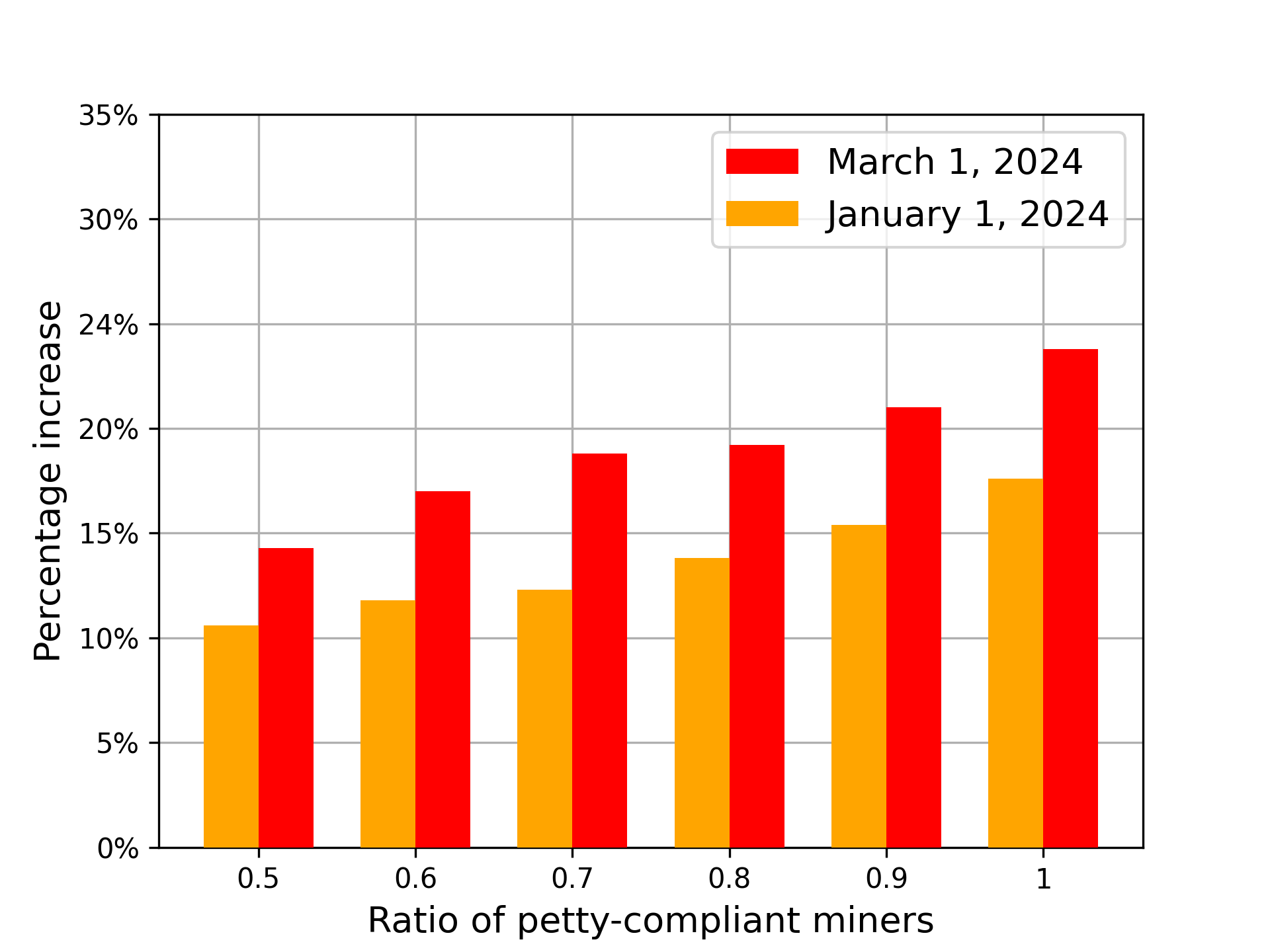}
    \caption{The percentage increase in time-averaged profit.}
    \label{fig:percentage_increase_reward_january_1}
\end{figure} 
The greater percentage increase observed under the mempool pattern of March 1, 2024, can be attributed to a higher normalized rate of growth in fee rewards during this period. Specifically, the ratio of the average fee reward for a block generated in 20 minutes to that for a block generated in 10 minutes is 1.29 under the mempool pattern of January 1, 2024, and 1.4 under that of March 1, 2024. This indicates that if the adversary succeeds in orphaning a preceding block and thereby extends the block generation time for its own block, it can accumulate a relatively higher fee reward under the mempool pattern of March 1, 2024.

\end{document}